\pdfoutput=1
\documentclass[11pt]{article}
\usepackage[margin=1in]{geometry}
\usepackage{amsmath,amssymb,amsthm}
\usepackage{xcolor}
\usepackage{array}
\usepackage{graphicx}
\usepackage{float}
\usepackage{placeins} %
\usepackage{tikz}
\usetikzlibrary{arrows.meta,fit,positioning}
\usepackage{pgfplots}
\pgfplotsset{compat=1.18}
\usepgfplotslibrary{fillbetween}

\usepackage[numbers,sort]{natbib}
\usepackage{hyperref}

\usepackage{pifont}
\newcommand{\cmark}{\ding{51}}
\newcommand{\xmark}{\ding{55}}

\newcommand{\R}{\mathbb{R}}
\newcommand{\Hist}{\mathcal{H}}
\newcommand{\Id}{\mathcal{I}}

\theoremstyle{plain}
\newtheorem{theorem}{Theorem}
\newtheorem{proposition}{Proposition}
\newtheorem{corollary}[proposition]{Corollary}
\newtheorem{lemma}[proposition]{Lemma}

\theoremstyle{definition}
\newtheorem{definition}{Definition}
\newtheorem{assumption}{Assumption}

\theoremstyle{remark}
\newtheorem{remark}{Remark}

\title{Who Moved the Frontier?\\ Paying Open Frontier Challenges with Free Identities}
\hypersetup{pdftitle={Who Moved the Frontier? Paying Open Frontier Challenges with Free Identities},
  pdfauthor={Luca Zanolini and Duc V. Le}}
\author{Luca Zanolini\thanks{Independent researcher. \texttt{lucazanolini1992@gmail.com}}
  \and Duc V. Le\thanks{Circle Research. \texttt{levduc112@gmail.com}}}
\date{November 2026}

\begin{document}
\maketitle

\begin{abstract}
In an open frontier challenge, anyone, person or AI agent, submits changes to a shared problem, and
a change that beats the best measured score becomes the public baseline. How should it pay when
identities are free? Rank prizes can be gamed: on a replay of one track of a recent challenge, one
solver's steps under two identities would have taken first and second prize. We consider a
prediction-market rule, proposed for machine-learning contests, that pays each accepted step for
the progress it adds; with exact scores, among rules that see only a step's start and end, only these resist
splitting and merging, and extra identities gain nothing. With noisy scores and independent normal
runs, we prove that any payment that never takes money back and pays real improvements also pays
something, on average, to a change that does nothing. We propose a fee per submission, at
least what luck pays per attempt, that removes each attempt's expected profit if verifiers are
honest and no submission is noisier than planned; nothing enforces that bound, and under noise the
rule does not cap the total paid. If crediting another and faking credit to one's second account look the same, a rule
that neither rewards fakes nor punishes honesty pays for credit at most what lying costs. We introduce a royalty on how much a later gain falls without the earlier change, which
needs no claims but needs removable changes. For five public challenges, we name the rule our
results would support.
\end{abstract}

\section{Introduction}\label{sec:intro}

An open frontier challenge is a contest with a public scoreboard and no gate. A sponsor posts a
problem, a scorer and a budget. Anyone, person or AI agent, submits a candidate under any identity.
If the candidate beats the current best, it is accepted, published at once, and becomes the base
that everyone builds on next. \emph{Autoresearch} is such a loop run by an AI agent, which
edits a piece of code, runs it, measures a score, and keeps the edit only if the score
improves~\cite{karpathy2026autoresearch}.
We study five public challenges with this shape or part of it: training a language model faster, proving
soundness bounds, speeding up a quantum-attack defence, optimising GPU
kernels~\cite{jordan2026speedrun,ef2026bettercodes,yukon2026qsbleaderboard,paradigm2026okc}, and
the one that coined the name \emph{open autoresearch} for the loop they
share~\cite{long2026ecdsafail}.

The loop is settled. How to pay the people in it is not. Three facts make the usual answer, prizes
for rank, fail. Identities are free, so one operator can be two solvers. Every accepted candidate
is public, so anyone can copy the best and add a little, which winner-take-all rewards. And nobody
referees: no human says what a
candidate is worth or whom it built on. A rule that these facts can game pays for identities,
copies or luck, and it takes that pay from those who made the progress.

Each failure is small enough to check by hand (Figure~\ref{fig:tricks}). Under winner-take-all, a
copier who adds $0.01$ to the leader's score takes the whole budget. Under ranked prizes or a
raffle, a solver with two large steps submits them under two identities and takes two prizes, or
two shares. Under a fixed bonus per accepted step, a solver cuts one step in two and earns two bonuses; ten
pieces earn nine extra.
And a check with one run a side, independent normal noise and no margin accepts a candidate that changes nothing half the time; on
our own training runs it accepted a renamed variable 8 times in 20 (Section~\ref{sec:noise}).

\begin{figure}[t]
\centering
\begin{tikzpicture}[font=\footnotesize,
  cell/.style={minimum width=2.15cm, minimum height=0.66cm, align=center, inner sep=2pt},
  bad/.style={cell, fill=red!14},
  ok/.style={cell, fill=green!16},
  head/.style={cell, font=\scriptsize\bfseries, minimum height=0.6cm},
  rowh/.style={text width=4.6cm, align=right, anchor=east, font=\scriptsize, inner sep=2pt}]
  \def\cw{2.2}\def\rh{0.72}
  \foreach \j/\t in {0/winner-take-all, 1/top-2 prizes, 2/raffle $\ge3$, 3/bonus per step, 4/\textbf{per-step}} {
    \node[head] at (\j*\cw, 0) {\t};
  }
  \node[rowh] at (-0.5*\cw-0.15, -1*\rh) {A copier takes Alice's code\\and adds $0.01$};
  \node[bad] at (0*\cw,-1*\rh) {$-\$1{,}000$};
  \node[ok]  at (1*\cw,-1*\rh) {$0$};
  \node[ok]  at (2*\cw,-1*\rh) {$0$};
  \node[ok]  at (3*\cw,-1*\rh) {$0$};
  \node[ok]  at (4*\cw,-1*\rh) {$0$};
  \node[rowh] at (-0.5*\cw-0.15, -2*\rh) {Alice submits her two steps\\under two names};
  \node[ok]  at (0*\cw,-2*\rh) {$0$};
  \node[bad] at (1*\cw,-2*\rh) {$+\$400$};
  \node[bad] at (2*\cw,-2*\rh) {$+\$100$};
  \node[ok]  at (3*\cw,-2*\rh) {$0$};
  \node[ok]  at (4*\cw,-2*\rh) {$0$};
  \node[rowh] at (-0.5*\cw-0.15, -3*\rh) {Alice cuts one of her steps\\in two};
  \node[ok]  at (0*\cw,-3*\rh) {$0$};
  \node[ok]  at (1*\cw,-3*\rh) {$0$};
  \node[ok]  at (2*\cw,-3*\rh) {$0$};
  \node[bad] at (3*\cw,-3*\rh) {$+\$50$};
  \node[ok]  at (4*\cw,-3*\rh) {$0$};
  \node[rowh] at (-0.5*\cw-0.15, -4*\rh) {A useless change, resubmitted\\until a noisy check accepts it};
  \node[bad] at (0*\cw,-4*\rh) {paid};
  \node[bad] at (1*\cw,-4*\rh) {paid};
  \node[bad] at (2*\cw,-4*\rh) {paid};
  \node[bad] at (3*\cw,-4*\rh) {paid};
  \node[ok]  at (4*\cw,-4*\rh) {$\le0$ on average\\ \scriptsize fee $b$, fresh runs};
  \foreach \i in {0,...,4} { \draw[black!25] (-0.5*\cw,-\i*\rh-0.5*\rh) -- (4.5*\cw,-\i*\rh-0.5*\rh); }
  
\end{tikzpicture}
\caption{Four tricks against five rules, on one example. Carol raised the frontier from $0$ to $4$; then Alice raised it from $4$ to $14$ in two steps of $5$. In rows 1--3 each cell is the change in Alice's pay compared with the same record without the trick ($+$: extra money Alice makes by cheating; $-$: money she loses to a cheat): under winner-take-all the copier takes the \$1,000 Alice would otherwise win. Under top-2 with two names, Alice takes \$600 + \$400 instead of \$600, and Carol loses her \$400. Row 4 asks whether a change that does nothing gets paid. Winner-take-all pays \$1,000; top-2 pays \$600 and \$400; the raffle shares \$600 among identities with gain at least $3$; the bonus pays \$100 per unit plus \$50 per step; the per-step rule pays \$100 per unit. Red: the trick pays, or costs an honest earlier contributor. Green: it changes nothing. Under the per-step rule the copier earns only the \$1 that its $0.01$ is worth (Proposition~\ref{prop:potential}); in the last row a fee, with pay on fresh runs, removes the expected profit (Proposition~\ref{prop:fee}).}
\label{fig:tricks}
\end{figure}
 
The fix is as simple as the failures. Pay each accepted step the rise it made on a price curve
fixed in advance, $\Phi(b)-\Phi(a)$. Then cutting a step in two pays $\Phi(b)-\Phi(a)$ in total,
the same; a second identity earns the same as the first; and a copier earns exactly the little it
added, while the leader keeps everything it earned. This rule is a market scoring
rule~\cite{hanson2003combinatorial}, and \citet{abernethy2011collaborative} proposed it for
machine-learning contests. We ask what else an open challenge needs.

\paragraph{Contributions.}
\begin{itemize}\setlength{\itemsep}{2pt}
\item \textbf{A model of the incentive layer} (Section~\ref{sec:formal}, Figure~\ref{fig:wrapper}).
  A problem is a solution space with a scorer, exact or noisy. A challenge adds identities, a fee,
  an acceptance test and a payment rule. It is open if identities are free, the frontier is public
  and nobody referees.
\item \textbf{Which rule for which challenge} (Section~\ref{sec:which}). We sort five real
  challenges by their scorer and name the rule our results support for each, and what it would
  need. On QSB's published prize schedule, replayed on one track, a solver's own steps under two
  identities would have taken first and second prize.
\item \textbf{Exact scores} (Section~\ref{sec:rule}). Among rules that pay a step by its two end
  points, the per-step rules are exactly the ones that splitting and merging cannot game; all are
  immune to fake identities and copying, and the per-step rule fits a budget $B$ iff
  $\sup\Phi-\Phi(s_0)\le B$ (Proposition~\ref{prop:potential}).
\item \textbf{Noisy scores} (Section~\ref{sec:noise}). With independent normal runs, every payment
  that is never negative pays a useless candidate something on average, or pays nobody
  (Proposition~\ref{prop:nonneg}). A fee per candidate of at least what luck pays per attempt, with
  pay on fresh runs, removes the expected profit of every useless attempt, under honest verifiers
  and a bound on the noise (Proposition~\ref{prop:fee}); a higher fee, not refunded, also stops splitting into
  equally noisy pieces (part (c)), at about twice the first in our simulation. The bound is the weak point: the operator's code sets its own noise.
\item \textbf{Enabling} (Section~\ref{sec:enabling}). If honestly crediting another person's step and
  falsely crediting one's own second account look the same to the rule, a rule that neither rewards
  fakes nor punishes honest credit pays for credit at most what lying costs (Theorem~\ref{thm:limit}). A royalty on measured interactions needs no claims, and exactly one,
  half of each interaction, resists cutting and timing (Theorem~\ref{thm:shapley}). But it needs
  removable changes: on the 24 pairs of one challenge where we tested removal, no later proof built
  without the earlier solver's work.
\end{itemize}
Each result holds under the conditions stated with it. Under noise the rule does not cap the total
paid, and the pot ended below zero in $35$--$42\%$ of our simulated challenges. Our evidence is four public
records and the published account of a fifth, none shown to meet all three openness conditions as run (Section~\ref{sec:which}), and the
training runs of one small model on one machine; the replays show what a rule would have paid, not
what anyone did.
\section{The model}\label{sec:framework}\label{sec:formal}

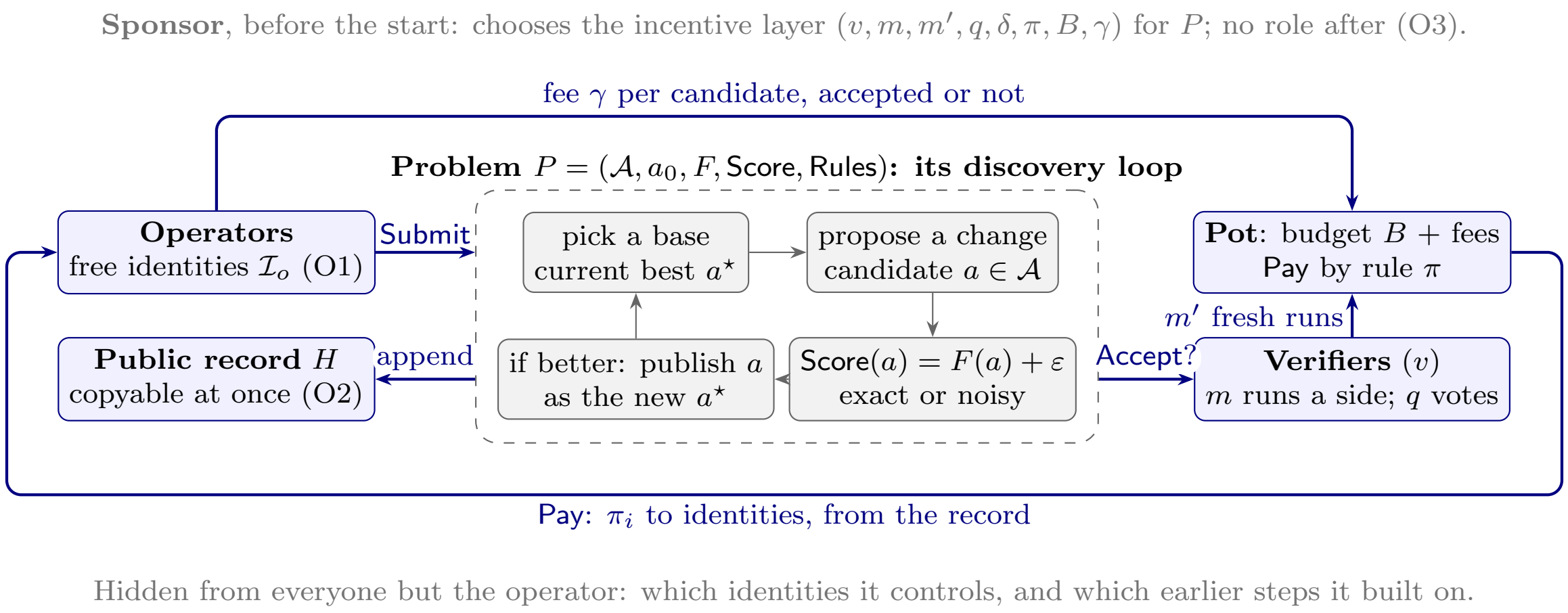
\begin{figure}[t]
\centering
\begin{tikzpicture}[font=\scriptsize, >={Stealth[length=4.5pt]},
  core/.style={draw=black!60, rounded corners=3pt, align=center, inner sep=3pt, minimum width=2.0cm, minimum height=0.78cm, fill=black!5},
  layer/.style={draw=blue!50!black, rounded corners=3pt, align=center, inner sep=3pt, minimum width=2.4cm, minimum height=0.78cm, fill=blue!6},
  flow/.style={->, blue!50!black, thick, rounded corners=4pt},
  lab/.style={font=\scriptsize, text=blue!50!black, fill=white, inner sep=1pt, align=center},
  hid/.style={align=center, font=\scriptsize, text=black!55}]
  \node[core] (base)    at (-1.45, 0.62) {pick a base\\ current best $a^\star$};
  \node[core] (change)  at ( 1.45, 0.62) {propose a change\\ candidate $a\in\mathcal{A}$};
  \node[core] (score)   at ( 1.45,-0.62) {$\mathsf{Score}(a)=F(a)+\varepsilon$\\ exact or noisy};
  \node[core] (promote) at (-1.45,-0.62) {if better: publish $a$\\ as the new $a^\star$};
  \node[draw=black!60, dashed, rounded corners=6pt, inner sep=6pt, fit={(base) (change) (score) (promote)}] (core) {};
  \node[font=\scriptsize\bfseries, anchor=south, inner sep=1.5pt] at (core.north) {Problem $P=(\mathcal{A},a_0,F,\mathsf{Score},\mathsf{Rules})$: its discovery loop};
  \draw[->, black!60] (base) -- (change);
  \draw[->, black!60] (change) -- (score);
  \draw[->, black!60] (score) -- (promote);
  \draw[->, black!60] (promote) -- (base);
  \node[layer] (ops) at (-5.55, 0.62) {\textbf{Operators}\\ free identities $\Id_o$ (O1)};
  \node[layer] (rec) at (-5.55,-0.62) {\textbf{Public record} $H$\\ copyable at once (O2)};
  \node[layer] (pot) at ( 5.55, 0.62) {\textbf{Pot}: budget $B$ + fees\\ $\mathsf{Pay}$ by rule $\pi$};
  \node[layer] (ver) at ( 5.55,-0.62) {\textbf{Verifiers} ($v$)\\ $m$ runs a side; $q$ votes};
  \draw[flow] (ops.east) -- node[lab, above=1pt]{$\mathsf{Submit}$} (core.west |- ops.east);
  \draw[flow] (ops.north) -- (-5.55,1.95) -- node[lab, above=1pt]{fee $\gamma$ per candidate, accepted or not} (5.55,1.95) -- (pot.north);
  \draw[flow] (core.east |- ver.west) -- node[lab, above=1pt]{$\mathsf{Accept}$?} (ver.west);
  \draw[flow] (core.west |- rec.east) -- node[lab, above=1pt]{append} (rec.east);
  \draw[flow] (ver.north) -- node[lab, left=1pt]{$m'$ fresh runs} (pot.south);
  \draw[flow] (pot.east) -- (7.6,0.62) -- (7.6,-1.75) -- node[lab, below=1pt]{$\mathsf{Pay}$: $\pi_i$ to identities, from the record} (-7.6,-1.75) -- (-7.6,0.62) -- (ops.west);
  \node[hid, anchor=north] at (0,-2.45) {Hidden from everyone but the operator: which identities it controls, and which earlier steps it built on.};
  \node[hid, anchor=south] at (0,2.55) {\textbf{Sponsor}, before the start: chooses the incentive layer $(v,m,m',q,\delta,\pi,B,\gamma)$ for $P$; no role after (O3).};
\end{tikzpicture}
\caption{The incentive layer (blue) wrapped around a problem's discovery loop (grey; the open-autoresearch loop of \citet[Fig.~5]{long2026ecdsafail}). The layer adds who submits, what a candidate costs, how acceptance is decided and how identities are paid. The paper chooses $\pi$ by the kind of scorer: the per-step rule when $\mathsf{Score}$ is exact; with a fee and pay on fresh runs when it is noisy, under honest verifiers, independent normal errors and a noise bound fixed in advance; a royalty only where changes can be removed and re-scored.}
\label{fig:wrapper}\label{fig:system}
\end{figure}
 
Figure~\ref{fig:wrapper} shows the two layers. The inner loop is the problem; it exists without us,
and it is the open-autoresearch loop of \citet[Fig.~5]{long2026ecdsafail}. The outer ring is the
incentive layer: who may submit, what a candidate costs, how acceptance is decided, and how
identities are paid. The paper is about the outer ring.

\subsection{Parties and the challenge}\label{sec:parties}

The \emph{sponsor} fixes everything before the start and does nothing after. An \emph{operator}
(a person or organisation, possibly running AI agents) controls a set $\Id_o$ of \emph{identities}
from a countable set $\Id$, disjoint across operators. The record shows identities; operators are
hidden. \emph{Verifiers} run the scorer and decide acceptance.

\begin{definition}[Problem]\label{def:problem}
A \emph{problem} is a tuple $P=(\mathcal{A},a_0,F,\mathsf{Score},\mathsf{Rules})$:
a \emph{solution space} $\mathcal{A}$, the candidate solutions of one problem instance, with a
public \emph{baseline} $a_0\in\mathcal{A}$;
a \emph{true score} $F:\mathcal{A}\to\R$, higher is better, which nobody observes;
a \emph{scorer}: one run of $a$ returns $F(a)+\varepsilon$, where $\varepsilon$ has mean $0$ and
standard deviation $\sigma(a)\ge0$, which may depend on $a$ because the operator wrote $a$; the
scorer is \emph{exact} if $\sigma\equiv0$ and \emph{noisy} otherwise;
and \emph{rules} $\mathsf{Rules}:\mathcal{A}\times\mathcal{A}\to\{0,1\}$, which say whether a
candidate may replace the current best.
\end{definition}

\begin{definition}[Frontier-reward rule]\label{def:frr}
A \emph{frontier-reward rule} $\pi$ maps a record $H$ to a payment $\pi_i(H)\in\R$ for each
identity $i$, with finitely many non-zero entries, paid from the budget $B$
(Definition~\ref{def:arp} defines both). It is
\emph{budget-feasible} if $\sum_i\pi_i(H)\le B$ for every record. An operator's \emph{payoff} is
$U_o(H)=\sum_{i\in\Id_o}\pi_i(H)$. For a randomised rule, $\pi_i$ is the expected payment.
\end{definition}

\begin{definition}[Frontier challenge]\label{def:arp}
A \emph{frontier challenge} on a problem $P$ adds an \emph{incentive layer}, fixed by the sponsor
before the first candidate arrives: $v\ge1$ verifiers; $m\ge1$ runs a side for acceptance and
$m'\ge1$ for payment; $q$ votes needed, $1\le q\le v$; a margin $\delta\ge0$; a reward rule $\pi$;
a budget $B>0$; and a fee $\gamma\ge0$ per candidate ($\gamma=0$ unless stated). It runs four
algorithms; $a^\star$ is the current best, initially $a_0$.
\begin{enumerate}\setlength{\itemsep}{0pt}\setlength{\parskip}{0pt}
\item $\mathsf{Submit}(i,a)$: an operator submits $a\in\mathcal{A}$ under an identity $i$ it
  controls and pays $\gamma$ into the pot; every candidate enters a \emph{log}.
\item $\mathsf{Measure}(a,a^\star)\to D$: $m$ runs of $a$ and of $a^\star$; $D$ is the difference
  of the two means.
\item $\mathsf{Accept}(i,a)$: verifier $j$ votes yes if $\mathsf{Rules}(a^\star,a)$ holds and its
  own $D_j\ge\delta$ (strictly when $\delta=0$). With $q$ yes votes the candidate is
  \emph{accepted}: its \emph{recorded gain} $g$ is the median of the $D_j$, $(i,a,g)$ is appended
  to the record, and $a$ becomes $a^\star$.
\item $\mathsf{Pay}(H)\to(\pi_i)_{i\in\Id}$: the rule $\pi$ (Definition~\ref{def:frr}) pays identities from the
  record; it may re-measure each accepted step on $m'$ fresh runs (why: Section~\ref{sec:noise}).
\end{enumerate}
\end{definition}

\noindent The vote and the median are simulated only; our results are for one verifier, or for
verifiers pooled by their mean (Section~\ref{sec:noise}). The rules, the record and every accepted
candidate are public. Hidden are which operator controls which identities (\emph{ownership}), which
earlier steps a step truly built on (\emph{ancestry}), and what any candidate cost.

\begin{definition}[Open frontier challenge]\label{def:oac}
A frontier challenge is \emph{open} if three conditions hold.
\begin{enumerate}\setlength{\itemsep}{0pt}
\item[(O1)] \textbf{Open entry.} Any operator may submit, under identities that nobody can link to
  it. A fresh identity costs nothing.
\item[(O2)] \textbf{Public frontier.} Every accepted candidate is published at once, with its
  identity and recorded gain, and becomes the baseline for everyone.
\item[(O3)] \textbf{No referee.} After the start, acceptance and payment follow from measurements
  and the record alone. Nobody rules on what a candidate is worth or on whom it built.
\end{enumerate}
\end{definition}

The three conditions are the threat model: (O1) makes fake identities free, (O2) makes copying
free, and (O3) removes the human who would otherwise catch both. They are also governance choices.
A sponsor that can verify people, keep results private, or judge cases by hand faces a different
problem; we study the one where it cannot.

\subsection{The score model: a record is a list of scores}\label{sec:model}

Scores live in $\mathcal{X}=[s_0,s^*]\subseteq\R$, with $s_0=F(a_0)$ and $s^*\le\infty$ the
target. A cost to be minimised becomes a score by a decreasing map such as minus its logarithm, under
which equal relative improvements are equal differences.

\paragraph{Exact scorer.} Every run of $a$ returns $F(a)$, so all verifiers agree and
$D=F(a)-F(a^\star)$. A record is then a list of identities and scores,
\[ H=\big((i_1,x_1),\dots,(i_n,x_n)\big),\qquad s_0=x_0<x_1<\dots<x_n\le s^*, \]
where step $k$ is owned by identity $i_k$ and raises the \emph{frontier} from $x_{k-1}$ to $x_k$.
A candidate is accepted iff its score beats the frontier ($\delta=0$; Section~\ref{sec:rule}
treats $\delta>0$). The start point $x_{k-1}$ is set by the record, not by the operator: a copy of
the frontier plus a little is a short step, whatever it cost. $\Hist$ is the set of
all records; a \emph{fresh} identity appears in none of $H$ (Figure~\ref{fig:record},
Appendix~\ref{app:score}).

\paragraph{Noisy scorer.} The same candidate scores differently on every run. Runs are
\emph{independent normal} when every run's error is drawn independently and follows a normal law
with mean $0$. One honest verifier with $m$ runs a side then reports $D$ with mean
$\mu=F(a)-F(a^\star)$, the true gain, and standard deviation
$\sqrt{(\sigma(a)^2+\sigma(a^\star)^2)/m}$; when both sides have the same standard deviation $\sigma$,
this is $s_m=\sigma\sqrt{2/m}$. Pooling $v$
verifiers by their mean is one verifier with $vm$ runs a side.

\subsection{What a rule should resist}\label{sec:manipulations}\label{sec:properties}\label{sec:properties-plain}

Each move below is one an operator can make alone. A rule \emph{resists} a move if the move never
raises the operator's payoff.
\begin{itemize}\setlength{\itemsep}{1pt}\setlength{\parskip}{0pt}
\item \textbf{Copying.} Copy the frontier, add a little, submit. The rule must not let this take
  pay from those who came before (Definition~\ref{def:nonexp}).
\item \textbf{Fake identities.} Move some of one's steps to a fresh identity
  (Definition~\ref{def:sybil}).
\item \textbf{Splitting and merging.} Cut one step in two at an intermediate score, or join two
  (Definition~\ref{def:split}).
\item \textbf{Luck-fishing.} Resubmit a useless candidate until a noisy check accepts it. Here the
  payoff is net of fees (Section~\ref{sec:noise}).
\item \textbf{Lying about ancestry, and re-cutting changes.} Hide a step one built on, or name one
  falsely; or cut a change into pieces and time them. These need a finer view of a step
  (Section~\ref{sec:enabling}).
\end{itemize}
A sponsor also wants the rule \emph{budget-feasible}, and may want it to pay for \emph{enabling}:
earlier work that made later progress possible. All of these compare payoffs on a fixed record of
others' steps; we do not analyse equilibrium. Splitting needs intermediate scores to exist.
\begin{assumption}[Divisibility]\label{ass:divisible}
An operator that can raise the frontier from $y$ to $z$ can also produce, at no cost, a verified
candidate with any score in $(y,z)$.
\end{assumption}
It holds where an improvement can be weakened at will, as a proved bound can. Only results on
other rules need it (that splitting games them, and the converse in Proposition~\ref{prop:potential}(a)); the
per-step rule's guarantees hold without it (Appendix~\ref{app:score}).

\begin{definition}[Split and merge]\label{def:split}
For a record $H$ of length $n$, $1\le k\le n$ and $y\in(x_{k-1},x_k)$, the \emph{split} of $H$ at
$(k,y)$ inserts the entry $(i_k,y)$ before entry $k$. A \emph{merge} replaces two consecutive
entries of one identity by the later one. A rule is \emph{split-proof} if
$U_o(\mathrm{split}(H))\le U_o(H)$ for every $H,k,y$, with $o$ the operator of $i_k$; it is
\emph{merge-proof} if the reverse inequality holds for the same $H,k,y$.
\end{definition}

\begin{definition}[Sybil-proofness]\label{def:sybil}
Let $K$ be a set of entries of $H$ owned by identities of $o$, and $i'\in\Id_o$ fresh. The
\emph{relabelling} $\rho_{K,i'}(H)$ gives every entry in $K$ the owner $i'$. A rule is
\emph{Sybil-proof} if $U_o(\rho_{K,i'}(H))\le U_o(H)$ for all $o,K,i',H$.
\end{definition}
This is false-name-proofness~\cite{yokoo2004falsename} on records.

\begin{definition}[Non-expropriation]\label{def:nonexp}
An \emph{extension of $H$ by others of $o$} appends entries owned outside $\Id_o$. A rule is
\emph{non-expropriating} if $U_o(H')\ge U_o(H)$ for every such extension $H'$ of every $H$.
\end{definition}
Non-expropriation is the copying property: whatever comes later, an operator keeps what it earned,
as protection against front-running~\cite{daian2020flash} would keep an earlier trade.
\begin{table}[!tb]
\centering\footnotesize
\setlength{\tabcolsep}{4pt}
\newcolumntype{L}[1]{>{\raggedright\arraybackslash}p{#1}}
\begin{tabular}{L{2.4cm}L{2.9cm}L{2.0cm}L{2.6cm}L{4.5cm}}
\hline
Challenge & Scorer; who draws the noise & Open as run? & Rule in use & Rule our results would support \\
\hline
better.codes~\cite{bettercodes2026leaderboard} & exact: a Lean proof checker & O1 ?, O2 yes, O3 no & discretionary awards & once pay is committed: per-step, no fee (Prop.~\ref{prop:potential}); a target bound or a capped $\Phi$ \\
NanoGPT speedrun~\cite{jordan2026speedrun} & noisy training time; the operator posts its own run logs & O1 ?, O2 yes, O3 no & none documented & if verifiers rerun and score each candidate: per-step on fresh runs with a fee, log price; the noisy loss check is outside our model \\
QSB pinning~\cite{yukon2026qsbleaderboard} & noisy; one fresh organiser run per candidate, against a stored best & O1 ?, O2 yes, O3 ? (payment fixed) & two top-3 periods + 3\% raffle, \$20,000 & with fresh runs of the best: per-step on the log score with a fee \\
Kernel challenge~\cite{paradigm2026okc} & noisy; serving track: organiser reruns the top 25 on held-out data & O1 no, O2 no, O3 no & serving track: top-10 prizes, \$100,000 & if opened: per-step on the mean of fresh reruns with a fee \\
ECDSA.Fail~\cite{long2026ecdsafail} & deterministic; the operator picks the test inputs by a 48-bit nonce & O1 ?, O2 yes, O3 ? & none documented in its paper & if inputs are drawn after submission: per-step with a fee; the sampled correctness check is outside our model \\
\hline
\end{tabular}
\caption{Five frontier challenges, which openness conditions of Definition~\ref{def:oac} each meets as run (? means our sources do not establish it; evidence for the first four in Table~\ref{tab:evidence}), and the reward rule our results would support for each. Every fee recommendation needs what Proposition~\ref{prop:fee} assumes: honest verifiers who draw fresh runs, independent normal errors, and a noise bound fixed in advance. None of the five establishes these, and under noise the per-step rule does not cap the total paid.}
\label{tab:which}
\end{table}

\section{Which rule for which challenge}\label{sec:which}\label{sec:evidence}

Three facts about a challenge pick the rule: is its scorer exact; if not, who draws the noise; and
can a change be removed from the current best and the rest re-scored. Table~\ref{tab:which} sorts
five public challenges by these facts. None is shown to be open in full as run, so each
recommendation says what it would need (records: Appendix~\ref{app:records}).

\paragraph{Exact scorer: pay each step its gain, and nothing else is needed.} On better.codes a
proof checker accepts or rejects every commit, so the record holds true scores and the per-step
rule applies as proved (Proposition~\ref{prop:potential}). Splitting is not hypothetical there: a
proved bound can be weakened at will, and one solver made five consecutive steps of $0.01$ bits.
Under the per-step rule they pay what one step of $0.05$ bits pays. The page already shows each
step's gain; it lacks a committed price, and its awards are discretionary, which (O3) excludes.
With no target a linear price in bits is unbounded, so the sponsor fixes a target bound or caps
$\Phi$ (Proposition~\ref{prop:potential}(b)).

\paragraph{Noisy scorer: first ask who draws the noise.} With a noisy scorer a useless candidate
sometimes scores well by luck. The remedy of Proposition~\ref{prop:fee} is a fee per candidate,
paid into the pot, with pay measured on fresh runs. It assumes that honest verifiers draw the runs,
that run errors are independent and normal on the score used, and that a bound on the noise is
fixed in advance. The fee is paid at $\mathsf{Submit}$, so it prices nothing that happens before.
The four noisy challenges answer it differently (evidence: Appendix~\ref{app:records}).
\begin{itemize}\setlength{\itemsep}{1pt}
\item \emph{The kernel challenge draws its own noise}: on the track that pays, it reruns the top 25
  entries at least three times on held-out data (rules of 23 October 2026; Appendix~\ref{app:records}). That is closest to our hypotheses, but it is not
  open as run: one account per individual (O1), a board with each account's best result and no
  artifact (O2), and scoring ``subject to Sponsor's discretion'' (O3). Our results are for means,
  not its median of at least three reruns.
\item \emph{The speedrun leaves the noise to the operator}, who chooses how many runs to make and
  which logs to post, before submission; no fee prices that choice. Its rules also list
  ``discretionary reasons why a PR may not be accepted'', so (O3) fails as run. Even with verifiers
  who rerun each candidate, its loss threshold stays a noisy pass/fail check, which
  Proposition~\ref{prop:fee} does not model. The price curve then decides who is paid: on the 91
  recorded steps, a linear price per second pays the first 10 steps 85\% of a budget, and a price in
  the logarithm of time pays them 43\% (Appendix~\ref{app:speedruncf}).
\item \emph{QSB runs each candidate once, fresh, against the best's stored score.} The organiser
  measures on its own GPU and promotes automatically above a set margin; one score's relative noise
  is about $1/\sqrt{K}$ for $K$ verified hits~\cite{layr2026qsbharness}. A stored score is where
  Proposition~\ref{prop:fee}(d) fails (Appendix~\ref{app:noiseproofs}).
\item \emph{ECDSA.Fail lets the operator pick the sample.} Its score is deterministic, but the
  9,024 test inputs are a hash of the submitted circuit, and a 48-bit nonce changes the hash
  without changing the circuit. Operators search the nonce offline; 19 of 400 accepted commits were
  such rerolls, ``not a circuit optimization''~\cite{long2026ecdsafail}. The first fix is in
  $\mathsf{Measure}$, not in $\mathsf{Pay}$: draw the inputs after submission, from randomness the
  operator cannot grind. Its correctness check is sampled on the same inputs, which our exact
  $\mathsf{Rules}$ does not model.
\end{itemize}

\paragraph{Rank prizes: a published schedule, replayed.} QSB's schedule is the one published
schedule we have a record to replay: top-3 prizes in two periods plus a \$6,000 raffle shared by
everyone who improved the frontier by $3\%$~\cite{starkware2026qsbrewards}; staff of the two
sponsoring firms are not eligible~\cite{starkware2026qsbleaderboard}. We replayed it on the one
track we could save, pinning, with the same \$20,000 and the 24 eligible solvers
(Table~\ref{tab:replay}), taking the recorded throughputs as exact scores. Three tricks pay, and
none involves new work. The winner of weeks 2--3 on this track would take that period's first
\emph{and} second prize under two identities: \$9,600 instead of \$6,000. Five solvers
could split their steps into groups that each clear the raffle's $3\%$; the largest gain in
Table~\ref{tab:replay}, \$4,167, combines three identities with the raffle. Two solvers who pool their steps under one identity would take second prize; this is
collusion between operators, which our definitions do not cover, but the per-step rule adds pay
over steps and so pays the pair the same. Under the per-step rule, with every price curve we tried,
each re-packaging pays exactly what the honest record pays. Nobody is claimed to have done any of
this, and on the published ranking, which adds QSB's two tracks, the same solver placed second.

\begin{table}[!tb]
\centering\footnotesize
\begin{tabular}{lrrrrr}
\hline
 & & & \multicolumn{3}{c}{Most a solver gains, in \$, by} \\
Rule (same \$20,000) & Gini & top-3 share & fake identities & pooling & splitting a step \\
\hline
QSB's published rule, replayed & 0.687 & 0.670 & $+4{,}167$ & $+3{,}286$ & 0 \\
Per-step rule (one price) & 0.574 & 0.483 & 0 & 0 & 0 \\
Winner-take-all & 0.958 & 1.000 & 0 & 0 & 0 \\
Fixed bonus per accepted step (\$487.80) & 0.349 & 0.390 & 0 & 0 & $+488$ \\
\hline
\end{tabular}
\caption{The same \$20,000 paid on the real QSB pinning record (44 steps by 27 solvers) under four
rules, to its 24 eligible solvers; the 3 staff members' steps count in the record but are not paid.
The last three columns give the
most one solver could gain by one trick with its own real steps: re-submitting them under 2 to 4
identities, two solvers pooling their steps under one identity (gain of the pair), or cutting one
step in two at the midpoint of its gain. Prices and pools are fixed in advance from the honest
record (Table~\ref{tab:replayfull}, Appendix~\ref{app:records}).}
\label{tab:replay}
\end{table}

The price curve moves the money: with one price the per-step rule is less concentrated than the
schedule (Gini $0.574$ against $0.687$); priced per period, as QSB's reset intends, it is more
concentrated (Appendix~\ref{app:records}). A copier who takes the final best and adds the smallest
real step wins the whole budget under winner-take-all and \$2.48 under the per-step rule
(Appendix~\ref{app:copyreplay}).

\paragraph{Removable changes: where we tested, removal failed.} Paying for enabling with a royalty
(Section~\ref{sec:enabling}) needs the current best to be scorable with one change removed. On
better.codes we rebuilt later proofs with an earlier solver's declarations deleted: in the 24 pairs
we could attempt, each built with them, and without them 22 failed to build, 2 ran past the time
limit, and none built
(Appendix~\ref{app:enablingbuild}). The later proof calls the earlier lemma; there is nothing to
switch off. On the speedrun we counted edits to earlier records' lines but did not test removal
(Appendix~\ref{app:speedruncensus}); on the other three we did not test it at all. Where removal
does work, as in our own training experiment, measuring an interaction of $1\%$ of the score costs about 132 runs,
by a planning calculation (Appendix~\ref{app:noise}). So the royalty's precondition is not established on any of the
five, and paying for enabling remains open for all of them.
\section{Theory I: exact scores}\label{sec:rule}\label{sec:results}\label{sec:rules}

With an exact scorer a record is a list of scores $s_0=x_0<x_1<\dots<x_n$, and step $k$ is owned by
identity $i_k$ (Section~\ref{sec:model}). A reward rule (Definition~\ref{def:frr}) pays no negative amounts here,
$\pi:\Hist\to\R_{\ge0}^{\Id}$; operators are risk-neutral, submitting costs nothing, and the budget
is $B>0$.

\begin{definition}[Endpoint rules]\label{def:step}
A reward rule $\pi$ is an \emph{endpoint rule} if there is a function $r(y,z)\ge0$ of pairs $y<z$ in
$\mathcal{X}$ with
\[ \pi_i(H)=\sum_{k\le n:\ i_k=i} r(x_{k-1},x_k)\qquad\text{for all } H\in\Hist,\ i\in\Id . \]
A step's pay is fixed by its two end points, known at acceptance and never revised.
\end{definition}

\begin{definition}[Per-step rule]\label{def:potential}
An endpoint rule is a \emph{per-step rule} if $r(y,z)=\Phi(z)-\Phi(y)$ for a non-decreasing
\emph{price curve} $\Phi:\mathcal{X}\to\R$ fixed in advance. ``Pay $\lambda$ per unit of score'' is
the per-step rule with $\Phi(x)=\lambda\,(x-s_0)$.
\end{definition}

\noindent Chunks $y=y_0<y_1<\dots<y_n=z$ pay
$\sum_k\big(\Phi(y_k)-\Phi(y_{k-1})\big)=\Phi(z)-\Phi(y)$, what one step pays, whoever submits
them. A linear $\Phi$ is a fixed price per unit of progress; a bounded $\Phi$ bounds the spend.
Other endpoint rules differ: $r(y,z)=(z-y)^2$ pays one step of 10 the amount 100 and ten steps of 1
the amount 10, so merging pays.

\begin{proposition}[Endpoint rules, exact scores]\label{prop:potential}\label{prop:robust}\label{prop:steprobust}\label{cor:all}\label{cor:fourprops}
Let scores be exact, the margin $\delta=0$, and $\pi$ an endpoint rule with step reward $r$.
\begin{enumerate}
\item[(a)] If $r(y,z)=\Phi(z)-\Phi(y)$ for some $\Phi$, then $\pi$ is split- and merge-proof, for
  every split, possible or not. Conversely, if $\pi$ resists every split and merge of
  Definition~\ref{def:split} on every record of $\Hist$, then $r(y,z)=\Phi(z)-\Phi(y)$ for a
  non-decreasing $\Phi$ with $\Phi(s_0)=0$, so $\pi$ is a per-step rule.
\item[(b)] A per-step rule pays $\sum_i\pi_i(H)=\Phi(x_n)-\Phi(s_0)$ on a record with frontier
  $x_n$. It is budget-feasible if and only if $\sup_{x\in\mathcal{X}}\Phi(x)-\Phi(s_0)\le B$.
\item[(c)] Moving steps among one operator's identities, and appending steps of others, leave that
  operator's payoff unchanged. So every endpoint rule is Sybil-proof and non-expropriating, and has all four properties
  if and only if it is a per-step rule.
\end{enumerate}
\end{proposition}

\noindent Proofs are in Appendix~\ref{app:score}. The converse in (a) needs every split to be
possible (Assumption~\ref{ass:divisible}); where fewer are, more rules resist them. A margin
$\delta>0$ weakens the monotonicity of $\Phi$ (Remark~\ref{rem:threshold}).

\paragraph{Four baseline rules fail where the per-step rule does not.} \emph{Winner-take-all} (WTA) pays $B$ to the
owner of the last step. \emph{Top-$k$} pays prizes $P_1\ge\dots\ge P_k$, with $P_1>0$, to the identities with the
$k$ largest total gains. The \emph{raffle} shares a pool $P$ equally among identities with total
gain at least $\theta$. The \emph{bonus} is $r(y,z)=\lambda(z-y)+\beta$ with $\beta>0$
(formal rules and Proposition~\ref{prop:failures}: Appendix~\ref{app:score}; top-$k$ and the raffle
are stylised from QSB's schedule, Appendix~\ref{app:rulesinuse}). The first three read only the last
owner or each identity's total, so they are split- and merge-proof too. Each fails elsewhere, on the
example of Figure~\ref{fig:tricks}: under WTA a copier who adds $0.01$ takes Alice's \$1,000; a
second identity gains her \$400 under top-2 and \$100 under the raffle; under the bonus each cut
adds $\beta=\$50$.
\emph{The per-step rule passes every score-model property}, and every baseline rule fails
at least one, except a raffle whose threshold nobody can reach, which pays nothing (Table~\ref{tab:results}); re-cutting is in Section~\ref{sec:enabling}.
\section{Theory II: noisy scores}\label{sec:noise}\label{sec:noisetheory}\label{sec:noiseresults}\label{sec:noisemeasured}\label{sec:noiserule}

With noise, Section~\ref{sec:rule} fails for two reasons. A candidate cut into $k$ pieces gets $k$
chances to be paid for luck. And a gain is recorded only if it cleared the bar, so it overstates the
true gain on average.

\paragraph{A one-run check paid \$8,480 for a renamed variable.} On a small language model scored by
validation bits per byte (1~mbpb $=0.001$), a one-run check paying \$100 per mbpb accepted a
candidate that only renames a variable 8 times in 20, against fresh baseline runs
(Appendix~\ref{app:renamedetail}); at the $\$31.47$ per mbpb of our simulation below, the same runs pay \$2,669.

\begin{assumption}[Noise]\label{ass:noise}
\begin{enumerate}\setlength{\itemsep}{0pt}\setlength{\parskip}{0pt}
\item[(N1)] Runs are independent, and each returns the true score plus normal noise with mean $0$
  and a positive standard deviation.
\item[(N2)] Verifiers are honest. A check is made by one verifier, or decided and paid on the mean
  of $v$ verifiers' differences, which is one verifier with $vm$ runs a side.
\item[(N3)] No run's standard deviation exceeds a bound $\bar\sigma$ fixed by the sponsor.
\item[(N4)] An accepted step is paid a fixed price $\lambda>0$ per unit of measured gain.
\end{enumerate}
\end{assumption}
\noindent $N$ and $N'$ are the standard normal cdf and density; $D$ has mean $\mu$, the true gain,
and standard deviation $s$ ($=\sigma\sqrt{2/m}$ when both sides have noise $\sigma$,
Section~\ref{sec:model}).

\begin{proposition}[Paying on the deciding runs]\label{prop:deciding}\label{prop:luck}
Assume (N1), (N2) and (N4). Let a candidate be accepted if $D\ge\delta$ and paid $\lambda D$ on
acceptance.
\begin{enumerate}
\item[(a)] \textbf{Luck is paid.} A candidate with $\mu=0$ is accepted with probability
  $1-N(\delta/s)$ and earns $\lambda s\,N'(\delta/s)>0$ per attempt in expectation; at $\delta=0$
  this is $\lambda s/\sqrt{2\pi}$.
\item[(b)] \textbf{Winner's curse.} For every $\mu$ and $\delta$, an accepted candidate is overpaid in
  expectation: $\mathbb{E}[D\mid D\ge\delta]=\mu+s\,N'(z)/(1-N(z))>\mu$ with $z=(\delta-\mu)/s$.
\end{enumerate}
\end{proposition}

\noindent Part (b) is the winner's curse~\cite{smith2006optimizer,lee2018winners}: in a simulated
stream, pay on the deciding runs counted $6{,}974$~mbpb against a true $3{,}194$
(Appendix~\ref{app:noisyscorer}); splitting also pays there (Proposition~\ref{prop:splitzero}).

\paragraph{No payment that is never negative avoids paying for luck.} A useless candidate reaches
every range of run scores a real one does, so it is paid wherever the real one is, as for
estimators~\cite{somekhbaruch2018nonexistence}.

\begin{proposition}[Payments that are never negative]\label{prop:nonneg}
Let $M$ be the vector of the run scores of a candidate and of its reference, those that decide
acceptance and those that set the pay alike. Suppose that some useless candidate's $M$ has a density that is positive everywhere,
and that every candidate's $M$ has a density. Let a candidate be paid $f(M,U)\ge0$, for a
measurable $f$ and randomness $U$ independent of $M$. If that useless candidate is paid $0$ in
expectation, then every candidate is paid $0$ with probability one.
\end{proposition}

\noindent (N1) gives both hypotheses; normality is not otherwise used, and noise of bounded support
breaks the first. So a sponsor that pays real work pays for luck or takes money back. Taking it
back needs a deposit, which is costly and which an operator escapes by walking away from a fresh
identity (Appendix~\ref{app:noisyscorer}); we take the money before, as a fee.

\paragraph{A fee per candidate takes the profit out of luck.} A useless candidate is a free lottery
ticket, and free identities buy any number. So sell each ticket for its average prize: every
candidate pays a fee $\gamma$ into the pot, accepted or not (in a variant, refunded on acceptance),
and pay is measured on fresh runs, which are not lucky by selection~\cite{cox1975datasplitting}.

\begin{proposition}[Fresh runs and a fee]\label{prop:fee}
A candidate is accepted if $D_1\ge\delta$, with $\delta\ge0$, where $D_1$ is the difference of the
means of $m$ runs of the candidate and $m$ runs of the current best. If accepted, it is paid
$\lambda\max\{0,D_2\}$, where $D_2$ is the difference of the means of $m'$ fresh runs of the
candidate and of the same reference, independent of $D_1$. Every candidate pays a fee $\gamma\ge0$
into the pot. Let $\bar s_1=\bar\sigma\sqrt{2/m}$, $\bar s_2=\bar\sigma\sqrt{2/m'}$,
$p_0=1-N(\delta/\bar s_1)$ and $b=p_0\,\lambda\,\bar s_2/\sqrt{2\pi}$.
\begin{enumerate}
\item[(a)] \textbf{Luck is bounded per attempt.} Under Assumption~\ref{ass:noise}, a candidate whose
  true gain against the current best is at most $0$ is accepted with probability at most $p_0$ and
  paid at most $b$ in expectation per attempt, before the fee.
\item[(b)] \textbf{A fee covers it.} Under Assumption~\ref{ass:noise}, let $\gamma\ge b$, or
  $\gamma\ge b/(1-p_0)$ if the fee is refunded on acceptance. Then a campaign of at most $K$
  candidates, $K$ fixed in advance, each with true gain at most $0$ against the current best at the
  time it is submitted, under any identities and deciding from what it has seen whether to try
  again, earns at most $0$ in expectation, net of fees. With $\gamma>b$, or $\gamma>b/(1-p_0)$ if
  refunded, every attempt made loses in expectation.
\item[(c)] \textbf{Splitting does not pay.} Assume (N1), (N2) and (N4), and let all runs, of the
  pieces and of their references, have one standard deviation $\sigma$. Let $A(x)$ be the acceptance
  probability of a piece with gain $x$, $G_2(x)=\mathbb{E}[\max\{0,D_2\}]$ its expected measured
  gain, and $b^*=\sup_{x\ge0}\lambda\,A(x)\,(G_2(x)-x)$. If $\gamma\ge2b^*$ and the fee is not
  refunded, then in the \emph{splitting experiment}, which cuts a candidate of true gain $\mu\ge0$
  into $k$ pieces with fixed non-negative gains summing to $\mu$, each tried once on fresh runs and
  dropped if rejected, splitting never raises its expected pay net of fees.
\item[(d)] \textbf{Copies do not change honest pay.} Let runs be independent, with any noise law;
  let verifiers be honest, each check draw fresh runs of its reference, and the rule have no cap and
  no deadline. Fix the other candidates, their order and their true gains. Add useless candidates
  that are copies of the current best, whose runs have its law. Then, with or without a fee, the
  acceptances and payments of the other candidates have the same joint law with or without the
  copies.
\end{enumerate}
\end{proposition}

\noindent Part (b) is the fee principle, Lemma~\ref{lem:fee} in Appendix~\ref{app:noiseproofs},
with $\beta=b$ from (a). The lemma needs only a bound on
what a useless candidate is paid; normality and fresh runs set the size of that bound. A refunded
fee must be higher, since it is kept only from rejected candidates. Part (c) does not cover pieces
that interact, or are timed against others' steps (Proposition~\ref{prop:sandwich}). Fees go into the pot, so on
average they put back at least what luck takes out. Private prediction markets charge such a fee
against bets on noise~\cite{frongillo2018bounded}. Figure~\ref{fig:noise} shows the mechanism and
the remedy; proofs are in Appendix~\ref{app:noise}.

\begin{figure}[t]
\centering
\begin{tikzpicture}
\begin{axis}[name=left, width=0.55\textwidth, height=4.6cm, font=\scriptsize,
  xlabel={measured gain $D$ (mbpb), one run a side}, ylabel={density},
  xmin=-170, xmax=240, ymin=0, ymax=0.0118, clip=false, ytick=\empty, axis lines=left,
  domain=-170:240, samples=120]
  \addplot[name path=f0, thick, black!70] {exp(-0.5*(x/53.6)^2)/(53.6*sqrt(2*pi))};
  \node[black!70, font=\scriptsize, anchor=east] at (axis cs:-58,0.0062) {useless, $\mu=0$};
  \addplot[name path=f1, thick, blue!60!black] {exp(-0.5*((x-60)/53.6)^2)/(53.6*sqrt(2*pi))};
  \node[blue!60!black, font=\scriptsize, anchor=west] at (axis cs:112,0.0055) {real, $\mu=60$};
  \path[name path=axis] (axis cs:0,0) -- (axis cs:240,0);
  \addplot[red!60, fill opacity=0.35] fill between[of=f0 and axis, soft clip={domain=0:240}];
  \draw[red!70!black, dashed, thick] (axis cs:0,0) -- (axis cs:0,0.0082);
  \node[red!70!black, font=\scriptsize, anchor=south east, align=right] at (axis cs:-3,0.0083) {accept if $D\ge0$};
  \node[red!70!black, font=\scriptsize, anchor=north west, align=left] at (axis cs:8,0.0117) {half the useless get in;\\ $21.4$~mbpb per attempt\\ on average};
\end{axis}
\begin{axis}[at={(left.east)}, anchor=west, xshift=2.6cm, width=0.38\textwidth, height=4.6cm, font=\scriptsize,
  ybar, bar width=14pt, scaled y ticks=false, ylabel={luck-fisher's mean take (\$)}, ymin=-900, ymax=18500,
  xmin=0.4, xmax=3.6, xtick={1,2,3},
  xticklabels={{deciding\\runs},{fresh\\runs},{fresh +\\fee $b$}},
  x tick label style={align=center, font=\scriptsize}, axis lines=left,
  nodes near coords, nodes near coords style={font=\scriptsize, /pgf/number format/fixed, /pgf/number format/1000 sep={{,}}},
  ytick={0,5000,10000,15000}, yticklabels={0,{5,000},{10,000},{15,000}}]
  \addplot[fill=red!40, draw=red!70!black] coordinates {(1,16707) (2,6172) (3,-5)};
\end{axis}
\end{tikzpicture}
\caption{Why a noisy scorer pays for luck, and what removes it. Left: the measured gain of a candidate that changes nothing is noise around zero (grey), so a one-run check at zero margin accepts half of them and pays $21.4$~mbpb per attempt on average, counting the rejected half as zero ($42.8$~mbpb per accepted one) (Proposition~\ref{prop:deciding}(a), at the pooled $\sigma=37.9$~mbpb); the real candidate (blue) overlaps it, so no never-negative payment can pay one and not the other (Proposition~\ref{prop:nonneg}). Right: in the simulated challenge of Section~\ref{sec:noise}, a luck-fisher with 100 useless candidates takes \$16,707 when paid on the runs that decided acceptance, \$6,172 when paid on fresh runs, and about zero once every candidate pays the fee $b=\$61.78$ (Proposition~\ref{prop:fee}; Appendix~\ref{app:noisyscorer}).}
\label{fig:noise}
\end{figure}
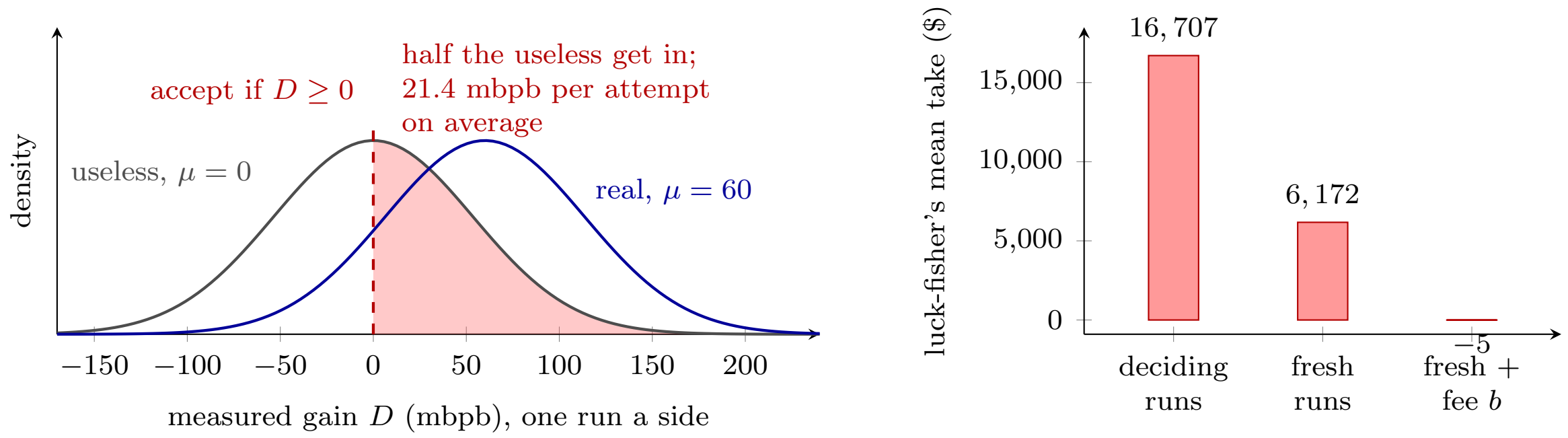
 
\paragraph{In simulation, a fee of $b$ took a luck-fisher's profit to about zero.} Our simulated
check uses three runs a side and a margin of $30.9$~mbpb, at $\sigma=37.9$~mbpb, the pooled noise of
our recipes, and a price of $\$31.47$ per mbpb. A useless candidate is then accepted with
probability $p_0=0.159$, and $b=\$61.78$ ($\$73.46$ if refunded). Paying on the deciding runs would
need a fee of $\$235.99$ at the same price, 3.8 times more. Each simulated challenge has 40 honest candidates from 20
identities and a budget of $\$20{,}000$. A luck-fisher adds 100 copies of the current best, at the
best of eight tested timings and identity spreads. Paid on fresh runs with no fee, it took
$\$6{,}172$ on average. With a fee of $b$ it netted $-\$5.48$, within two standard errors of zero. Copies of the current
best attain $b$ exactly, so under normal noise the fees paid equal the expected take by
construction: the near-zero net checks the closed form, not independent evidence.
The same fee does not save the rank rules, where luck is worth far more per attempt
(Appendix~\ref{app:noisyscorer}). With each run's noise resampled from our measured runs, luck paid
$3.2\%$ more than $b$ per attempt at $\sigma=16.26$ and $2.0\%$ less at $\sigma=37.9$. At a fee of
$1.5\,b$ the luck-fisher lost money at both levels. Splitting needs $2b^*=\$126.89$, about $2.05\,b$.
At a fee of $b$, splitting can pay when the pay is measured on far fewer runs than acceptance
(Appendix~\ref{app:noise}); with our parameters it did not. We take the fee that is not refunded
as the default, because part (c) is proved only for it. It charges honest work: each honest
identity paid \$123.55 in fees on \$1,005 of pay. The refunded variant, at \$73.46, charges only
rejected candidates, and we have no splitting result for it.

\paragraph{Open.} Nothing enforces (N3), and the operator controls its code's noise: useless
candidates twice as noisy netted \$10,036 against a fee set for the recipe. Our own recipes' runs
have standard deviations from $11.1$ to $111.9$~mbpb (Appendix~\ref{app:interactionreal}). At
$\bar\sigma=111.9$, with the design above, $b$ would be about \$422, and an honest identity's two
fees about \$843, some $84\%$ of its mean pay of \$1{,}005 with the stream and price unchanged. So
(N3) is the binding assumption, and not only against useless candidates: with the design above, a
candidate with a true gain of $20$~mbpb expects $42\%$ more pay if its own runs are twice as noisy,
and one of $60$~mbpb $9\%$ less (closed form, as in the proof of part (a)). Operators whose gains
are below or near the margin want noise. Normality (N1) is an approximation too: the standardised
residuals of our two pools of measured runs have excess kurtosis $-1.12$ and $0.53$
(Appendix~\ref{app:noisyscorer}). Honest work pays the fee as well: at
a fee of $1.5\,b$, real candidates below $4.7$~mbpb lose money at $\sigma=37.9$. And the guarantee
holds in expectation: with no hard cap, and a price set to spend the budget on average, the pot
ended below zero in about $35\%$ of simulated challenges at $\sigma=37.9$ and $42\%$ at
$\sigma=16.26$ (Appendix~\ref{app:noisyscorer}).
Appendix~\ref{app:noise} gives the measurements, the proofs and campaigns outside
Proposition~\ref{prop:fee}. Appendix~\ref{app:noisyscorer} gives the 192 check designs, a
$q$-of-$v$ vote and a median gain (simulated only), the luck-fisher under every rule, and why we
chose a fee over a deposit.
\section{Theory III: paying for enabling}\label{sec:enabling}

The per-step rule pays a step only for its own gain. Take a change $c$ worth 1~mbpb alone. Three
later changes by others are worth 1 each without $c$ and 4 each with it. Of the 12~mbpb that follow
$c$, 9 exist because of $c$. At $\lambda=\$100$ per mbpb, the owner of $c$ gets \$100 and nothing
for enabling. We test two ways to pay for it.

\subsection{Paying for being named}\label{sec:impossibility}

\paragraph{The ancestry model.} Each step $k$ names a set $\mathrm{Pa}_k$ of earlier steps as its
\emph{parents}, chosen by its operator. A rule $\pi$ maps the record and the named parents to each
identity's total payment $\pi_i$, an expectation if the rule is randomised
(Appendix~\ref{app:ancestry} gives the conventions). Two things are hidden from the rule:
\emph{ownership}, which identities belong to one operator, and \emph{true ancestry}, the set $T_k$
of earlier steps that step $k$ built on.
\begin{itemize}\setlength{\itemsep}{1pt}
\item[(D)] \textbf{Domain.} For every record, every ownership partition and every choice of the sets
  $T_k$ is possible.
\end{itemize}
Fix a record, an earlier step $j$, a later step $k$ (the \emph{follower}) of identities
$i_j\ne i_k$, and a set
$Q\not\ni j$ of other parents named by $k$. Let $\Delta_j$ and $\Delta_k$ be the changes in the
payments of $i_j$ and $i_k$ when $k$ names $Q\cup\{j\}$ instead of $Q$. Lying may cost extra: at
most $\ell_{\mathrm{hide}}\ge0$ for hiding a true parent (an expected penalty for detected reuse),
and $\ell_{\mathrm{fake}}\ge0$ for naming a false one (a forfeited stake per claim). Costs paid
either way cancel.
\begin{itemize}\setlength{\itemsep}{1pt}
\item[(NH)] \textbf{No hiding.} If the operator of $i_k$ controls no other identity and $j\in T_k$,
  then $\Delta_k\ge-\ell_{\mathrm{hide}}$.
\item[(NF)] \textbf{No faking.} If one operator controls exactly $i_j$ and $i_k$ and $j\notin T_k$,
  then $\Delta_j+\Delta_k\le\ell_{\mathrm{fake}}$.
\end{itemize}

\begin{theorem}\label{thm:limit}
Assume (D), or only its instance: the record is consistent both with $i_k$'s operator controlling
no other identity and $j\in T_k$, and with one operator controlling exactly $i_j$ and $i_k$ with $j\notin T_k$. Then
every rule satisfying (NH) and (NF) has $\Delta_j\le\ell_{\mathrm{hide}}+\ell_{\mathrm{fake}}$. In
particular, if lying costs nothing extra, being named by a later step of another identity never
raises a step's payment.
\end{theorem}
\noindent In words: call being named as a parent by a later step of another identity \emph{credit}.
If honest credit to another operator and fake credit to one's own second identity look the same, a
rule that neither rewards fakes (NF) nor punishes honest credit (NH) pays for credit at most what
lying costs.

\noindent\emph{Proof sketch.} The rule sees one record in two worlds: in one, $i_k$'s operator is
alone and $j\in T_k$, so (NH) gives $\Delta_k\ge-\ell_{\mathrm{hide}}$; in the other, one operator
holds exactly $i_j$ and $i_k$ and $j\notin T_k$, so (NF) gives $\Delta_j+\Delta_k\le\ell_{\mathrm{fake}}$.
$\Delta_j$ and $\Delta_k$ are the same in both, which gives the bound (Appendix~\ref{app:royaltyproofs}).

The bound is attained by a rule that moves $\ell_{\mathrm{hide}}$ from the follower to each named
step and adds $\ell_{\mathrm{fake}}$ from the sponsor (Appendix~\ref{app:ancestry}). Four ways
around it remain: penalise detected reuse, stake each claimed parent, certify ancestry, or pay for
something observable instead.

\subsection{Paying for a measured interaction}\label{sec:royalty}\label{sec:royaltytheory}

A \emph{royalty} pays an earlier change for what a later change draws from it. Verifiers measure
this by \emph{ablation}: they re-score the current best with one or two changes removed
(Figure~\ref{fig:ablation}). Nothing is declared. The \emph{change model} has five parts
(Appendix~\ref{app:royaltymodel}).
\begin{itemize}\setlength{\itemsep}{1pt}\setlength{\parskip}{0pt}
\item \textbf{Additive records.} Steps add changes and remove none; the operator chooses the cut.
\item \textbf{Pairwise landscape.} Progress is the sum of own gains $g(c)$ and one
  \emph{interaction} $h(c,d)$ per pair; no three-way terms.
\item \textbf{Full separability, exact measurement.} Every subset of changes builds and scores
  exactly, so Figure~\ref{fig:ablation} gives $h$.
\item \textbf{Royalty.} A step earns $\lambda>0$ times its gain. For each pair with $c$ added before
  $d$ by a different step, $c$'s owner receives $\lambda\varphi(h(c,d))$, $\varphi(0)=0$, paid by
  $d$'s owner, the follower (\emph{follower-financed}), or by the sponsor (\emph{sponsor-financed}).
  \emph{Clamped}: $\varphi(h)=\alpha\max\{0,h\}$; \emph{signed}: $\varphi(h)=\alpha h$;
  $\alpha\in(0,1]$. The zero royalty is the per-step rule.
\item \textbf{Settlement.} Once, at the end, with no window and no cap. Net pay can be negative, and
  a stake covers it: we assume it here; Section~\ref{sec:noise} explains why a deposit is costly,
  and we leave the gap open.
\end{itemize}
In the opening example, at $\alpha=\tfrac12$, each follower passes \$150 to the owner of $c$, who
receives \$550 in all. Each follower keeps \$250.
\begin{figure}[H]
\centering
\begin{tikzpicture}[x=1cm,y=1cm,
  cell/.style={draw,minimum width=2.6cm,minimum height=1.1cm,align=center,font=\small},
  lab/.style={font=\small}]
  \node[lab] at (1.3,2.9) {with $d$};
  \node[lab] at (4.0,2.9) {without $d$};
  \node[lab,anchor=east] at (-0.3,1.75) {with $c$};
  \node[lab,anchor=east] at (-0.3,0.45) {without $c$};
  \node[cell,fill=black!6] at (1.3,1.75) {$p(S)$\\$=5$};
  \node[cell] at (4.0,1.75) {$p(S\setminus\{d\})$\\$=1$};
  \node[cell] at (1.3,0.45) {$p(S\setminus\{c\})$\\$=1$};
  \node[cell] at (4.0,0.45) {$p(S\setminus\{c,d\})$\\$=0$};
  \node[lab,align=left,anchor=north west] at (6.0,2.4)
    {$\hat h(c,d)=5-1-1+0=3$~mbpb\\[2pt]
     $d$ owes $c$: $\alpha\lambda\hat h=\tfrac12\cdot100\cdot3=\$150$\\[2pt]
     Nothing is declared: the four cells\\are four runs of the scorer.};
\end{tikzpicture}
\caption{How the royalty measures an interaction. Take a current best $S$ made of two changes, $c$
and $d$, and score four versions of it: with both, without $d$, without $c$, and without either (the
baseline, scored 0). Here $c$ alone adds 1~mbpb, $d$ alone adds 1, and both together add 5. The
interaction is $5-1-1+0=3$~mbpb: what $c$ and $d$ give together beyond what each gives alone. If $c$
and $d$ were added by different steps, $c$ first, the owner of $d$ pays the owner of $c$ the royalty
of Definition~\ref{def:pairroyalty}: at rate $\alpha=\tfrac12$ and price $\lambda=\$100$ per mbpb,
$\tfrac12\cdot100\cdot3=\$150$, under the clamped and the signed royalty alike, since the
interaction is positive. The four values are runs of
the scorer, not anyone's claim about who built on whom.}
\label{fig:ablation}
\end{figure}
 
\paragraph{Results.} All rest on one computation (Lemma~\ref{lem:payoff}). Under follower
financing, an operator earns $\lambda$ times the progress of its own changes, plus $\lambda\varphi(h)$
on each interaction with others where its change came first, and $\lambda(h-\varphi(h))$ where it
came second (proofs: Appendix~\ref{app:royaltyproofs}).
\begin{itemize}\setlength{\itemsep}{2pt}
\item \textbf{Packaging} (Theorem~\ref{thm:royalty}). Under follower financing, with any $\varphi$,
  the sponsor pays $\lambda x_n$. Regrouping one's changes into other accepted steps or identities,
  keeping which come before or after each change of others, leaves the payoff unchanged: extra
  identities, splits, merges and complementary pieces gain nothing. A later step of others never
  lowers a payoff iff $\varphi\ge0$, so the clamped royalty is non-expropriating and the signed one
  is not. Under sponsor financing, every $\varphi\not\equiv0$ pays for splitting or merging one's
  own changes.
\item \textbf{The sandwich} (Proposition~\ref{prop:sandwich}). A \emph{refinement} replaces a change
  by pieces that together do the same; it is \emph{sign-preserving} if each piece's interaction with
  every other change has the original's sign or is zero, and \emph{sign-changing} otherwise. A
  \emph{timing} places the pieces in the record of others, fixed and known. A \emph{sandwich} cuts a
  change that conflicts with an earlier change of others ($h=-u$) and complements a later one
  ($h=+w$), placing the conflicting piece first and the complementing piece last; it is
  sign-preserving. If direct gains are large enough for every step to be accepted, it gains
  $\lambda\min\{u,w\}$ under the per-step rule, paid by another owner: with $\lambda=1$,
  $g(c)=5$, $u=3$ and $w=2$, the whole change earns at most 5, the sandwich 7.
\item \textbf{Exactly one royalty resists it} (Theorem~\ref{thm:shapley}). Under follower financing,
  with the accepted changes of others fixed and known, three statements are equivalent: no
  sign-changing refinement with any timing beats the best timing of the unrefined change; every
  payoff is independent of the order of steps; $\varphi(h)=h/2$. This royalty pays each operator
  $\lambda$ times the Shapley values~\cite{shapley1953value} of its changes,
  $g(c)+\tfrac12\sum_{d\ne c}h(c,d)$, and resists every refinement, with any timing.
\item \textbf{A trade-off} (Corollary~\ref{cor:tradeoff}). No follower-financed pairwise royalty,
  the per-step rule included, is both non-expropriating and refinement-proof on sign-changing
  refinements. The clamped royalty with $\alpha\ge\tfrac12$ resists sign-preserving refinements by an
  operator $o$ if some accepted record places all of $o$'s changes before every change of others.
  With $\alpha<\tfrac12$, or without such a record, sign-preserving refinements can gain.
\end{itemize}
Table~\ref{tab:results} collects the marks.

\begin{table}[tb]
\centering\footnotesize
\setlength{\tabcolsep}{3pt}
\resizebox{\textwidth}{!}{%
\begin{tabular}{lccccccc}
\hline
Rule & Non-exp. & Sybil & Split/merge & Packaging & Refinement & Credit & Total paid \\
\hline
\multicolumn{8}{l}{\emph{Baseline rules}} \\
WTA & \xmark\ A.a & \cmark & \cmark\ A.e & -- & -- & none & $B$ \\
Top-$k$, $k\ge2$, $P_2>0$ & \xmark\ A.f & \xmark\ A.d & \cmark\ A.e & -- & -- & none & $\le\sum P_j$ \\
Raffle, $\theta\le L$ & \xmark\ A.f$^\dagger$ & \xmark\ A.c & \cmark\ A.e & -- & -- & none & $P$ or $0$ \\
Bonus per step & \cmark\ A & \cmark\ A & \xmark\ A.b & -- & -- & none & unbounded \\
\hline
\multicolumn{8}{l}{\emph{The per-step rule, alone and with a follower-financed royalty (Section~\ref{sec:enabling})}} \\
\textbf{Per-step} ($\Phi$) & \cmark\ A & \cmark\ A & \cmark\ A (iff) & \cmark & \xmark\ P\ref{prop:sandwich} & none & $\Phi(x_n)-\Phi(s_0)$ \\
\textbf{Per-step + clamped royalty} & \cmark\ T\ref{thm:royalty} & \cmark & \cmark & \cmark\ T\ref{thm:royalty} & sign-pres.\ C\ref{cor:tradeoff} & $\alpha\lambda\sum h^+$ & $\lambda x_n$ \\
\textbf{Per-step + Shapley royalty} & \xmark\ T\ref{thm:royalty} & \cmark & \cmark & \cmark\ T\ref{thm:royalty} & \cmark\ T\ref{thm:shapley} & $\tfrac{\lambda}{2}\sum h$ & $\lambda x_n$ \\
\hline
\multicolumn{8}{l}{\emph{Other rules we analyse}} \\
Per-step + any royalty, sponsor-paid & iff $\varphi\ge0$ & \cmark & \xmark\ T\ref{thm:royalty} & \xmark\ T\ref{thm:royalty} & \xmark & $\lambda\sum\varphi(h)$ & $\lambda x_n+\lambda\sum\varphi$ \\
Declared ancestry, (NH)(NF), (D) & -- & -- & -- & -- & -- & $\le \ell_{\mathrm{hide}}+\ell_{\mathrm{fake}}$ T\ref{thm:limit} & -- \\
\hline
\end{tabular}}
\caption{Which rule has which property, and where it is proved. A alone = Proposition~\ref{prop:potential} (Section~\ref{sec:rule}, proof in
Appendix~\ref{app:score}); A.a--A.f = the parts of Proposition~\ref{prop:failures} (Appendix~\ref{app:score}).
T, P, C = theorem, proposition, corollary of this section and
Appendix~\ref{app:royaltyproofs}; a dash means not analysed. Score-model columns are for exact
scores. Change-model columns are for the idealised royalty (every pair measured, uncapped, no
window, settled once) on fully separable pairwise landscapes with exact measurement and additive
records, linear base $\Phi(x)=\lambda x$. The sponsor-paid row assumes $\varphi\not\equiv0$.
``iff'': among endpoint rules, exactly the per-step rules, when every split of
Definition~\ref{def:split} can be made. ``sign-pres.'': sign-preserving refinements only,
$\alpha\ge\tfrac12$, operator able to place all its changes first. $L=s^*-s_0$ is the length of the
score interval; $^\dagger$if $L\ge2\theta$; $h^+=\max\{0,h\}$; (D) is the domain assumption of
Section~\ref{sec:impossibility}. Regimes, witnesses and the remaining marks:
Appendix~\ref{app:notes}. The noisy-scorer guarantees are not in the table; they are in
Section~\ref{sec:noise}.}
\label{tab:results}
\end{table}

\paragraph{Copies defeat both royalties.} A follower can bundle a renamed copy of $c$ with its own
change. Removing $c$ leaves the copy, so the measured interaction is negative. Under the Shapley
royalty, the owner of $c$ then pays the follower half of $c$'s value. The clamped royalty pays
nothing for the conflict, but the copy also hides a real dependence on $c$
(Appendix~\ref{app:copyattack}). Without copy detection, neither royalty pays for enabling. Shapley
payment in data markets has the same weakness~\cite{agarwal2019marketplace,ohrimenko2019collaborative}.

\subsection{Can the interaction be measured?}\label{sec:measurable}

Every result above needs the four versions of Figure~\ref{fig:ablation} to exist and score. If $d$
calls $c$, or rewrites it in place, removing $c$ can leave nothing that runs.
Definition~\ref{def:pairroyalty} then says nothing about the pair. Paying nothing there can lose
non-expropriation: a later step of another operator that calls $c$ voids the ablation, and the owner
of $c$ loses its royalty. A follower can also void an ablation on purpose. On better.codes, no later
proof built without the earlier one in the 24 pairs we attempted (Section~\ref{sec:which}); Appendix~\ref{app:removalelsewhere} lists
where it worked. Where it works, an interaction is a second difference of four noisy means: it
needs many runs, the one test we fixed in advance failed, and noise biases the clamped royalty upward
but not the signed one (Appendix~\ref{app:interactionreal},
Proposition~\ref{prop:royaltynoise}). Counting a failed build as full dependence is a weak fallback
(Appendix~\ref{app:enablingroyalty}).
\section{Related work}\label{sec:related}

\paragraph{Open autoresearch defines the setting; we add the incentive layer.}
\citet{long2026ecdsafail} define \emph{open autoresearch}: humans and AI agents publish
evaluator-verified improvements to a public leaderboard, and each valid one becomes a shared base
for everyone. Our problem with its scorer and rules (Definition~\ref{def:problem}) is that loop. Their paper gives no payment rule;
it notes that ``evaluators and incentives therefore influence which discoveries are pursued and preserved''. Their ECDSA.Fail is one
of the challenges we place in Section~\ref{sec:which}. Gauntlet~\cite{lidin2025gauntlet} is a
deployed, permissionless system that pays peers for training a language model. It scores each
contribution by how much it lowers the loss and, since single scores are noisy, pays by a rating
built from their ranks. It names peer copying and registering many times as attacks. We study the
payment rule itself.

\paragraph{The per-step rule is a market scoring rule, proposed for machine-learning contests.} A
market scoring rule~\cite{hanson2003combinatorial} pays each report the difference of a fixed
function between the new report and the previous one, so the total depends only on the first and
last reports~\cite{hanson2007logarithmic}. \citet{abernethy2011optimization} prove that a market
maker whose price does not depend on how a trader splits a purchase prices by a cost function;
\citet{frongillo2018axiomatic} characterise such markets by axioms.
\citet{abernethy2011collaborative} turn a prediction market into a contest: each update to a public
model earns by how much it lowered the loss on test labels revealed at the end, against collateral.
In this design and its extensions~\cite{waggoner2015market,harris2019decentralized}, an update is
scored by its loss on the revealed labels.

\paragraph{False-name manipulations.} Our Sybil, split and merge manipulations are the false-name
manipulations of combinatorial auctions~\cite{yokoo2004falsename} and of reputation and accounting
mechanisms~\cite{cheng2005sybilproof,seuken2014sybilproof,stannat2021sybilproofness}.
\citet{ju2007nonmanipulable} characterise the division rules under which no group gains by
transferring its characteristics among its members; Proposition~\ref{prop:potential} is a statement
of that kind for rules that read a step's two end points. \citet{knudsen2012merging} give the
(im)possibilities for merging and splitting in cooperative games. Shapley sharing over identities is
not false-name-proof~\cite{yokoo2005coalitional,burnat2026quotient}. \citet{ohta2008anonymity}
repair it by computing a Shapley-based value over skills, which the designer knows in advance. Our
royalty makes the same move with changes (Theorem~\ref{thm:royalty}). Replicas defeat Shapley
payment in data markets; the remedies penalise similar inputs~\cite{agarwal2019marketplace} or
divide payments so that replicating does not pay~\cite{ohrimenko2019collaborative}.

\paragraph{Credit for enabling.} \citet{green1995division} ask how to divide profit between
successive innovators. \citet{banerjee2014reincentivizing} pay for sharing partial progress on a
known network of subtasks. \citet{kleinberg2022credit} show that misallocated credit can steer
project choice. Referral rewards~\cite{emek2011multilevel}, lottery and incentive
trees~\cite{douceur2007lottery,lv2016incentivetree} and propagation
rewards~\cite{babaioff2012redballoons} face the same tension: a reward for enabling invites fake
enablers. \citet{babaioff2012redballoons} use incentive inequalities over observationally identical
chains with different ownership, in propagation networks with clones; Theorem~\ref{thm:limit} uses
the same two-world method.

\paragraph{The fee's precedent.} The closest fee to ours is in private prediction markets, where
noise is added to the published prices and a trader can bet against it.
\citet{waggoner2015market} note that a small fee deters such bets. \citet{frongillo2018bounded}
choose a fee per trade that exactly balances the expected profit from them; covering it in
expectation is enough ``because a trader must pay the fee to trade before the random noise is
drawn''. Our fee applies the same principle to noise in measurement: it is paid before the runs and
sized to what luck pays per attempt (Proposition~\ref{prop:fee}).

Volunteer computing, contests, trustworthy measurement,
charging per use, blockchain incentives, personhood credentials and audit contests are in
Appendix~\ref{app:relatedextra}.
\section{Conclusion}\label{sec:limits}\label{sec:discussion}

\paragraph{The open problem: enabling where changes cannot be removed.} The royalty of
Section~\ref{sec:royalty} pays for enabling only where each change can be removed and the rest
scored. When a later change calls or rewrites an earlier one, a version without the earlier change
may not run, or may not exist until someone writes it; better.codes is such a case
(Section~\ref{sec:which}). The royalty is not defined there. Values for cooperative games with precedence constraints or permission structures fit
this case, but they take the dependence as known (Appendix~\ref{app:enablingroyalty}). Declared
ancestry does not help (Theorem~\ref{thm:limit}).
A rule that pays for such enabling and resists manipulation is open.

\paragraph{Five limits.} First, the fee needs honest verifiers and a bound $\bar\sigma$ on every
run's noise, and we do not say how to enforce either; a candidate noisier than planned, or whose
randomness its operator controls, falls outside its guarantee. Second, under noise the rule does
not cap the total paid: the fee removes the profit from luck only in expectation, per attempt, for
campaigns bounded in advance, so a fixed pot can end below zero. Scaling payments down to the pot
keeps the fee result but loses non-expropriation (Appendix~\ref{app:noisyscorer}). Third, a pending
candidate can be copied if the log is public and a later submission is processed first; we do not
model the order. Fourth, nobody is paid for the runs: with one verifier an accepted candidate costs
$2m+2m'$ runs, twelve in our simulation, and a rejected one $2m$. Fifth, our evidence is narrow: four
public records, the published account of a fifth, and training runs of one small model on one
machine. The replays show what a rule would have paid, not what anyone did.

\paragraph{What stays with people.} The model takes the verifiers as honest, the eligibility rules
as given, the noise bound as planned and the price curve as fixed (paying equal relative
improvements equally forces a fixed price per unit of log score, Remark~\ref{rem:pricecurve}). Each is a governance choice that
a sponsor makes outside the rule, and each moves who is paid: the price curve alone moves the
first ten speedrun steps from 85\% of a budget to 43\% (Section~\ref{sec:which}), and a fee weighs
most on small contributors (Section~\ref{sec:noise}). The rule
removes the pay for identities, copies and luck; it does not decide what progress is worth.
 
\section*{Use of generative AI}
The research question, the model, the direction of the work and the claims are the authors'.
We used AI assistants to draft and edit the text and proofs, to search the literature, and to write
and run the code of the simulations. The authors are responsible for the content.

{\small\setlength{\bibsep}{2pt plus 1pt}
\bibliographystyle{plainnat}
}

\appendix
\section{Proofs and baseline rules for Section~\ref{sec:rule}}\label{app:score}

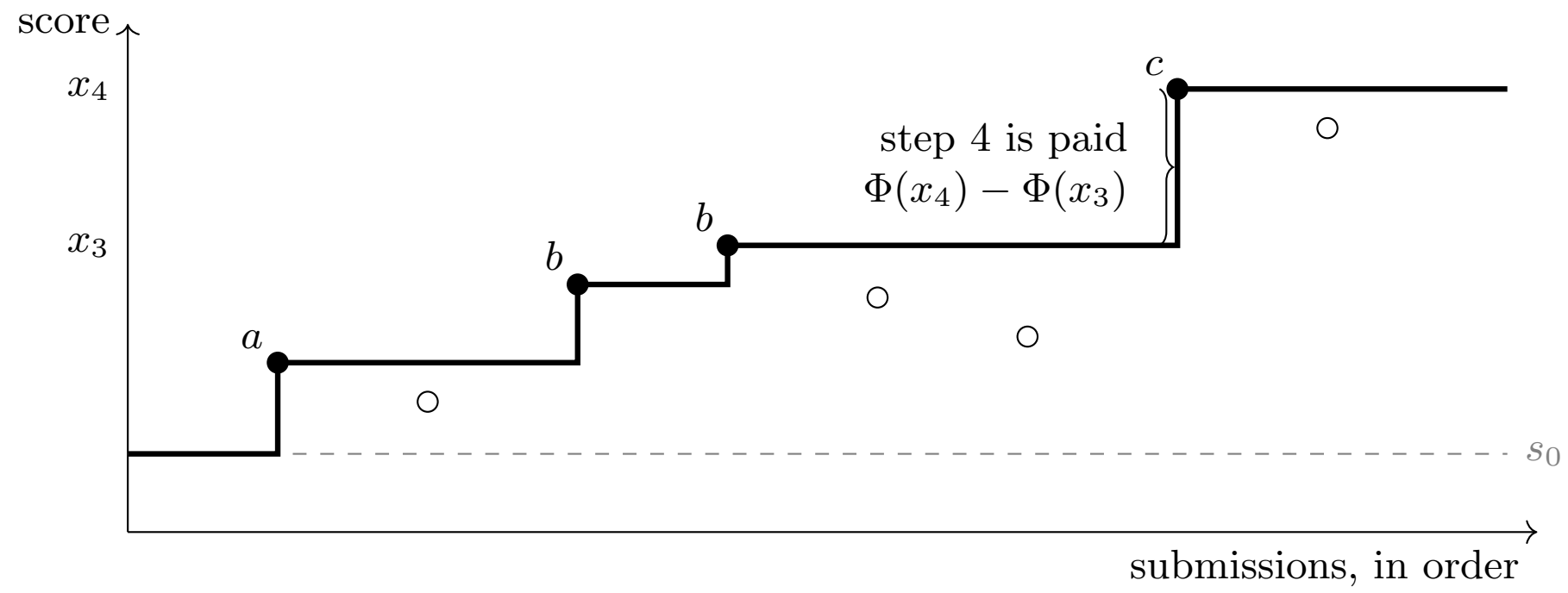
\begin{figure}[t]
\centering
\begin{tikzpicture}[x=1.15cm,y=1.0cm,font=\footnotesize]
  \draw[->] (0,0) -- (9.4,0) node[below left] {submissions, in order};
  \draw[->] (0,0) -- (0,3.9) node[left] {score};
  \draw[dashed,black!50] (0,0.6) -- (9.2,0.6) node[right] {$s_0$};
  \draw[very thick] (0,0.6) -- (1,0.6) -- (1,1.3) -- (3,1.3) -- (3,1.9) -- (4,1.9) -- (4,2.2) -- (7,2.2)
    -- (7,3.4) -- (9.2,3.4);
  \foreach \x/\y/\id in {1/1.3/a, 3/1.9/b, 4/2.2/b, 7/3.4/c}
    \filldraw (\x,\y) circle (2.2pt) node[above left=-1pt] {$\id$};
  \foreach \x/\y in {2/1.0, 5/1.8, 6/1.5, 8/3.1}
    \draw[fill=white] (\x,\y) circle (2.2pt);
  \draw[decorate,decoration={brace,amplitude=3pt,mirror}] (6.88,2.2) -- (6.88,3.4)
    node[midway,left=3pt,align=right] {step 4 is paid\\ $\Phi(x_4)-\Phi(x_3)$};
  \node[left] at (0,2.2) {$x_3$};
  \node[left] at (0,3.4) {$x_4$};
\end{tikzpicture}
\caption{A record with an exact scorer. Candidates arrive in order; a filled dot is accepted, an open dot rejected. Each accepted step raises the frontier (thick line) and is recorded with its identity ($a$, $b$, $c$). Identity $b$ made two steps; one operator may control several identities. Under the per-step rule of Section~\ref{sec:rule}, step $k$ is paid $\Phi(x_k)-\Phi(x_{k-1})$: its start point is the frontier when it is accepted, set by the record.}
\label{fig:record}
\end{figure}
 
\paragraph{Where the rule comes from.} The per-step rule is not new in form: market scoring rules pay
each forecast this way~\cite{hanson2003combinatorial}, and \citet{abernethy2011collaborative}
brought it to machine-learning contests. Proposition~\ref{prop:potential}(a) is the analogue, for
endpoint rules, of Theorem~1 of \citet{abernethy2011optimization}: a market maker whose price for a
bundle does not depend on how a trader splits the purchase prices by a cost function.

\paragraph{Divisibility.} Assumption~\ref{ass:divisible} is what makes splitting possible. It holds when an improvement can
be weakened by any amount at no cost, for example a proved bound that can be stated weaker by any
amount. A bundle of independent changes allows only some splits, at the scores of its subsets: with
gains $1$ and $3$, the scores $1$ and $3$, but not $2$. It fails when a weaker version costs work, as
on a training problem, where each version needs its own runs. The guarantees for the per-step rule
hold for every split, possible or not, so they do not need it. Only the failures of other rules
that use a split do (Proposition~\ref{prop:failures}(b), (c) and (d), their proofs below, and the matching entries of Table~\ref{tab:results}), and so does the
replay of Section~\ref{sec:evidence}, which cuts one step in two at the midpoint of its gain; where
such cuts cannot be made, those rules are harder to game. The replay's first and second prize needs
no cut, since it only moves existing steps to a second identity, and Sections~\ref{sec:noise}
and~\ref{sec:enabling} do not use the assumption.
The ``only'' direction of Proposition~\ref{prop:potential}(a) is a statement about rules that resist
every split; it uses splits at every intermediate score, which is Assumption~\ref{ass:divisible} in
substance. Where fewer splits can be made, a rule need resist fewer, and more rules do.

\subsection{Proofs of Proposition~\ref{prop:potential} and Remark~\ref{rem:threshold}}

\begin{proof}[Proof of Proposition~\ref{prop:potential}]
(a) Take a record $H$ with an entry $k$ owned by $i$, and $y\in(x_{k-1},x_k)$. Splitting at $(k,y)$
replaces the term $r(x_{k-1},x_k)$ in $\pi_i$ by $r(x_{k-1},y)+r(y,x_k)$. No other term of any
identity's payment changes, because the end points of all other entries are unchanged. So the payoff
of the operator $o$ controlling $i$ changes by exactly $r(x_{k-1},y)+r(y,x_k)-r(x_{k-1},x_k)$.
Every triple $y<z<w$ in $\mathcal{X}$ arises in this way: take the record $\big((i,y),(i,w)\big)$ if $y>s_0$, or
$\big((i,w)\big)$ if $y=s_0$, and split the entry ending at $w$ at $z$. Hence split-proofness and
merge-proofness together are equivalent to
\begin{equation}\label{eq:additive}
r(y,w)=r(y,z)+r(z,w)\qquad\text{for all } s_0\le y<z<w \text{ in } \mathcal{X} .
\end{equation}
If $r(y,z)=\Phi(z)-\Phi(y)$, then \eqref{eq:additive} holds by telescoping; this direction uses no
particular record, so it holds for every split, possible or not. Conversely, assume
\eqref{eq:additive}. Define $\Phi(s_0)=0$ and $\Phi(x)=r(s_0,x)$ for $x>s_0$. For $y=s_0<z$ we have
$r(y,z)=\Phi(z)-\Phi(y)$ by definition. For $s_0<y<z$, applying \eqref{eq:additive} to the triple
$s_0<y<z$ gives $r(s_0,z)=r(s_0,y)+r(y,z)$, that is, $r(y,z)=\Phi(z)-\Phi(y)$. Finally, $r\ge0$
gives $\Phi(z)\ge\Phi(y)$ whenever $y<z$.

(b) The total payment is $\sum_{k=1}^n(\Phi(x_k)-\Phi(x_{k-1}))=\Phi(x_n)-\Phi(x_0)$. Every
$x\in \mathcal{X}$ with $x>s_0$ is the frontier of the record $\big((i,x)\big)$, and $\Phi$ is non-decreasing,
so the supremum of the total payment over $\Hist$ is $\sup_{x\in \mathcal{X}}\Phi(x)-\Phi(s_0)$.

(c) For an endpoint rule, $U_o(H)=\sum_{k:\,i_k\in\Id_o}r(x_{k-1},x_k)$. A relabelling $\rho_{K,i'}$ with
$i'\in\Id_o$ and $K$ a set of entries owned by $\Id_o$ changes no end point and keeps every entry of
$K$ under an identity of $o$. So the set of entries counted in $U_o$ and their rewards are unchanged,
and $U_o(\rho_{K,i'}(H))=U_o(H)$. An extension $H'$ of $H$ by others of $o$ leaves the first $n$
entries, and therefore their rewards, unchanged. It adds only entries owned outside $\Id_o$, which
do not count in $U_o$. Hence $U_o(H')=U_o(H)$. Both are equalities, so Sybil-proofness and
non-expropriation hold.

By (c), every endpoint rule is Sybil-proof and non-expropriating. So an endpoint rule has all
four properties if and only if it is split-proof and merge-proof, which by (a) holds if and only if
it is a per-step rule.
\end{proof}

\begin{remark}[A principle for the price curve]\label{rem:pricecurve}
Proposition~\ref{prop:potential} leaves $\Phi$ free. One principle fixes it up to a rate. Let the
score be the logarithm of the quantity improved (minus the logarithm, for a cost), so that a step's
gain $z-y$ is its relative improvement, and let $L=s^*-s_0$, with $0<L\le\infty$. Suppose a per-step rule pays
equal relative improvements equally: $\Phi(z)-\Phi(y)=\rho(z-y)$ for all $y<z$ in $\mathcal{X}$,
for some function $\rho$. Then $\Phi(x)=\Phi(s_0)+\lambda\,(x-s_0)$ for a rate $\lambda\ge0$: a
fixed price per unit of log score. Indeed, for $u,w>0$ with $s_0+u+w\in\mathcal{X}$, the points $s_0$, $s_0+u$
and $s_0+u+w$ give $\rho(u+w)=\rho(u)+\rho(w)$, and $\rho(u)=\Phi(s_0+u)-\Phi(s_0)$ is
non-decreasing because $\Phi$ is; a non-decreasing function that is additive in this sense on
the gains $\{u>0:\ s_0+u\in\mathcal{X}\}$ is linear there. By Proposition~\ref{prop:potential}(b) the rule is then budget-feasible if and only if
$\lambda L\le B$ when $L<\infty$. So a positive rate needs a finite target, and $\lambda=B/L$
spends the budget exactly when the target is reached; with no target, only $\lambda=0$ is
budget-feasible. The principle is a
choice, not a consequence of our properties: it prices a late $1\%$ like an early $1\%$, whatever
each cost to find.
\end{remark}

\begin{remark}[A margin]\label{rem:threshold}
Let the margin be $\delta>0$, so that a step must gain at least $\delta$, and let a split be legal if
both pieces gain at least $\delta$. An endpoint rule resists every legal split and merge, on every
record the margin allows, if and only if $r(y,z)=\Phi(z)-\Phi(y)$ on every step such a record can
contain, for some $\Phi$ with $\Phi(s_0)=0$. Then $\Phi(z)\ge\Phi(y)$ whenever $z-y\ge\delta$, but
$\Phi$ need not be non-decreasing. Proposition~\ref{prop:potential}(b) holds with the supremum over
the frontiers the margin allows.
\end{remark}

\begin{proof}
A step on a record the margin allows starts at $s_0$ or at an earlier frontier, which is at least
$s_0+\delta$. So every such step $(y,z)$ has $z-y\ge\delta$ and either $y=s_0$ or $y\ge s_0+\delta$.
If $r(y,z)=\Phi(z)-\Phi(y)$ on these steps, a legal split of a legal step into legal pieces leaves the
total unchanged by telescoping. Conversely, define $\Phi(s_0)=0$ and $\Phi(x)=r(s_0,x)$ for
$x\ge s_0+\delta$. For a step $(y,z)$ with $y\ge s_0+\delta$, the record $\big((i,y),(i,z)\big)$ is
allowed, and it is a legal split of the allowed record $\big((i,z)\big)$. Resisting this split and its
merge gives $r(s_0,z)=r(s_0,y)+r(y,z)$, that is, $r(y,z)=\Phi(z)-\Phi(y)$. For $y=s_0$ this holds by
definition. Since $r\ge0$, $\Phi(z)\ge\Phi(y)$ on every allowed step, that is, whenever
$z-y\ge\delta$. Values of $\Phi$ on $(s_0,s_0+\delta)$ never enter a payment. The constraints still
bind steps that cannot themselves be split: for $\delta=1$, $r(0,2)=r(0,1)+r(1,2)$ fixes the step
$(1,2)$, of length in $[\delta,2\delta)$.
\end{proof}

\subsection{Baseline rules and their failures}

We compare the per-step rule with four \emph{baseline rules}: top-$k$ prizes and an equal-share
raffle, stylised from real schedules (QSB uses both; Appendix~\ref{app:rulesinuse}), and
winner-take-all and a fixed bonus per step, the simplest prize and the simplest payment per step.
Let $T_i(H)=\sum_{k:\,i_k=i}(x_k-x_{k-1})$ be the total gain owned by identity $i$.
\begin{itemize}\setlength{\itemsep}{0pt}\setlength{\parskip}{0pt}
\item \emph{Winner-take-all} (WTA): $\pi_{i_n}(H)=B$ if $n\ge1$, and all other payments are $0$.
\item \emph{Top-$k$ by total gain}: only \emph{active} identities, those with $T_i(H)>0$,
  are eligible. Prizes $P_1\ge P_2\ge\dots\ge P_k\ge 0$, with $P_1>0$, go to the active identities with the $k$
  largest values of $T_i(H)$. Ties are broken by any fixed rule. If fewer than $k$ identities are
  active, the remaining prizes are not paid.
\item \emph{Equal-share qualification} (a stylised raffle): a pool $P>0$ is split equally among the
  qualifying identities $Q(H)=\{i: T_i(H)\ge\theta\}$, for a threshold $\theta>0$. Nothing is paid if
  $Q(H)=\emptyset$.
\item \emph{Per-step bonus}: the endpoint rule $r(y,z)=\lambda\,(z-y)+\beta$ with $\lambda\ge0$ and $\beta>0$.
\end{itemize}

\begin{proposition}[Failures]\label{prop:failures}
Let scores be exact, and let $L=s^*-s_0\in(0,\infty]$ be the length of the score interval.
\begin{enumerate}
\item[(a)] WTA is expropriating. If entry $n$ of $H$ is owned by operator $o$, then any extension of
  $H$ by one entry of another operator lowers $U_o$ from $B$ to $0$.
\item[(b)] The per-step bonus is not split-proof. Each split raises the splitter's payoff by
  exactly $\beta$.
\item[(c)] Equal-share qualification is Sybil-proof if $\theta>L$, because then no identity ever
  qualifies and nothing is paid. It is not Sybil-proof if $\theta\le L$. Moreover, it admits the
  following \emph{splitting attack}, which needs $L\ge(c+j)\theta$: suppose operator $o$ holds all its
  steps under one identity, with total gain $T_o\ge j\theta$ for an integer $j\ge2$, and
  $c\ge1$ identities outside $\Id_o$ qualify. Then by splitting and relabelling, $o$ raises its
  payoff from $P/(c+1)$ to $jP/(c+j)$, which is strictly larger.
\item[(d)] Top-$k$ with $k\ge2$ and $P_2>0$ is not Sybil-proof. Specifically, suppose $o$ holds all
  its steps under one identity, and $T_o>2g$, where $g\ge0$ is the largest total gain of any
  identity outside $\Id_o$ (with $g=0$ if there is none). Then by splitting and relabelling, $o$ can
  hold the two largest totals, raising its payoff from $P_1$ to $P_1+P_2$. Such records exist whenever
  $L>0$.
\item[(e)] WTA, top-$k$ and equal-share qualification are split-proof and merge-proof.
\item[(f)] Top-$k$ is expropriating. Equal-share qualification is expropriating if
  $L\ge2\theta$.
\end{enumerate}
\end{proposition}

\paragraph{Four worked failures.}
\begin{itemize}\setlength{\itemsep}{1pt}\setlength{\parskip}{0pt}
\item \textbf{WTA and a copier.} Alice raises the frontier from 0 to 10 and holds $B=\$1{,}000$.
  Bob copies the frontier, adds $0.01$, and is accepted. Bob now holds $\$1{,}000$ and Alice $0$.
\item \textbf{Top-2 and a fake identity.} Prizes are $\$600$ and $\$400$. An operator's one identity
  has total gain 10; the only rival has 4. Moving half its gain to a second identity gives totals
  5 and 5, both above 4, so it takes $\$1{,}000$ instead of $\$600$.
\item \textbf{The raffle and a split.} A pool of $\$600$ is shared by identities with total gain at
  least 3. An operator with gain 6 and one qualifying rival gets $\$300$. Cut into two identities
  of gain 3 each, it gets two shares of three, $\$400$.
\item \textbf{The bonus and a split.} With $\lambda=\$100$ and $\beta=\$50$, a step from 0 to 2 pays
  $\$250$. Cut at 1, it pays $\$150+\$150=\$300$: each cut adds $\beta$.
\end{itemize}

\begin{proof}
(a) Under WTA, $U_o(H)=B$ because $i_n\in\Id_o$. In any extension $H'$ of length $n+1$, the last
entry is owned outside $\Id_o$, so $U_o(H')=0$.

(b) By the computation in the proof of Proposition~\ref{prop:potential}(a), splitting changes the
splitter's payoff by $r(y,z)+r(z,w)-r(y,w)=\beta$.

(c) If $\theta>L$, every $T_i(H)\le x_n-s_0\le L<\theta$, so $Q(H)=\emptyset$ and all payments are
$0$ on every record. In particular, no relabelling changes any payoff. If $\theta\le L$, take
$H=\big((i_1,s_0+\theta/2),(i_2,s_0+\theta)\big)$ with $i_1,i_2\in\Id_o$ and no other entries.
Both identities have total $\theta/2<\theta$, so $Q(H)=\emptyset$ and $U_o(H)=0$. Relabel both entries
to a fresh $i'\in\Id_o$: now $T_{i'}=\theta$, $Q=\{i'\}$ and $U_o=P>0$. This violates
Definition~\ref{def:sybil}.

For the splitting attack: a record with the stated properties exists only if
$L\ge T_o+c\theta\ge(c+j)\theta$, since the totals of distinct identities cover disjoint parts of
$(s_0,x_n]$. Given such a record, the steps of $o$ are entries $(x_{k-1},x_k]$ of total length
$T_o\ge j\theta$. By Definition~\ref{def:split}, $o$ can split its entries at any interior points. Do so: walk through the steps of $o$ in record order, accumulating length, and cut
each time the accumulated length reaches $\theta,2\theta,\dots,(j-1)\theta$. This yields $j$ groups of
consecutive pieces. The first $j-1$ groups each have total length exactly $\theta$, and the last has
total length $T_o-(j-1)\theta\ge\theta$. Relabel each group to its own identity of $o$, using $j-1$
fresh identities. Splitting and relabelling change no other operator's entries, so exactly $c$
identities outside $\Id_o$ qualify before and after. Before, $|Q|=c+1$ and $o$ receives
$P/(c+1)$. After, $|Q|=c+j$ and $o$ receives $jP/(c+j)$. Finally,
$jP/(c+j)>P/(c+1)$ if and only if $j(c+1)>c+j$, that is, $(j-1)c>0$, which holds.
Splitting alone changes no payment (the rule reads only the totals $T_i$), so at least one of the
$j-1$ relabellings strictly raises $U_o$, violating Definition~\ref{def:sybil}.

(d) Prizes go only to identities with $T_i>0$, so the other identities of
$o$, which own no steps, receive nothing. Before, $o$'s single active identity has the largest total
($T_o>2g\ge g$, and $T_o>g$ strictly), so $o$ receives exactly $P_1$. As in (c), split and relabel the
steps of $o$ into two identities, each with total gain $T_o/2>g$. These two identities now hold
the two largest totals, so $o$ receives $P_1+P_2$. The split alone changes no payment, so the single
relabelling is a strict Sybil gain of $P_2>0$. A record satisfying the hypothesis exists whenever $L>0$:
take $H=\big((i,s^\dagger)\big)$ with $i\in\Id_o$, $s^\dagger\in(s_0,s^*]$, and $g=0$.

(e) A split at $(k,y)$ inserts an entry owned by $i_k$ before entry $k$, and a merge removes one;
neither changes the owner of the last entry or any identity's total $T_i$, since the two pieces'
gains add up to the gain of the step they replace. WTA reads only the owner of the last entry, and
top-$k$ and equal-share qualification read only the totals $T_i$, so no payment changes.

(f) Top-$k$: pick $\varepsilon>0$ with $(k+1)\varepsilon<L$ and let $o$ own the record
$H=\big((i,s_0+\varepsilon)\big)$. Its identity is the only active one, so $U_o(H)=P_1>0$.
Extend $H$ by $k$ entries of other operators, under $k$ distinct identities, each raising the
frontier by more than $\varepsilon$; this fits, since $L>(k+1)\varepsilon$ leaves room for gains
slightly above $\varepsilon$. Each new identity's total exceeds $T_i=\varepsilon$, so the $k$ prizes go
to them and $U_o$ falls to $0$. Equal-share qualification: let $o$ own $H=\big((i,s_0+\theta)\big)$, so
$Q(H)=\{i\}$ and $U_o(H)=P$. Since $L\ge2\theta$, another operator can append a step to
$s_0+2\theta$ under a fresh identity, which then qualifies, and $U_o$ falls to $P/2$.
\end{proof}
\section{The enabling results: statements and proofs}\label{app:royaltyproofs}

\subsection{The ancestry model and Theorem~\ref{thm:limit}}\label{app:ancestry}

\paragraph{Conventions.} Records are as in the score model, with the named parents $\mathrm{Pa}_k$
added. A rule $\pi$ can be any map from records to the total payment $\pi_i$ of each identity $i$.
Payments made at different times are valued under a fixed convention. For a randomised rule, $\pi_i$
is the expected payment, assumed finite. Operators are risk-neutral and declare before the rule's
randomness is revealed. Ownership is a partition of identities into operators. Under (D), looking at
the record, the rule cannot exclude any ownership partition or any choice of true-parent sets; the
proof of Theorem~\ref{thm:limit} uses only the instance stated in the theorem.

\paragraph{Tightness.} Let each naming move $\ell_{\mathrm{hide}}$ from the follower to the named
step, and let the sponsor add $\ell_{\mathrm{fake}}$. Then $\Delta_k=-\ell_{\mathrm{hide}}$ and
$\Delta_j+\Delta_k=\ell_{\mathrm{fake}}$. The rule satisfies (NH) and (NF), and being named raises a
step's pay by exactly $\ell_{\mathrm{hide}}+\ell_{\mathrm{fake}}$.

\paragraph{Scope.} Theorem~\ref{thm:limit} assumes nothing else about the rule: not a per-step rule,
a balanced budget, monotone pay, or when payments are made. It bounds the \emph{increment} from being
named, not all payments to earlier steps.

\paragraph{Four ways around it.} Raise $\ell_{\mathrm{hide}}$ by detecting and penalising un-named
reuse. Raise $\ell_{\mathrm{fake}}$ by a forfeitable stake per claimed parent. Certify ancestry. Or
reward something observable instead. The royalty of Section~\ref{sec:royalty} does the last: it pays
for a measured interaction between changes and does not infer $T_k$. That removes the declaration,
but not every manipulation. Operators can still manufacture real interactions, which is why the
follower pays (Theorem~\ref{thm:royalty} and the example after it).

\subsection{The change model}\label{app:royaltymodel}

Fix a finite set $C$ of changes and a \emph{landscape} $p:2^C\to\R$ with
$p(\emptyset)=0$, the progress of a set of changes, assumed \emph{pairwise}:
\begin{equation}\label{eq:pairwise}
p(S)=\sum_{c\in S}g(c)+\sum_{\{c,d\}\subseteq S}h(c,d),
\end{equation}
for a direct gain $g:C\to\R$ and a symmetric interaction $h$ on pairs, with $h(c,c)=0$. Our results
need this assumption. A record is a sequence of steps $k=1,\dots,n$. Step $k$, submitted by identity
$i_k$, \emph{adds} a non-empty set $A_k\subseteq C$ of changes not yet present. We set
$S_k=A_1\cup\dots\cup A_k$ and $x_k=p(S_k)$, so $x_0=p(\emptyset)=0$. The step is accepted only if
$x_k>x_{k-1}$.

\begin{definition}[Additive record]\label{def:additive}
A record is \emph{additive} if each step adds a non-empty set of changes not yet present and no
step removes one, so that $S_{k-1}\subseteq S_k$ for every $k$.
\end{definition}

All records of this appendix are additive; removals are outside it. The \emph{owner step} of a change
is the step that added it, and its \emph{owner identity} is that step's identity. $C_o$ is the set
of changes owned by operator $o$. ``$c$ precedes $d$'' means that $c$'s owner step comes first. At
settlement the verifiers hold $S_n$ and measure, for $c,d\in S_n$, the ablation
$\hat h(c,d)=p(S_n)-p(S_n\setminus\{c\})-p(S_n\setminus\{d\})+p(S_n\setminus\{c,d\})$. We assume
\emph{full separability}: every subset of $C$ can be built and scored. Then every pair is eligible
and $\hat h=h$. A rule that drops pairs whose ablations cannot be built has a different payoff
formula; Section~\ref{sec:measurable} discusses such rules.

\begin{definition}[Pairwise royalty rules]\label{def:pairroyalty}
Fix a price $\lambda>0$ per unit of progress and a \emph{royalty function} $\varphi:\R\to\R$ with
$\varphi(0)=0$. Each accepted step $k$ earns the per-step payment $r_k=\lambda(x_k-x_{k-1})$. For
every pair $c,d\in S_n$ with $c$ preceding $d$ and different owner steps, the owner identity of $d$
pays the owner identity of $c$ the \emph{royalty} $\lambda\,\varphi(\hat h(c,d))$
(\emph{follower-financed}). In the \emph{sponsor-financed} variant the sponsor pays it instead, and
the owner of $d$ keeps $r_k$. Royalties are settled once, at the end. We write $\varphi=\alpha\psi$
with a rate $\alpha\in(0,1]$ when the shape $\psi$ matters:
\begin{itemize}
\item the \emph{clamped} royalty, $\psi^+(h)=\max\{0,h\}$, pays only for complementarity;
\item the \emph{signed} royalty, $\psi^{\pm}(h)=h$, shares conflicts as well as complementarities.
\end{itemize}
The \emph{zero} royalty $\varphi\equiv0$ is the per-step rule. A follower-financed royalty can make
a step's net payment negative. We assume that operators post a stake that covers it, and we do not
cap liabilities here. An operator's payoff $U_o$ is the sum, over its identities, of per-step
payments plus royalties received minus royalties paid.
\end{definition}

In the terms of \citet{grabisch1997kadditive}, $p$ is \emph{at most 2-additive}: its M\"obius
transform vanishes on every set of more than two changes (we allow $h\equiv0$, and $p$ is a signed
set function, not a normalised measure), and $g(c)$, $h(c,d)$ are the M\"obius masses of $\{c\}$
and $\{c,d\}$. Hence $h(c,d)=p(S\cup\{c,d\})-p(S\cup\{c\})-p(S\cup\{d\})+p(S)$ for every $S$
disjoint from $\{c,d\}$, by \eqref{eq:pairwise} directly. This second difference is the quantity
the interaction indices of the cooperative-game literature average over $S$
\citep{grabisch1999interaction}, and on a 2-additive game the average is the constant.

\subsection{Statements}

\begin{lemma}[Operator payoff]\label{lem:payoff}
Fix a pairwise landscape, an additive record with exact measurement, and an operator $o$. Under the
follower-financed rule with royalty function $\varphi$,
\begin{equation}\label{eq:payoff}
U_o=\lambda\,p(C_o)
+\lambda\!\!\sum_{\substack{c\in C_o,\ d\notin C_o\\ c\ \mathrm{precedes}\ d}}\!\!\varphi(h(c,d))
+\lambda\!\!\sum_{\substack{c\notin C_o,\ d\in C_o\\ c\ \mathrm{precedes}\ d}}\!\!\big(h(c,d)-\varphi(h(c,d))\big).
\end{equation}
Under the sponsor-financed rule the third sum is replaced by $\sum h(c,d)$ over the same pairs, and
a fourth term $\lambda\sum\varphi(h(c,d))$ is added over the pairs $c,d\in C_o$ owned by different
steps of $o$. Under both financings, the owner of $c$ receives, beyond its per-step payments,
$\lambda\sum_d\varphi(h(c,d))$ over the later changes $d$ of other operators (credit for enabling).
With $\varphi=\alpha\psi^+$ this is $\alpha\lambda$ times the measured complementarity that later
work of others drew from $c$.
\end{lemma}

\begin{theorem}[Packaging]\label{thm:royalty}
Consider the follower-financed pairwise royalty rule with any $\varphi$, on fully separable
pairwise landscapes and additive records, with exact measurement and no cap on liabilities.
\begin{enumerate}
\item[(a)] \textbf{Budget.} The total paid is $\lambda(x_n-x_0)=\lambda x_n$, as under the per-step
  rule.
\item[(b)] \textbf{Sybil-, split- and merge-proofness; manufactured complementarity.} $U_o$ depends
  on the record only through $C_o$, the landscape, and which changes of $o$ precede or follow each
  change of other operators. So two records that differ only in how $o$'s changes are partitioned
  into accepted steps and identities, with the same precedences relative to the changes of others,
  give $o$ the same payoff. In particular, relabelling steps of $o$ to other identities of $o$,
  splitting a step of $o$ into consecutive accepted steps, merging consecutive steps of $o$, and
  submitting two complementary pieces under two identities one after the other, all accepted,
  change nothing.
\item[(c)] \textbf{Non-expropriation.} A later accepted step by another operator never lowers $U_o$
  if and only if $\varphi\ge0$. The clamped royalty is non-expropriating; the signed one is not:
  when $h(c,d)<0$, the owner of $c$ pays $\alpha\lambda|h(c,d)|$ of the conflict.
\end{enumerate}
The sponsor-financed rule pays in total $\lambda x_n+\lambda\sum\varphi(h(c,d))$ over the pairs
with different owner steps: more than $\lambda x_n$ when $\varphi\ge0$ and some royalty is
positive, possibly less when $\varphi$ takes negative values. It pays the credit of
Lemma~\ref{lem:payoff}, and it satisfies (c): a later accepted step by another operator never lowers
$U_o$ if and only if $\varphi\ge0$. With $\varphi\equiv0$ it is the per-step rule and satisfies
(b). For every other $\varphi$ it fails (b): an operator collects $\lambda\varphi(h)$ for each pair
of its own changes owned by different steps. So if $\varphi(h)>0$ for some $h$, it gains
$\lambda\varphi(h)$ by submitting two pieces with interaction $h$ as separate accepted steps rather
than as one. If $\varphi(h)<0$, it gains $-\lambda\varphi(h)$ by merging them. Relabelling alone
changes nothing under either financing.
\end{theorem}

\paragraph{Why the follower pays.} An operator whose change is worth 6 can rewrite it as two halves
worth 1 each alone and 6 together, and submit them under two identities. If the sponsor paid
royalties on top, at $\alpha=\tfrac12$ and $\lambda=100$, this would earn it $+200$ for an
interaction it manufactured. Under follower financing the royalty goes from one of its identities
to the other, and the gain is zero (Theorem~\ref{thm:royalty}(b)).

\paragraph{Refinement and timing.} A stronger manipulation reshapes the changes: it replaces one
change by pieces that together do the same thing, and submits the pieces at different times.
Theorem~\ref{thm:royalty} does not cover this, because the pieces can sit on different sides of a
change of another operator.

\begin{definition}[Refinement; timing]\label{def:refine}
A \emph{refinement} of a change $c$ replaces it by two changes $c_1,c_2$ with
$p(S\cup\{c_1,c_2\})=p(S\cup\{c\})$ for all $S\subseteq C\setminus\{c\}$, the landscape staying
pairwise; equivalently, $g(c_1)+g(c_2)+h(c_1,c_2)=g(c)$ and $h(c_1,d)+h(c_2,d)=h(c,d)$ for every
other change $d$. It is \emph{sign-preserving} if, for every $d$, $h(c_1,d)$ and $h(c_2,d)$ are both
of the sign of $h(c,d)$ or zero; otherwise it is \emph{sign-changing}. To define timing, fix the
\emph{external record}: the changes of the other operators, in their order, all of which are
accepted in every record compared, and known to $o$ when it chooses its timing. A \emph{timing} of
operator $o$ is an interleaving of $o$'s changes (after a refinement, of the pieces) into the
external record such that every step of the resulting complete record is accepted. The comparison
thus holds the final set of changes fixed, up to replacing $c$ by its pieces. An operator who can
make a rival's candidate fail acceptance is outside it. A rule is \emph{refinement-proof} (on a
class of refinements) if, whenever the unrefined change has at least one timing, no refinement in the
class, with any timing of its pieces, earns the operator more than its best timing of the unrefined
change. When the unrefined change has no timing, the definition is silent.
\end{definition}

The record, not the operator, sets each step's start point, so timing decides who is charged for an
interaction. The per-step rule itself is not refinement-proof. A \emph{sandwich} cuts a change that
conflicts with an earlier change of others and complements a later one into a conflicting piece
submitted first and a complementing piece submitted last.

\begin{proposition}[The per-step rule can be sandwiched]\label{prop:sandwich}
Under the per-step rule ($\varphi\equiv0$), an operator can gain by a sign-preserving refinement with
timing. Let $o$ hold $c$, and let other operators submit $d_1$ then $d_2$, with $h(c,d_1)=-u<0$ and
$h(c,d_2)=w>0$, and direct gains large enough that every record considered is accepted.
Submitting $c$ at any single time pays $o$ at most
$\lambda\big(g(c)+\max\{0,w-u\}\big)$. Refining $c$ into $c_1$ with $h(c_1,d_1)=-u$, $h(c_1,d_2)=0$
and $c_2$ with $h(c_2,d_1)=0$, $h(c_2,d_2)=w$, submitting $c_1$ before $d_1$ and $c_2$ after $d_2$,
pays $\lambda(g(c)+w)$, a gain of $\lambda\min\{u,w\}$ at the expense of one other owner (of $d_2$ if
$u\ge w$, of $d_1$ if $u<w$). For example, with $\lambda=1$, $g(c)=5$, $u=3$ and $w=2$, the best single timing pays $5$
and the sandwich pays $7$.
\end{proposition}

\begin{theorem}[Characterisation]\label{thm:shapley}
For a follower-financed pairwise royalty rule with royalty function $\varphi$, on all fully
separable pairwise landscapes and additive records with exact measurement, with timings as in
Definition~\ref{def:refine} (the accepted changes of others fixed and known), the following are
equivalent:
\begin{enumerate}
\item[(i)] the rule is refinement-proof on sign-changing refinements;
\item[(ii)] every operator's payoff is independent of the order of the steps;
\item[(iii)] $\varphi(h)=h/2$: the signed royalty at rate $\alpha=\tfrac12$.
\end{enumerate}
Under this rule every operator receives $\lambda$ times the sum of the Shapley
values~\cite{shapley1953value} of its changes in the game $p$ on $S_n$, which on a pairwise game
equal $g(c)+\tfrac12\sum_{d\ne c}h(c,d)$ and coincide with the Banzhaf values; and the rule is
refinement-proof on all refinements, with any timing.
\end{theorem}

So the signed royalty at $\alpha=\tfrac12$ is Shapley sharing of the landscape. It ignores
identities because it pays changes, not names (Theorem~\ref{thm:royalty}(b)); Shapley sharing over
identities is not false-name-proof in general~\cite{yokoo2005coalitional,burnat2026quotient}. It is
order-free and refinement-proof because every refinement of Definition~\ref{def:refine} keeps the
landscape pairwise. We do not cover refinements that create a three-way term (compare
\citet{knudsen2012merging} and the amalgamation property of \citet{lehrer1988banzhaf}). The price is
Theorem~\ref{thm:royalty}(c): an earlier operator pays half of every conflict that later work runs
into.

\begin{corollary}[A trade-off]\label{cor:tradeoff}
No follower-financed pairwise royalty rule, the per-step rule included, is both non-expropriating
and refinement-proof on sign-changing refinements. The clamped royalty with rate $\alpha\ge\tfrac12$
is non-expropriating, and it is refinement-proof on sign-preserving refinements for an operator $o$
whenever some record that places \emph{all} of $o$'s changes before every external change is
accepted. With $\alpha<\tfrac12$, or when no such record is accepted, sign-preserving refinements
can gain; in the latter case even when the refined change itself can be placed first, because
a piece of it can be placed early next to another own change where the whole change cannot.
\end{corollary}

The corollary is the designer's choice, among follower-financed pair royalties on the linear base,
and always among records with the same accepted changes of others. The clamped royalty keeps the
promise that nobody loses by what comes later. The Shapley royalty is safe against every refinement
and every timing.

\paragraph{The copy attack.}\label{app:copyattack} Let $c$, with $g(c)=4$, be followed by $d$, of
another operator, made of its own change $d'$ with $g(d')=2$ and a renamed copy of $c$ that adds
nothing once $c$ is present. Removing $c$ leaves the copy, so ablation measures $g(d)=6$ and
$h(c,d)=6-4-6=-4$, while the frontier gain of $d$ is $2$. At $\alpha=\tfrac12$ and $\lambda=1$, the
signed royalty moves $2$ from $c$'s owner to $d$'s: $d$'s owner is paid $4$ for a change that adds
$2$, and $c$'s owner $2$ for a change worth $4$. Under the clamped royalty, as under the per-step rule
alone, they are paid $2$ and $4$. In general the follower collects half of the copied change's
value. Theorem~\ref{thm:shapley} does not cover this, since $d$ is not a refinement of its owner's
own changes: it carries another operator's work, which in an open frontier challenge nobody can rule
out (Definition~\ref{def:oac}, O3). The same bundle also hides a real dependence. If the follower's
own change needs $c$, removing $c$ leaves the copy, so the measured interaction is not positive and
the follower owes nothing. Where copies cannot be detected, this favours the clamped royalty, but
only because it limits the loss: its known gains from refinement need particular interactions with
others' changes, while the copy needs only a public earlier change. Copies are a known weakness of
Shapley-based payment: in a data market, a seller raises its Shapley share by selling a replica of
its own data, and the remedies penalise similar inputs~\cite{agarwal2019marketplace} or divide
payments so that replicating does not pay~\cite{ohrimenko2019collaborative}.

\subsection{Proofs}

\begin{proof}[Proof of Theorem~\ref{thm:limit}]
By hypothesis, the rule sees the same record in two worlds. In world~1, the operator of $i_k$ is alone and
$j\in T_k$, so (NH) gives $\Delta_k\ge-\ell_{\mathrm{hide}}$. In world~2, one operator controls
exactly $i_j$ and $i_k$ and $j\notin T_k$, so (NF) gives
$\Delta_j+\Delta_k\le\ell_{\mathrm{fake}}$. The rule sees only the record, so $\Delta_j$ and
$\Delta_k$ are the same numbers in both worlds. Hence
$\Delta_j\le\ell_{\mathrm{fake}}-\Delta_k\le\ell_{\mathrm{fake}}+\ell_{\mathrm{hide}}$.
\end{proof}

\begin{proof}[Proof of Lemma~\ref{lem:payoff}]
By~\eqref{eq:pairwise}, the gain of step $k$ is
\[
x_k-x_{k-1}=\sum_{d\in A_k}g(d)+\sum_{\{c,d\}\subseteq A_k}h(c,d)+\sum_{c\in S_{k-1},\,d\in A_k}h(c,d),
\]
the direct gains of its changes, the interactions within the step, and the interactions with every
earlier change. Summing $r_k$ over the steps of $o$ gives
$\lambda\big[\sum_{c\in C_o}g(c)+\sum_{\{c,d\}\subseteq C_o}h(c,d)+\sum_{c\notin C_o,\,d\in C_o,\
c\text{ precedes }d}h(c,d)\big]$, whose first two terms are $\lambda p(C_o)$. Royalties: a pair with
both changes in $C_o$ moves money between two identities of $o$ under follower financing, so it adds
nothing to $U_o$; under sponsor financing it brings $o$ the amount $\lambda\varphi(h(c,d))$ from the
sponsor when the owner steps differ. A pair with $c\in C_o$ preceding $d\notin C_o$ brings $o$ the
amount $\lambda\varphi(h(c,d))$ under both financings; these pairs give the credit for enabling. A
pair with $c\notin C_o$ preceding $d\in C_o$ costs $o$ the amount $\lambda\varphi(h(c,d))$ under
follower financing and nothing under sponsor financing. Adding up gives the two formulas.
\end{proof}

\begin{proof}[Proof of Theorem~\ref{thm:royalty}]
(a) Royalties are transfers, so the total is $\sum_kr_k=\lambda(x_n-x_0)=\lambda x_n$.
(b) The right-hand side of~\eqref{eq:payoff} involves the record only through $C_o$ and the
precedences between changes of $o$ and changes of others. Relabelling, splitting into consecutive
steps and merging consecutive steps keep both, and so does every repartition of $o$'s changes into
accepted steps that keeps the precedences with others' changes, such as two complementary pieces
submitted one after the other.
(c) Appending accepted steps of other operators adds the terms $\lambda\varphi(h(c,d))$ for the new
changes $d$ and $c\in C_o$, and changes nothing else; every real value of $h(c,d)$ occurs on some
landscape. The same holds under sponsor financing: the new changes are not in $C_o$, so they add no
pair to the third sum or the fourth term, and $o$'s own per-step payments are unchanged. The other
claims about the sponsor-financed rule read off its fourth term.
\end{proof}

\begin{proof}[Proof of Proposition~\ref{prop:sandwich}]
Under $\varphi\equiv0$ the pair term in~\eqref{eq:payoff} is $0$ if $o$ precedes and $h$ if $o$
follows. With $c$ submitted whole the options are: before both ($0$), between ($-u$), after both
($w-u$). With the pieces as stated, $c_1$ precedes $d_1$ (term $0$) and $c_2$ follows $d_2$ (term
$w$), and $h(c_1,c_2)=0$ with $g(c_1)+g(c_2)=g(c)$ keeps $p(C_o)$; the direct gains can be chosen
large enough for every step to be accepted. The others' payments fall by the same total, since the
total is $\lambda x_n$.
\end{proof}

\begin{proof}[Proof of Theorem~\ref{thm:shapley}]
Write $\varphi(h)=h/2+\chi(h)$; then $\chi(0)=0$, and the pair term of $o$ in~\eqref{eq:payoff} is
$h/2+\chi(h)$ when $o$ precedes and $h/2-\chi(h)$ when $o$ follows.

(iii)$\Rightarrow$(ii): with $\chi\equiv0$ both terms are $h/2$, so
$U_o=\lambda\big[p(C_o)+\tfrac12\sum_{c\in C_o,\,d\notin C_o}h(c,d)\big]$ whatever the order.
(ii)$\Rightarrow$(iii): take one change $c$ of $o$ and one change $d$ of another operator, with
$h(c,d)=h$ and direct gains large enough for both orders to be accepted. Order-independence gives
$h/2+\chi(h)=h/2-\chi(h)$, so $\chi(h)=0$; $h$ was arbitrary.
(iii)$\Rightarrow$(i): under (iii), $U_o$ is the expression above. The changes of others are the
same accepted set in every record compared (Definition~\ref{def:refine}), a refinement keeps
$p(C_o)$ and replaces $h(c,d)$ by $h(c_1,d)+h(c_2,d)=h(c,d)$ in the sum, and timing does not enter.
So no refinement and no timing changes $U_o$.
(i)$\Rightarrow$(iii): take $c\in C_o$ and one change $d$ of another operator with $h(c,d)=0$, and
the sign-changing refinement of $c$ into pieces with $h(c_1,d)=x$ and $h(c_2,d)=-x$, $h(c_1,c_2)=0$,
with the direct gains of $d$ and of both pieces larger than $|x|$ plus the margin, and $g(c)$ their
sum, so that every record used below is accepted. Submitting $c$ unrefined pays $o$ the amount
$\lambda p(C_o)$ in either order, since $\chi(0)=0$. The pieces, $c_1$ before $d$ and $c_2$ after,
pay $\lambda p(C_o)+\lambda[\chi(x)-\chi(-x)]$; so refinement-proofness gives $\chi(x)\le\chi(-x)$,
and exchanging the roles of the pieces gives $\chi(-x)\le\chi(x)$: $\chi$ is even. Both pieces
before $d$ pay $\lambda p(C_o)+\lambda[\chi(x)+\chi(-x)]=\lambda p(C_o)+2\lambda\chi(x)$, so
$\chi(x)\le0$; both after pay $\lambda p(C_o)-2\lambda\chi(x)$, so $\chi(x)\ge0$. Hence
$\chi\equiv0$. Placing $c_1$ before $d$ uses the known external record.

The Shapley value of $c$ averages its marginal contribution $g(c)+\sum_{d\in W}h(c,d)$ over the
coalitions $W$ preceding $c$ in a uniformly random order, in which each $d\ne c$ precedes $c$ with
probability $\tfrac12$; the Banzhaf value averages it over all $W\subseteq S_n\setminus\{c\}$ with
equal weight, in which each $d$ belongs to $W$ with probability $\tfrac12$. Both give
$g(c)+\tfrac12\sum_{d\ne c}h(c,d)$, and summing over $c\in C_o$ gives
$p(C_o)+\tfrac12\sum_{c\in C_o,\,d\notin C_o}h(c,d)$, which is $U_o/\lambda$.
\end{proof}

\begin{proof}[Proof of Corollary~\ref{cor:tradeoff}]
Non-expropriation needs $\varphi\ge0$ (Theorem~\ref{thm:royalty}(c)); refinement-proofness on
sign-changing refinements needs $\varphi(h)=h/2$ (Theorem~\ref{thm:shapley}), which is negative for
$h<0$. For the clamped rule with $\alpha\ge\tfrac12$: the pair term of $o$ for a change of others
with interaction $h$ is $\alpha\max\{0,h\}$ when $o$ precedes and $h-\alpha\max\{0,h\}$ when it
follows, and the first is at least the second for every $h$ ($\alpha h\ge(1-\alpha)h$ for $h>0$;
$0\ge h$ for $h\le0$). So, in~\eqref{eq:payoff}, every external pair term of $o$ is at most its
value when the own change precedes, whatever the timing; and $p(C_o)$ does not depend on timing.
Hence, for any timing of all of $o$'s changes, $U_o\le\lambda[p(C_o)+\sum\alpha\max\{0,h(c,d)\}]$,
the sum over all pairs of an own change $c$ and an external $d$, with equality on any record that
places all of $o$'s changes first. A sign-preserving refinement keeps $p(C_o)$ (it preserves every
interaction, including those with $o$'s other changes) and splits each $h(c,d)$ into two values of
the same sign, on which $\max\{0,\cdot\}$ is additive, so the bound is the same number before and
after refining. By hypothesis an all-first record of the unrefined changes is accepted, so the bound
is attained by a timing of the unrefined changes, and no refinement with any timing beats it. For
the last sentences: with $\alpha<\tfrac12$ and $h>0$, following pays $(1-\alpha)h>\alpha h$, so a
piece carrying the positive interactions is best submitted late while a piece carrying the
negative ones is best submitted early, and the two together beat any single time for $c$ when $c$
has interactions of both signs with later changes and the direct gains are large, as in
Proposition~\ref{prop:sandwich}. When $c$ cannot be placed first, a piece can: with
$\alpha=\tfrac34$, $g(c)=-1$, $g(d)=10$ and $h(c,d)=4$, the only timing of $c$ is after $d$,
paying $o$ the amount $\lambda[-1+4-3]=0$; the sign-preserving refinement $g(c_1)=1$, $g(c_2)=-2$,
$h(c_1,c_2)=0$, $h(c_1,d)=1$, $h(c_2,d)=3$ admits the record $c_1,d,c_2$ (gains $1$, $11$, $1$)
and pays $o$ the amount $\lambda[-1+\tfrac34+\tfrac34]=\lambda/2$. And when $c$ can be placed
first, but not together with another own change $e$ (here with $\lambda=1$): with $\alpha=\tfrac34$, $g(c)=12$,
$g(e)=2$, $g(d)=40$, $h(c,e)=-16$, $h(c,d)=8$, $h(e,d)=16$, the accepted timings of $\{c,e\}$ are
$c,d,e$ ($U_o=8$), $e,d,c$ ($12$), $d,c,e$ and $d,e,c$ ($4$ each), so the best is $12$; the
sign-preserving refinement of $c$ into $a,b$ with $g(a)=1$, $g(b)=11$, $h(a,b)=0$, $h(a,e)=-1$,
$h(b,e)=-15$, $h(a,d)=2$, $h(b,d)=6$ admits $a,e,d,b$ (gains $1,1,58,2$) and pays $13$: refining $c$
made an early placement of a piece of $c$, next to $e$, acceptable.
\end{proof}
\section{Noise: measurements, proofs and the royalties under noise}\label{app:noise}

\subsection{The measurements}

\paragraph{Training scores are noisy.} On an Apple M4 Mac mini, we used the Apple-Silicon port of
autoresearch~\cite{trevin2026autoresearchmlx}, which trains a small language model. Its score is the
validation loss in bits per byte (\texttt{val\_bpb}, lower is better); 1~\emph{mbpb} is a decrease
of $0.001$. We trained every combination of three changes to the recipe (\emph{batch}, halve the
batch; \emph{lr}, double one learning rate; \emph{depth}, halve the depth setting, which in this port also narrows the model), with 3 seeds each and 5
for the unchanged recipe: 26 five-minute runs. In five minutes the unchanged recipe trains a model
of 50 million parameters for 20 optimizer steps on 2.6 million tokens; halving the batch, the depth
or both allows 38 to 133 steps. The pooled standard deviation of a recipe is $38$~mbpb; the gains of
depth and batch are about $9\times$ and $3\times$ that (Table~\ref{tab:real}). These are short
runs on one machine; we did not test whether longer training is less noisy. An independent study
on a benchmark modelled on the speedrun, with 693 runs of its unchanged code, estimates that a check
with one seed and its acceptance floor of 64 optimizer steps would credit about one useless change in
seven~\cite{ren2026outerloop}.

\begin{table}[H]
\centering\small
\begin{tabular}{>{\raggedright\arraybackslash}p{0.36\textwidth}rr>{\raggedright\arraybackslash}p{0.27\textwidth}}
\hline
Quantity & Estimate & 95\% interval & Compared with noise \\
\hline
Noise: standard deviation for the same recipe (pooled) & 0.038 & & \\
Gain of depth & 0.353 & $[0.179, 0.528]$ & about $9\times$ \\
Gain of batch & 0.121 & $[0.023, 0.219]$ & about $3\times$ \\
Gain of lr & $-0.002$ & $[-0.017, 0.013]$ & not resolved \\
Interaction of batch and depth & 0.094 & $[-0.192, 0.381]$ & not distinguishable from 0 \\
\hline
\end{tabular}
\caption{Real measurements in \texttt{val\_bpb} units (higher gain is better). The
seeds vary only the initialisation, with the data order fixed, on one machine, so the noise is an
estimate in a restricted setting, and it may omit sources of variability that verifiers would
face. Seeds are shared across configurations, so the intervals treat each seed as a block (t
intervals, about 2 degrees of freedom; 4 for lr). No measured interaction is distinguishable from zero.}\label{tab:real}
\end{table}

\subsection{The renamed variable and the data order}\label{app:renamedetail}

\paragraph{The renamed variable.} %
The one-run check of Section~\ref{sec:noise} accepted the rename 8 times in 20 (exact $95\%$
interval $19$--$64\%$). With three runs a side and a margin of $30.9$~mbpb, none of six pooled
decisions accepted it; at the measured $\sigma=16.3$~mbpb that design accepts a useless candidate
with probability $1\%$. The rename changes
\texttt{t\_start} to \texttt{t\_begin}, three lines. We scored its twenty candidates and their forty
runs under four payment designs fixed before the first run (the written plan is dated three minutes
before it; five of the reference runs are those of the measurements above, run two days earlier).
The one-run design's $\$8{,}480$ (Section~\ref{sec:noise}) is $424$ per candidate; on the 15 pairs
run in the same session, 5 were accepted, $436$ per candidate. One of the twenty renamed runs scored
below the best of the five earlier runs of the unchanged recipe, the number a leaderboard would show
as the record. The rename changes no computation; the runs do not resolve a difference between the two arms either (mean
difference $+4.5$~mbpb, Welch $95\%$ interval $-5.9$ to $15.0$, $p=0.38$), and the forty runs put this recipe's run-to-run standard
deviation at $16.3$~mbpb, between the baseline-only $12.2$ and the $37.9$ pooled over
the configurations of Table~\ref{tab:real}; the payment per candidate was correspondingly below the $\$2{,}138$ that
$\sigma=37.9$ predicts, and within sampling error of the $688$ that $\sigma=12.2$ predicts. An
exact scorer gives the renamed code the same score as the original and accepts none of the
twenty.

\paragraph{Data order.} The runs above vary only the initialisation seed, not the data order. We
also ran the unchanged recipe twenty times with the data order varied; these runs trained on
different documents, since a run reads only a few of the 827 row groups (one run made 19 steps, not 20). Their standard deviation was $18.5$~mbpb ($95\%$ interval $14.1$ to
$27.0$), against $14.0$ for twenty seed-only runs and $16.3$ pooled; the ratio to the former, $1.32$
($0.83$ to $2.10$), is not resolved, and the runs came from a later session. We do not rescale our
numbers. Under the noise model, a useless change's payment at zero margin grows in proportion to $\sigma$, and the planning run count for a fixed target
with $\sigma^2$; fixed-margin acceptance rates and payment shares do not scale this simply.
The single-recipe estimates of the standard deviation thus range from $11.1$ to $18.5$~mbpb,
depending on what varies between runs.

\subsection{One interaction, ten seeds per cell}\label{app:interactionreal}
The planning calculation behind Section~\ref{sec:measurable}: an interaction is a second difference of four
cell means, so with $m$ runs per cell its standard error is $2\sigma/\sqrt m$ under equal variance and
independence, and resolving an interaction $h$ at $z$ standard errors needs $m\ge4z^2\sigma^2/h^2$ runs
per cell, four cells per pair. At $\sigma=0.038$ and $z=1.96$, an interaction worth one percent of the
baseline score needs 33 runs per cell, 132 per pair. It uses the pooled $\sigma$ and is not a bound.
At the pooled $\sigma$ it gives 92 runs per pair for an interaction the size of the speedrun's median
step ($1.2\%$) and 8 for one the size of the batch change's gain.
With the four cells' own standard deviations ($0.008$ to $0.076$) it gives 57 runs per cell for the batch--depth design and 15 for batch--lr. With three seeds, the measured standard error of the batch--depth interaction
was $0.067$ against the formula's $0.044$.

To test the calculation we ran the batch--depth design again with ten fresh seeds per cell, the same seed in
all four cells (40 runs, about 3.9 hours). We predicted before the first run, in a plan dated
an hour and a half before it, that the interaction would be positive with a $95\%$ interval excluding
zero. The prediction failed. The
gains of batch and depth were resolved ($96.6$ and $373.7$~mbpb; intervals $67.7$ to $125.6$ and $342.8$ to
$404.6$). The interaction was $h=+70.7$~mbpb with a $95\%$ interval from $-13.5$ to $+154.8$ (blocked
$t$, nine degrees of freedom, $p=0.09$). Its standard error, $37.2$~mbpb, was $3.6$ times what the
formula gives with the $\sigma=16.3$ of the renamed-variable runs and $1.6$ times what it gives with
the pooled $37.9$. The reason is that the changed recipes are noisier than the unchanged one: the
cells' standard deviations were $11.1$ (unchanged), $34.8$ (batch), $44.9$ (depth) and $111.9$~mbpb (both).
Paid on the point estimate, the clamped royalty at $\lambda=100$ and $\alpha=\tfrac12$ would move
$\$3{,}532.53$ from the later change's owner to the earlier one's; over the interval, $0$ to
$7{,}741.89$.

\subsection{Proofs for Section~\ref{sec:noise}}\label{app:noiseproofs}

Throughout, $N$ and $N'$ are the standard normal distribution function and density, and for
$X$ normal with mean $\mu$ and standard deviation $s$, $G_s(\mu)=\mathbb{E}[\max\{0,X\}]
=\mu\,N(\mu/s)+s\,N'(\mu/s)$. It is increasing in $\mu$ (its derivative is $N(\mu/s)$), and
$G_s(0)=s/\sqrt{2\pi}$. The $G$ of Proposition~\ref{prop:splitzero} is $G_s$.

\begin{proposition}[Splitting pays at zero margin]\label{prop:splitzero}
Assume (N1), (N2) and (N4). Let a candidate be accepted if $D\ge\delta$ and paid $\lambda D$ on
acceptance, as in Proposition~\ref{prop:deciding}. Let $\delta=0$ and every piece have the same $s$.
With $G(\mu)=\mathbb{E}[\max\{0,D\}]=\mu\,N(\mu/s)+s\,N'(\mu/s)$, a candidate of true gain
$\mu\ge0$ submitted as $k$ equal pieces in the splitting experiment of Proposition~\ref{prop:fee}(c) earns $\lambda k\,G(\mu/k)$
instead of $\lambda G(\mu)$. This is larger if $k>1+\mu\sqrt{2\pi}/(2s)$. So for every $\mu$ some
number of pieces gains, and each extra piece adds about $\lambda s/\sqrt{2\pi}$.
\end{proposition}

\begin{proof}[Proof of Proposition~\ref{prop:deciding}]
(a) $D$ is normal with mean $0$ and standard deviation $s$, so $\Pr(D\ge\delta)=1-N(\delta/s)$ and
$\mathbb{E}[D\,\mathbf{1}\{D\ge\delta\}]=s\,N'(\delta/s)$, the standard upper-tail integral.
(b) The same integral with mean $\mu$: $\mathbb{E}[D\,\mathbf 1\{D\ge\delta\}]
=\mu(1-N(z))+s\,N'(z)$; divide by $1-N(z)>0$.
\end{proof}

\begin{proof}[Proof of Proposition~\ref{prop:splitzero}]
Payments are $\lambda$ times the score quantities, so compare $kG(\mu/k)$ with $G(\mu)$.
$G'(\mu)=N(\mu/s)\ge\tfrac12$ for $\mu\ge0$ and $G(0)=s\,N'(0)=s/\sqrt{2\pi}$, so
$G(\mu/k)\ge G(0)+\mu/(2k)$ and $k\,G(\mu/k)\ge k\,G(0)+\mu/2$. Also $G(\mu)\le\mu+G(0)$, since
$N\le1$ and $N'$ is maximal at $0$. Hence $k\,G(\mu/k)-G(\mu)\ge(k-1)G(0)-\mu/2$, positive
when $(k-1)s/\sqrt{2\pi}>\mu/2$. This condition is sufficient, not necessary.
\end{proof}

\begin{proof}[Proof of Proposition~\ref{prop:nonneg}]
\label{app:nonneg}A randomised payment $f(M,U)$, with $U$ independent of $M$, is paid
$\bar f(M)=\mathbb{E}[f(M,U)\mid M]\ge0$ in expectation given $M$, so it suffices to treat $\bar f$.
If the useless candidate of the hypothesis earns nothing in expectation, $\mathbb{E}[\bar f(M)]=0$
under its law; a non-negative random variable with mean zero is zero almost surely, so $\bar f(M)=0$
almost surely under that law. That law has a density that is positive everywhere, so $\bar f=0$
almost everywhere. Every other candidate's $M$ has a density, so $\bar f(M)=0$ almost surely there
too; then $f(M,U)$, which is non-negative with conditional mean $\bar f(M)$, is $0$ almost surely.
Normality enters only through these two hypotheses. Under (N1) the run vector, of the candidate and
of its reference, is a vector of independent normal variables with positive standard deviations, so
its law has a density that is positive everywhere, for every candidate. This is the argument of
\citet{somekhbaruch2018nonexistence} for constrained estimators, with the payment in place of the
estimator and $0$ as the edge of its range.
\end{proof}

\begin{lemma}[Fee principle]\label{lem:fee}
Suppose every candidate whose true gain against the current best at the time it is submitted is at
most $0$ is paid at most $\beta$ in expectation, and accepted with probability at most $p_0$, given
the history and whoever submits it. Let $\gamma\ge\beta$, or $\gamma\ge\beta/(1-p_0)$ if the fee is
refunded on acceptance. Then any campaign of at most $K$ such candidates, $K$ fixed in advance,
under any identities and any adaptive choices, earns at most $0$ in expectation, net of fees. With
strict inequality in the fee, every attempt made loses in expectation.
\end{lemma}

\begin{proof}[Proof of Lemma~\ref{lem:fee}]
Number the attempts $t=1,\dots,K$, with $K$ fixed in advance, and let $Y_t$ be the net result of
attempt $t$ ($0$ if it is not made). Whether attempt $t$ is made, and which candidate is submitted
under which identity, depend only on the history. Given the history and that the attempt is made,
its pay $X_t\ge0$ has $\mathbb{E}[X_t]\le\beta$ and it is accepted with probability $q_t\le p_0$, by
hypothesis. With the fee not refunded, $Y_t=X_t-\gamma$ and $\mathbb{E}[Y_t\mid\text{history,
made}]\le\beta-\gamma\le0$. With the fee refunded on acceptance, $Y_t=X_t+\mathbf 1\{\text{accepted}\}
\gamma-\gamma$, whose conditional expectation is at most $\beta+q_t\gamma-\gamma\le\beta-(1-p_0)\gamma
\le0$. Summing over $t$, $\mathbb{E}[\sum_tY_t]\le0$. With strict inequality in the fee, each
conditional expectation is negative.
\end{proof}

\begin{proof}[Proof of Proposition~\ref{prop:fee}]
Write $s_1$ and $s_2$ for the standard deviations of $D_1$ and $D_2$ for a given candidate. With
candidate and reference runs of standard deviations $\sigma_c,\sigma_r\le\bar\sigma$,
$s_1=\sqrt{(\sigma_c^2+\sigma_r^2)/m}\le\bar s_1$, and likewise $s_2\le\bar s_2$.
(a) By independence, the expected pay is $\Pr(D_1\ge\delta)\cdot\lambda G_{s_2}(\mu)$. For $\mu\le0$
and $\delta\ge0$, $\Pr(D_1\ge\delta)=1-N((\delta-\mu)/s_1)\le1-N(\delta/s_1)\le1-N(\delta/\bar s_1)
=p_0$, and $G_{s_2}(\mu)\le G_{s_2}(0)=s_2/\sqrt{2\pi}\le\bar s_2/\sqrt{2\pi}$. The product is at most $b$.
(b) Each attempt's runs are fresh and independent of the history by (N1), and its gain $\mu\le0$ is
measured against the current best at that time, so (a) holds given the history. The pay of a
candidate and its fee do not depend on who submits it. So the hypotheses of Lemma~\ref{lem:fee} hold
with $\beta=b$, and it gives (b).
(c) Let $e(x)=\lambda A(x)(G_2(x)-x)\ge0$, so that a piece of gain $x$ is paid
$\lambda A(x)G_2(x)=\lambda A(x)x+e(x)$ in expectation. The acceptance probability $A$ is
non-decreasing in the gain, so for pieces of gains $\mu_1,\dots,\mu_k\ge0$ summing to $\mu$, with
$k\ge2$, the pieces earn net at most $\sum_i(\lambda A(\mu)\mu_i+b^*)-k\gamma=\lambda A(\mu)\mu+kb^*-k\gamma$,
and the whole earns $\lambda A(\mu)\mu+e(\mu)-\gamma\ge\lambda A(\mu)\mu-\gamma$. The difference is at
most $kb^*-(k-1)\gamma\le0$, since $\gamma\ge2b^*\ge kb^*/(k-1)$.
(d) Condition on the other candidates, their order and their true gains; the rule has no cap and
no deadline, so the copies change nothing else. Couple the two worlds by giving each other
candidate, in both, the same runs for its check and its payment, of the candidate and of its
reference. A copy, accepted or not, leaves the current best's true score and the distribution of its
runs unchanged, so each other candidate's reference has the same run distribution in both worlds;
those runs are drawn for that check alone, independently of everything before, so the coupling is
valid. Each acceptance and payment is then the same function of the same runs in both worlds, and
the joint distribution is the same. No step uses normality.
\end{proof}

\paragraph{Where normal candidate runs come from.} Parts (a) and (b) assume that the candidate's
runs are normal with standard deviation at most $\bar\sigma$. Definition~\ref{def:problem} lets the
operator set its code's noise $\sigma(a)$, so this needs the verifier, not the operator, to draw the
runs' randomness, and a way to enforce $\bar\sigma$, which we do not give. The shape matters too: a
useless candidate whose runs are $\pm\bar\sigma$ with probability $\tfrac12$ each, against normal
reference runs, is paid $\$24.83$ and $\$63.71$ at the simulated design of
Appendix~\ref{app:noisyscorer}, $1.031\,b$ at both noise levels (computed in closed form).

\paragraph{Where Proposition~\ref{prop:fee}(d) fails.} The proof uses that no check or payment
reads runs drawn before it. A rule that accepts a candidate only if its run beats the stored score
that made the current best current fails it: with all changes useless and one run each, a last
useless candidate is accepted with probability $1/2$ alone, but with probability $1/(r+2)$ after
$r$ useless copies, since its run must then be the best of $r+2$. The same holds for a payment
that reads such a stored score. Under the rule of Proposition~\ref{prop:fee}, no run is stored.

\paragraph{What Proposition~\ref{prop:fee}(b) and (c) do not cover.} Part (b) is about each attempt
and about campaigns bounded in advance. Take a luck-fisher with unlimited funds and attempts that
submits copies of the current best, with runs on both sides as noisy as $\bar\sigma$, and pays a fee
of exactly $b$, not refunded: its net is then a random walk with mean zero, so if it stops as soon as
it is ahead, it ends ahead with probability one. Its number of attempts is not fixed in advance, so
Lemma~\ref{lem:fee} does not apply. A fee above $b$ removes this example. Part (b) also
covers only candidates whose true gain against the current best is not positive, and part (c) only
fees of at least $2b^*$. Two campaigns can profit at a
fee of $b$; both lie outside the hypotheses. Take $\lambda=1$, $\delta=0$, $\sigma=1/\sqrt2$.
\emph{Lower, then restore}: with $m=1$ and $m'=10{,}000$, so $\bar s_1=1$ and $\bar s_2=0.01$,
submit a step of gain $-1$ and, if it is accepted, a step of gain $+1$ against the lowered best. The
second step's gain is positive, so (b) does not apply; the campaign nets $0.1312$ in expectation at
$\gamma=b\approx0.0020$, and the final best is never better than the start. \emph{Splitting}: with
$m=10{,}000$ and $m'=1$, a change of gain $0.2$ nets $0.3074$ at $\gamma=b\approx0.1995$, and two pieces
of $0.1$ net $0.5029$; (c) needs $\gamma\ge2b^*$, which this fee is not. In both, one of the two
measurements is far noisier than the other. With the parameters of our simulated challenge, neither
campaign gained (Appendix~\ref{app:noisyscorer}).

\subsection{Royalties under noise}\label{app:royaltynoise}

The royalties of Section~\ref{sec:royaltytheory} are computed
from ablations, so under noise each measured interaction is $\hat h=h+\eta$, where $\eta$ has mean
zero and, with $m$ runs per ablated artifact and normal run noise, standard deviation
$s_\eta=2\sigma/\sqrt m$ (four measured scores enter $\hat h$). The two royalty functions react
differently.

\begin{proposition}[The clamped royalty is biased, the signed one is not]\label{prop:royaltynoise}
Fix an accepted additive record on a fully separable pairwise landscape, with its true scores and
its exact linear per-step payments, as in Section~\ref{sec:royaltytheory}; let only the royalties
be settled on measurements $\hat h=h+\eta$ taken after acceptance, where the $\eta$ have mean zero,
the same integrable distribution for every pair measured, and are independent across pairs and of
the record (the proof uses only each $\eta$'s law, not independence across pairs).
\begin{enumerate}
\item[(a)] Under the signed royalty at rate $\alpha$, the expected royalty on every pair is
  $\lambda\alpha h$. At $\alpha=\tfrac12$ this is $\lambda\varphi(h)$ with $\varphi(h)=h/2$, so every
  operator's expected payoff is its exact-measurement payoff of
  Theorem~\ref{thm:shapley} on that record, and the packaging and refinement invariances hold in
  expectation on the comparison domain of Definition~\ref{def:refine} and Theorem~\ref{thm:shapley}.
  This says nothing about per-step payments that are themselves set from noisy measurements: their
  bias (Proposition~\ref{prop:deciding}(b)) is untouched by the royalty.
\item[(b)] Under the clamped royalty, the expected royalty on a pair is at least
  $\alpha\lambda\max\{0,h\}$:
  \[
  \alpha\lambda\,\mathbb{E}\big[\max\{0,h+\eta\}\big]\ \ge\ \alpha\lambda\max\{0,h\},
  \]
  with strict inequality whenever $\eta$ has positive probability on both sides of $-h$. A pair
  with no interaction pays $\alpha\lambda\,\mathbb{E}[\eta^+]$ in expectation, which for normal
  $\eta$ is $\alpha\lambda s_\eta/\sqrt{2\pi}$.
\item[(c)] So the clamped royalty is not refinement-proof in expectation, not even on
  sign-preserving refinements. Refining a change into $k$ pieces with the same total interaction
  with a later change $d$ of another operator, each pair measured separately with the same noise
  law, raises the expected royalty received from $d$'s owner by a non-negative amount, and by
  exactly $(k-1)\,\alpha\lambda\,\mathbb{E}[\eta^+]$ when the pieces have no interaction with $d$.
\end{enumerate}
\end{proposition}
\begin{proof}
(a) Linearity of expectation: $\mathbb{E}[\alpha\hat h]=\alpha h$, which is $h/2$ at $\alpha=\tfrac12$, and the per-step payments are exact by
assumption, so the expected payoff is the expression of Lemma~\ref{lem:payoff} with
$\varphi(h)=h/2$, on which Theorem~\ref{thm:shapley} rests. (b) $\max\{0,\cdot\}$ is convex, so Jensen gives the inequality; strictness is the
standard condition for Jensen with a function that is affine on each side of $0$ and not across it.
At $h=0$, $\mathbb{E}[\max\{0,\eta\}]=\mathbb{E}[\eta^+]$, which for $\eta\sim N(0,s_\eta^2)$ is
$s_\eta/\sqrt{2\pi}$. (c) Let $f(t)=\mathbb{E}[\max\{0,t+\eta\}]$ for the common law of $\eta$.
Then $f\ge0$, $f$ is non-decreasing and $1$-Lipschitz (both because $\max\{0,\cdot\}$ is), and
$f(t)\ge t$. Hence $f$ is subadditive: for $y\ge0$, $f(x+y)\le f(x)+y\le f(x)+f(y)$; for $y<0$,
$f(x+y)\le f(x)\le f(x)+f(y)$. The pieces' pairs with $d$ have interactions $h_1,\dots,h_k$ summing
to $h(c,d)$, each measured with its own $\eta_i$ of the same law, so the expected royalty is
$\alpha\lambda\sum_if(h_i)$ against $\alpha\lambda f(h(c,d))$ unrefined, and
$\sum_if(h_i)\ge f(\sum_ih_i)$ by induction on subadditivity. When all $h_i=0$ the refined amount
is $k\,f(0)=k\,\mathbb{E}[\eta^+]$ against $f(0)$.
\end{proof}

The bias is not small at the measured noise: with $m=1$, $s_\eta=75.8$~mbpb and
$\mathbb{E}[\eta^+]=30$~mbpb, so at $\alpha=\tfrac12$ and $\lambda=100$ every additional piece of
an operator's work collects about $\$1{,}500$ per later change of others, from nothing ($870$
with $m=3$). Under noise the trade-off of Corollary~\ref{cor:tradeoff} tilts: the clamped royalty
pays for noise in proportion to how finely operators slice their work, while the signed royalty
keeps its guarantees in expectation.
\section{Rules in use, and notes to Table~\ref{tab:results}}\label{app:notes}

\subsection{Rules in use}\label{app:rulesinuse}

Two baseline rules of Section~\ref{sec:rule}, top-$k$ and the equal-share raffle, are stylised from
the published reward schedules below, read from the challenge pages on 2026-10-23; winner-take-all
and the fixed bonus are not taken from them. We describe what the pages offer, not what was paid.
\begin{itemize}\setlength{\itemsep}{1pt}\setlength{\parskip}{0pt}
\item \emph{QSB} (Yukon, \$20{,}000): two separately ranked periods with top-3 prizes
  (1{,}000/600/400 in week~1; 6{,}000/3{,}600/2{,}400 in weeks~2--3), the baseline reset after week~1
  so that the gains of ``early entrants'' would not make ``the later rounds unwinnable'', and a
  \$6{,}000 raffle shared equally by every solver who improved the frontier by at least 3\% over the
  whole challenge~\cite{starkware2026qsbrewards}. We stylise the periods as two top-$k$ rules on
  cumulative gain, each on its own record, and the raffle as equal-share qualification on the whole
  record. Proposition~\ref{prop:failures}(c),(d) concern these stylised rules; whether relabelling
  and splitting are feasible on the deployed platform is not established here.
\item \emph{ARC Prize 2024} (Kaggle): five ranked progress prizes and a conditional grand prize, with
  ``one account per participant''~\cite{arcprize2024rules}; the grand prize went
  unclaimed~\cite{knoop2024arcwinners}. Top-$k$, with fake identities forbidden by the terms, which
  open entry (O1) rules out.
\item \emph{Muse Spark Open Kernel Challenge} (Paradigm, Meta): \$100{,}000 on a serving track as
  ranked prizes for places 1 to 10, decided by the median of at least three reruns of the top 25
  entries on held-out data; one account per individual, no teams~\cite{paradigm2026okc}. Top-$k$,
  with fake identities forbidden by the terms. The kernel board we inspected shows each account's
  best result and no artifact, so on our reading it has no public frontier (O2).
\item \emph{better.codes} shows every accepted submission's gain, the per-step rule's accounting, but
  makes awards ``experimental and discretionary'' and reserves discretion over how a contribution
  ``should be measured where improvements build upon or interact with other
  submissions''~\cite{ef2026programterms}: no committed rule.
\item \emph{autoresearch@home} specifies result sharing and a global best, and no payment
  rule~\cite{mutablestate2026athome}.
\end{itemize}
None of these schedules commits to a payment per unit of progress on acceptance, and none prescribes a rule that pays for
measured interactions.

\subsection{The rules compared, and notes to Table~\ref{tab:results}}\label{app:tablenotes}

Table~\ref{tab:results} is in Section~\ref{sec:enabling}. These notes give the regimes and witnesses
behind its marks.

Score-model columns (non-expropriation, Sybil, split/merge) are for exact scores. Change-model
columns (packaging, refinement, credit) are for the idealised royalty (every pair measured, uncapped,
no window, settled once) on fully separable pairwise landscapes with exact measurement and additive
records (Definition~\ref{def:additive}). Packaging keeps the order relative to others' changes, and
refinement compares records with the same accepted changes of others. The royalty rows are for the
linear base $\Phi(x)=\lambda x$ only. The per-step packaging \cmark\ holds for every $\Phi$: with
precedences relative to others' changes kept, the others' start and end scores are unchanged, and
the operator's total telescopes.

\paragraph{Baseline rules.} The marks of the baseline rules are those of
Proposition~\ref{prop:failures}, in the parameter regimes stated there. The \cmark\ entries for WTA
and the bonus follow from their definitions: WTA pays only the last identity, so relabelling and
splitting one's own steps change nothing; the bonus is an endpoint rule
(Proposition~\ref{prop:potential}). Top-$k$ and the raffle are split- and merge-proof as well
(Proposition~\ref{prop:failures}(e)): both read each identity's cumulative gain, which splitting a
step into consecutive steps of the same identity, or merging such steps, leaves unchanged. So split- and merge-proofness alone separates the per-step rule only among endpoint rules;
WTA, top-$k$ and the raffle fail on copying or extra identities instead. Top-$k$ is
also expropriating when $L>0$: later identities with larger totals push an earlier one down, out of
the prizes (Proposition~\ref{prop:failures}(f)). The equal-share raffle is expropriating if $L\ge2\theta$: a later qualifier lowers every
share from $P/N$ to $P/(N+1)$ (Proposition~\ref{prop:failures}(f)). With $\theta\le L<2\theta$ no
second identity can qualify, so no share is diluted. A raffle nobody can
qualify for pays nothing and fails nothing.

\paragraph{Royalty rows.} Sponsor-paid row: for $\varphi\not\equiv0$, splitting (if $\varphi(h)>0$
for some $h$) or merging (if $\varphi(h)<0$) two own changes into different steps pays
$\lambda|\varphi(h)|$ from the sponsor; $\varphi\equiv0$ is the per-step rule. Its non-expropriation
mark, iff $\varphi\ge0$, is stated with Theorem~\ref{thm:royalty}. Its refinement \xmark: if
$\varphi(h)>0$ for some $h$, refining an own change into two pieces with internal interaction $h$,
submitted as separate steps, collects $\lambda\varphi(h)$; if $\varphi\le0$ everywhere, a timing
witness: with $g(c)=2$, $g(d)=g(e)=10$, $h(c,d)=-4$, $h(c,e)=1$ and external order $d,e$, the only
accepted timing of $c$ is first, paying $o$ the amount $\lambda[2+\varphi(-4)+\varphi(1)]$; refining
$c$ into unit-gain pieces $a,b$ with $h(a,d)=-4$, $h(b,e)=1$ and all other new interactions zero,
the record $a,d,e,b$ is accepted and pays $\lambda[2+\varphi(-4)+1]$, since under sponsor financing
a follower keeps the whole interaction, a gain of $\lambda(1-\varphi(1))>0$. The declared-ancestry
row needs the domain assumption (D) of Section~\ref{sec:impossibility}. Credit for enabling is the
gross royalty income of the owner of a change, summed over later changes of others, with
$h^+=\max\{0,h\}$; the sponsor's total also includes royalties between an operator's own steps.

\paragraph{Total paid.} On a given record. A rule is budget-feasible when the supremum of this over
records is at most $B$, which for the per-step rule is $\sup\Phi-\Phi(s_0)\le B$
(Proposition~\ref{prop:potential}).
\section{Paying on a noisy scorer}\label{app:noisyscorer}

Section~\ref{sec:noise} states the results on paying under noise; their proofs are in
Appendix~\ref{app:noise}. This appendix gives the simulated checks, the baseline rules and the fee
under the same noise, and why we do not use a deposit.

\paragraph{Cheap checks pay for luck, and careful checks cost many runs.} We simulated independent
normal runs at $\sigma=37.9$~mbpb, pooled over all recipes, the value a sponsor who does not know
what will be submitted plans with (and, for sensitivity, at $12.2$ and $76$~mbpb). The figures below
are for $\sigma=37.9$. A
candidate is checked as in Definition~\ref{def:arp}: each of $v$ verifiers runs the candidate and the
current best $m$ times each and says yes if the candidate's average is better by at least a margin
$\delta$; it is accepted if $q$ say yes, and paid the median measured difference. With $v=4$, $m=3$
and $q=3$, a candidate costs 24 runs.
\begin{itemize}
\item \textbf{A single check pays for luck.} With one verifier who runs each side once ($v=m=1$)
  and no margin ($\delta=0$), a candidate that does nothing is accepted half the time (probability
  $0.50$). In a stream of 200 candidates, 60\% useless and 25\%, 10\% and 5\% with true gains of 20, 60
  and 150~mbpb, 36.6\% of all payments go to useless candidates.
\item \textbf{Winner's curse.} Paying on the measurements that decided acceptance overstates gains
  (Proposition~\ref{prop:deciding}(b)): in the same stream, paid progress is $6{,}974$~mbpb against a
  true $3{,}194$.
\item \textbf{No cheap design does it all.} Of 192 designs, none keeps the luck share below 5\%,
  accepts 80\% of real 60~mbpb candidates and costs under one machine-hour per candidate. The cheapest
  meeting the first two uses 24 runs ($\delta=30.9$~mbpb, $m=3$, $v=4$, $q=3$; luck share 2.4\%). At
  the single-recipe noise of $12.2$~mbpb, 16 of the 192 designs meet all three.
\end{itemize}
Without a fee, reruns and a margin thus keep luck to $2.4\%$ of payments in this stream, but nothing
stops a determined luck-fisher.

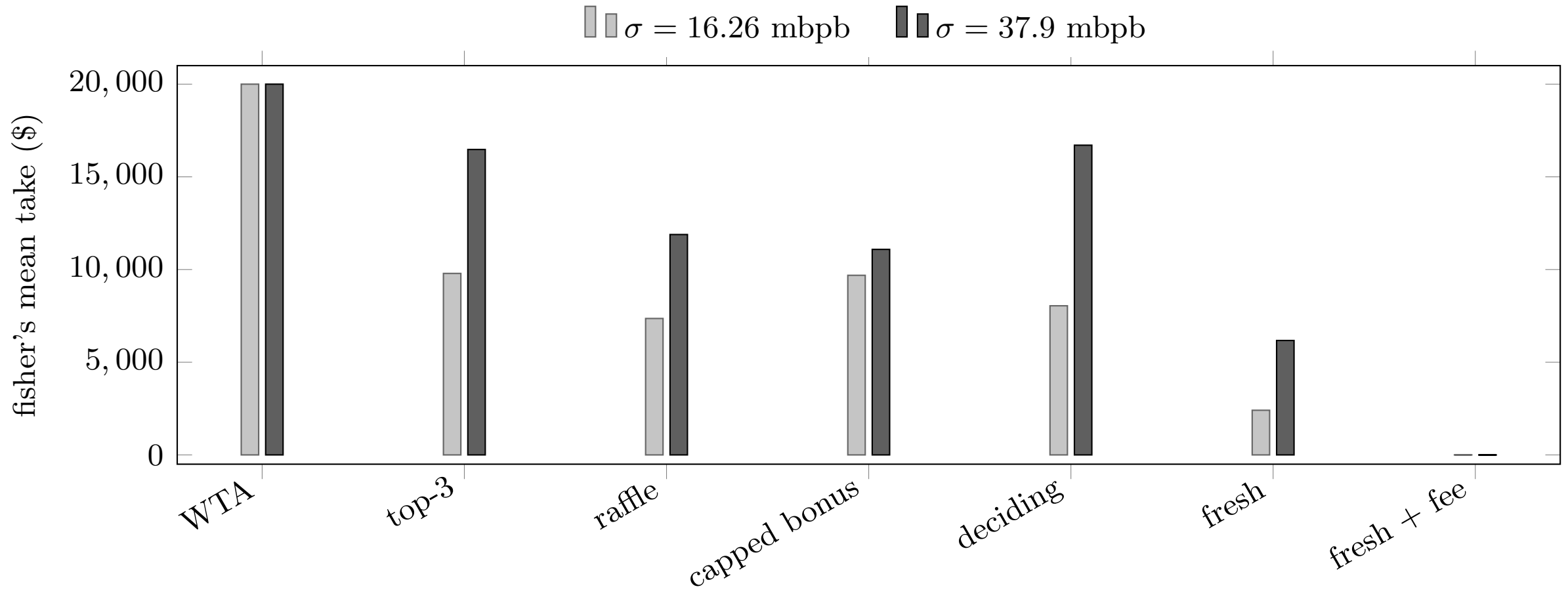
\begin{figure}[t]
\centering
\begin{tikzpicture}
\begin{axis}[
  ybar, bar width=5pt, width=0.94\linewidth, height=5.6cm,
  ymin=-500, ymax=21000, ytick={0,5000,10000,15000,20000},
  scaled y ticks=false,
  yticklabel style={font=\footnotesize, /pgf/number format/fixed, /pgf/number format/1000 sep={,}},
  ylabel={fisher's mean take (\$)}, ylabel style={font=\footnotesize},
  symbolic x coords={WTA, top-3, raffle, capped bonus, deciding, fresh, fresh + fee},
  xtick=data, x tick label style={font=\footnotesize, rotate=30, anchor=east},
  enlarge x limits=0.07,
  legend style={font=\footnotesize, at={(0.5,1.02)}, anchor=south, draw=none, legend columns=2, /tikz/every even column/.append style={column sep=1em}},
  legend cell align=left,
  every axis plot/.append style={fill opacity=0.9},
]
\addplot[fill=black!25, draw=black!60] coordinates {
  (WTA,20000.00) (top-3,9790.20) (raffle,7360.10) (capped bonus,9688.18)
  (deciding,8044.16) (fresh,2408.79) (fresh + fee,1.12)};
\addplot[fill=black!70, draw=black] coordinates {
  (WTA,20000.00) (top-3,16476.20) (raffle,11888.26) (capped bonus,11087.00)
  (deciding,16707.12) (fresh,6172.27) (fresh + fee,-5.48)};
\legend{$\sigma=16.26$~mbpb, $\sigma=37.9$~mbpb}
\end{axis}
\end{tikzpicture}
\caption{What a luck-fisher takes. A luck-fisher submits 100 changes that do nothing and hopes that
a noisy check accepts some. Each bar is its average take in $20{,}000$ simulated challenges, at the
best of the eight tested choices of timing and identities (setup in Section~\ref{sec:noiserule};
Table~\ref{tab:noiserules}).
Each of the seven labels is one rule and has two bars, one per noise level. Here $\sigma$ is the
standard deviation of the score of one training run of the same code. Light bars:
$\sigma=16.26$~mbpb, the unchanged recipe alone. Dark bars: $\sigma=37.9$~mbpb, pooled over the
recipes of Table~\ref{tab:real}. The first four labels are the baseline rules; the bonus stops when
the budget is spent. The last three are the per-step rule under noise, each with no hard cap.
\textbf{deciding}: pay on the runs that decided acceptance (Proposition~\ref{prop:luck}).
\textbf{fresh}: pay $\lambda\max\{0,D_2\}$ on fresh runs, never a negative amount.
\textbf{fresh + fee}: the same, with a fee of $b$ per submission ($\$24.08$ and $\$61.78$) paid into
the pot by every submission; the bar is the take net of fees (Proposition~\ref{prop:fee}). The fee is
set for useless changes as noisy as the recipe; noisier ones beat it (text).}
\label{fig:fisher}
\end{figure}
 
\paragraph{The rules under the same noise.} We simulated all rules under one noise model (normal,
independent runs, $\sigma=16.26$, the unchanged recipe alone, and $37.9$~mbpb, 20{,}000 challenges per cell): 40 honest
candidates by 20 identities and a luck-fisher with $K$ useless ones, budget $B=\$20{,}000$. Each
per-step rule's price is set so that its mean total with no luck-fisher is $B$.
Table~\ref{tab:noiserules} gives the luck-fisher's take at the best of its eight tested timing and
identity choices, for a verifier with three runs a side and a margin.
The luck-fisher's useless candidates are copies of the current best, placed at random among the
honest candidates or all at the end, and sent from a fresh identity each, from one identity, or in
turn from three or from ten identities; we did not test other timings, adaptive play or cutting
candidates differently. Winner-take-all, top-3 and the raffle are as in Section~\ref{sec:rules}. The
bonus pays a fixed amount per accepted step, the budget divided by the average number of accepted
steps without a luck-fisher, in order of acceptance, and stops when the budget is spent; the last
paid step gets what remains. This cap can only lower the luck-fisher's take.
The per-step rules have no hard cap: with no luck-fisher and no fee, their realised total exceeds $B$
in 47\% to 48\% of simulated challenges, and its 97.5th percentile is $\$31{,}972$ to $\$36{,}555$.
On an exact scorer, a bounded price curve caps the total (Section~\ref{sec:rule}); on a noisy scorer
with fresh measurements it does not, because the measured steps need not add up to one total. A rank
rule has a hard cap, with its manipulations. The rule we propose has none. A variant that scales
every payment at settlement by $\kappa=\min\{1,\text{pot}/\text{total owed}\}$, the pot being the
budget plus the fees, and $\kappa=1$ when nothing is owed, would cap the total and keep part of Proposition~\ref{prop:fee}: every payment only shrinks, so the bound $b$ of
part (a) still holds, and with it the fee result of part (b). It makes the budget a shared pool
again: an accepted step of others, lucky or real, can lower $\kappa$, so the rule is no longer
non-expropriating, and the guarantee of part (d) is lost, since copies can change $\kappa$ and with it
the other candidates' pay; we have not checked part (c), since
splitting changes the total owed. We have not simulated this variant, and we leave a hard cap without
these losses open.

\begin{table}[H]
\centering\small
\begin{tabular}{lrrrr}
\hline
 & \multicolumn{2}{c}{$\sigma=16.26$} & \multicolumn{2}{c}{$\sigma=37.9$} \\
Rule & fisher take & unpaid progress & fisher take & unpaid progress \\
\hline
Winner-take-all & $20{,}000.00$ & $100\%$ & $20{,}000.00$ & $100\%$ \\
Top-3 by total gain & $9{,}790.20$ & $58\%$ & $16{,}476.20$ & $80\%$ \\
Raffle & $7{,}360.10$ & $4\%$ & $11{,}888.26$ & $0\%$ \\
Bonus, capped at the budget & $9{,}688.18$ & $37\%$ & $11{,}087.00$ & $46\%$ \\
Per-step, deciding measurement & $8{,}044.16$ & $0\%$ & $16{,}707.12$ & $0\%$ \\
Per-step, fresh runs, no fee & $2{,}408.79$ & $1\%$ & $6{,}172.27$ & $3\%$ \\
Per-step, fresh runs, fee $b$ (net) & $1.12$ & $1\%^*$ & $-5.48$ & $3\%^*$ \\
\hline
\end{tabular}
\caption{Mean amount, in dollars, taken by a luck-fisher with $K=100$ useless candidates at the best of its eight
tested timing and identity choices, and the share of accepted honest true progress that belongs to
honest identities with no positive net pay (``unpaid progress''), under the verifier with three runs
a side and a margin; budget $20{,}000$; 20{,}000 simulated challenges per cell. The fresh rows pay $\lambda\max\{0,D_2\}$ on three fresh runs a side; with the fee, every
candidate pays $b$ ($\$24.08$ and $\$61.78$) into the pot, and the take is net of fees. $^*$Unpaid
progress is computed on pay before fees.}
\label{tab:noiserules}
\end{table}

\paragraph{The fee in simulation.} The fresh rule's price is $\$28.60$ and $\$31.47$ per mbpb at
$\sigma=16.26$ and $37.9$, set so that its mean total with no luck-fisher is $B$; a useless change
is accepted with probability $p_0=0.159$, and $b$ is $\$24.08$ and $\$61.78$. Paying on the deciding
runs instead, a useless candidate earns $\lambda s_1N'(\delta/s_1)$ per attempt
(Proposition~\ref{prop:deciding}(a)), $\$91.98$ and $\$235.99$ at the fresh rule's price: a fee would have to be $3.82$ times
larger. In the simulation the deciding rule has its own calibrated price, $\$24.99$ and $\$22.26$ per mbpb. When the useless changes' runs are as noisy as
the recipe's, the simulated pay per useless candidate matched $b$, the bound of
Proposition~\ref{prop:fee}(a), which such copies attain, in all 8 checks (two noise levels, $K=20$
and $100$, two timings), within three standard errors. When their runs had twice the standard
deviation, with the reference's unchanged, it matched in all 8 checks the exact expected pay
$\Pr(D_1\ge\delta)\,\lambda s_2/\sqrt{2\pi}$ from the proof of Proposition~\ref{prop:fee}(a), with
$s_1=s_2=\sigma\sqrt{5/3}$: $\$63.18$ and $\$162.08$. This exceeds $b$; the bound of (a) with
$\bar\sigma=2\sigma$ on both sides is $93.54$ and $239.94$. The pay on the deciding runs matched
Proposition~\ref{prop:deciding}(a) in all 16 corresponding checks.
With a fee of $b$ not refunded, or of $b/(1-p_0)$ refunded on acceptance ($\$28.63$ and $\$73.46$),
the luck-fisher's net was within two standard errors of zero ($1.12$ and $-5.48$, standard errors
$7.38$ and $18.85$, at a fee of $b$ not refunded); it ended ahead in $46\%$ of challenges at a fee of
$b$ not refunded, and in $47\%$ with the refunded fee; with $1.5\,b$ and $2b$
not refunded, it lost money on average ($1{,}202.72$ and $3{,}094.36$ at $1.5\,b$) and ended ahead in
$13\%$ and $2\%$ of challenges. On average the pot did not fall: at $\sigma=37.9$, with a fee of $b$,
its mean balance was $\$2{,}378$ without a luck-fisher and $2{,}408$ with one at its best choice;
without the fee, $-93$ and $-6{,}240$. The balance is random: with the fee, it ended below zero in
$35\%$ of challenges without the luck-fisher and $36\%$ with it ($42\%$ and $42\%$ at $\sigma=16.26$). Honest pay before fees did not
change with the luck-fisher: $\$1{,}004.65$ per honest identity without it, $1{,}004.69$ with 100
useless candidates spread through the stream. The fee must go into the pot: a fee paid elsewhere,
such as a blockchain's transaction fee, would also make attempts unprofitable, but it would not
refill the pot, and someone willing to lose money could still drain it. The splitting threshold
$2b^*$ of Proposition~\ref{prop:fee}(c) is $\$49.46$ and $\$126.89$. Each honest
identity paid $\$48.15$ and $\$123.55$ of fees for its two candidates, from $1{,}006.77$ and
$1{,}004.65$ of pay; at $\sigma=37.9$ it netted $\$881$ instead of $\$1{,}005$. With the useless changes' runs twice as noisy,
a fee set at $b$ left the luck-fisher $3{,}904$ and $10{,}036$, and a fee of $2b$ still $1{,}497$ and
$3{,}859$; in this experiment an accepted noisy change is measured later as a reference with the
recipe's noise, so it does not test noisy changes that stay noisy. Under the rank rules, the same fee
of $b$ left the luck-fisher $\$5{,}711$ to $\$13{,}822$ at $\sigma=37.9$, and $4{,}909$ under the capped
bonus.

\paragraph{The fee on real noise.} Proposition~\ref{prop:fee} assumes normal noise. We reran the
per-attempt quantities and the fee cells with each run's noise $\sigma z$, where $z$ is drawn with
replacement from the standardised residuals of our measured runs (each run's score minus the mean of
its group, rescaled to mean $0$ and variance $1$): 40 runs of the unchanged recipe, in three groups
(the renamed runs, that session's reference runs, the five earlier ones), for
$\sigma=16.26$ (standard deviation $15.98$~mbpb before rescaling) and the 26 runs of the eight recipes
for $\sigma=37.9$ ($38.52$). Neither pool is normal in shape: the standardised residuals
have skewness $-0.11$ and excess kurtosis $-1.12$ in the first pool, flatter than normal, and $0.67$
and $0.53$ in the second. Prices and fees are those of the normal model. With 2 million draws per
cell, a useless change was accepted with probability $0.163$ and $0.157$ (normal: $0.159$), and luck
paid $\$24.84$ and $\$60.55$ per attempt against $b=24.08$ and $61.78$ (Monte Carlo standard errors
$\$0.07$ and $\$0.19$), ratios $1.032$ and $0.980$ (standard errors about $0.003$). In the stream with 100 useless candidates and a fee
of $b$, the luck-fisher netted $+\$77.41$ and $-\$115.06$ (standard errors $7.48$ and $19.01$); with
$1.5\,b$, $-1{,}126$ and $-3{,}204$. In run order, the residuals' lag-one autocorrelations
($-0.20$, $-0.20$ and $-0.08$ for the candidate, reference and unchanged-recipe runs) and linear
trends were not distinguishable from zero. The standard errors are Monte Carlo errors only, not the
sampling error of 40 and 26 runs; resampling cannot produce noise beyond the largest measured run, and
the second pool mixes eight recipes.

\paragraph{Campaigns outside Proposition~\ref{prop:fee}(b) and (c).} With the simulated
challenge's parameters (three runs on each side, margins $13.26$ and $30.9$~mbpb, prices $\$28.60$ and
$\$31.47$ per mbpb), we computed two quantities in closed form. \emph{Lowering, then restoring}: the campaign's expected
net, for a lowering of $x$ from $0$ to $200$~mbpb in steps of $0.05$. \emph{Splitting}: the
expected net of $k=2,\dots,10$ equal pieces minus that of the unsplit change, for a gain $\mu$ from
$0.1$ to $300$~mbpb in steps of $0.1$. At a fee of $b$ not refunded, and at $b/(1-p_0)$ refunded on
acceptance, neither quantity was positive on these grids (the largest value was zero up to
rounding, at $x=0$ or for large $\mu$). At $1.5\,b$ not refunded, the campaign lost $\$13.95$ and
$\$35.80$ at best, and splitting lowered the expected net by at least $12.04$ and $30.89$. A fee
of only $b$, refunded on acceptance, is below the threshold of Proposition~\ref{prop:fee}(b): a
single useless candidate then nets $p_0b$ in expectation, $\$3.83$ and $\$9.82$. The two profitable
examples of Appendix~\ref{app:noise} need one measurement far noisier than the other.

\paragraph{Why not a deposit.} Proposition~\ref{prop:nonneg} leaves two ways to stop paying for luck:
take money before the measurement, as a fee, or after it, as a negative payment. A negative payment
$X=\lambda D_2$ on fresh runs pays a useless change nothing in expectation if it is always collected,
but with free identities it can be collected only from a deposit $B_s$ posted in advance; against a
finite deposit an operator walks away, and receives $\max\{X,-B_s\}$, whose expectation
$\tau N'(B_s/\tau)-B_s(1-N(B_s/\tau))$, with $\tau=\lambda s_{m'}$ for $m'$ payment runs, is positive. As an illustration, in the
simulated challenge at $\sigma=37.9$~mbpb with three fresh runs on each side, at the price of
$\$35.12$ per mbpb that makes such signed payments total $20{,}000$ on average, a deposit of
$3\tau$, our choice of coverage, is about $\$3{,}260$ per accepted step, about twice the $1{,}570$
that an accepted step earns on average at that price (its average true gain, $44.7$~mbpb, times the
price), and still leaves about $\$0.42$ of luck per accepted useless step. The ratio depends on
the design: with ten times as many payment runs, the same deposit coverage would be about two thirds
of what an accepted step earns. Honest operators would also be charged after unlucky
measurements. We judged the fee the simpler choice.
\section{Enabling on a real record: better.codes}\label{app:enabling}

Does one solver's accepted work enable another's on a real record? better.codes is the cleanest
place to ask: every accepted submission is a commit in a public Lean
repository~\cite{proximityprize2026repo}, and a proof checker decides acceptance, so scores have no
noise. We saved the repository (head \texttt{aec85b65}, 2026-10-24). It has 101 submissions on four
tracks (Lower~81, Upper~11, BinaryUpper~5, BinaryLower~4). Two submissions are by \emph{different
solvers} when their author sets (the commits' co-author trailers) are disjoint.

\subsection{Dependence, by rebuilding}\label{app:enablingbuild}\label{app:enablingreuse}

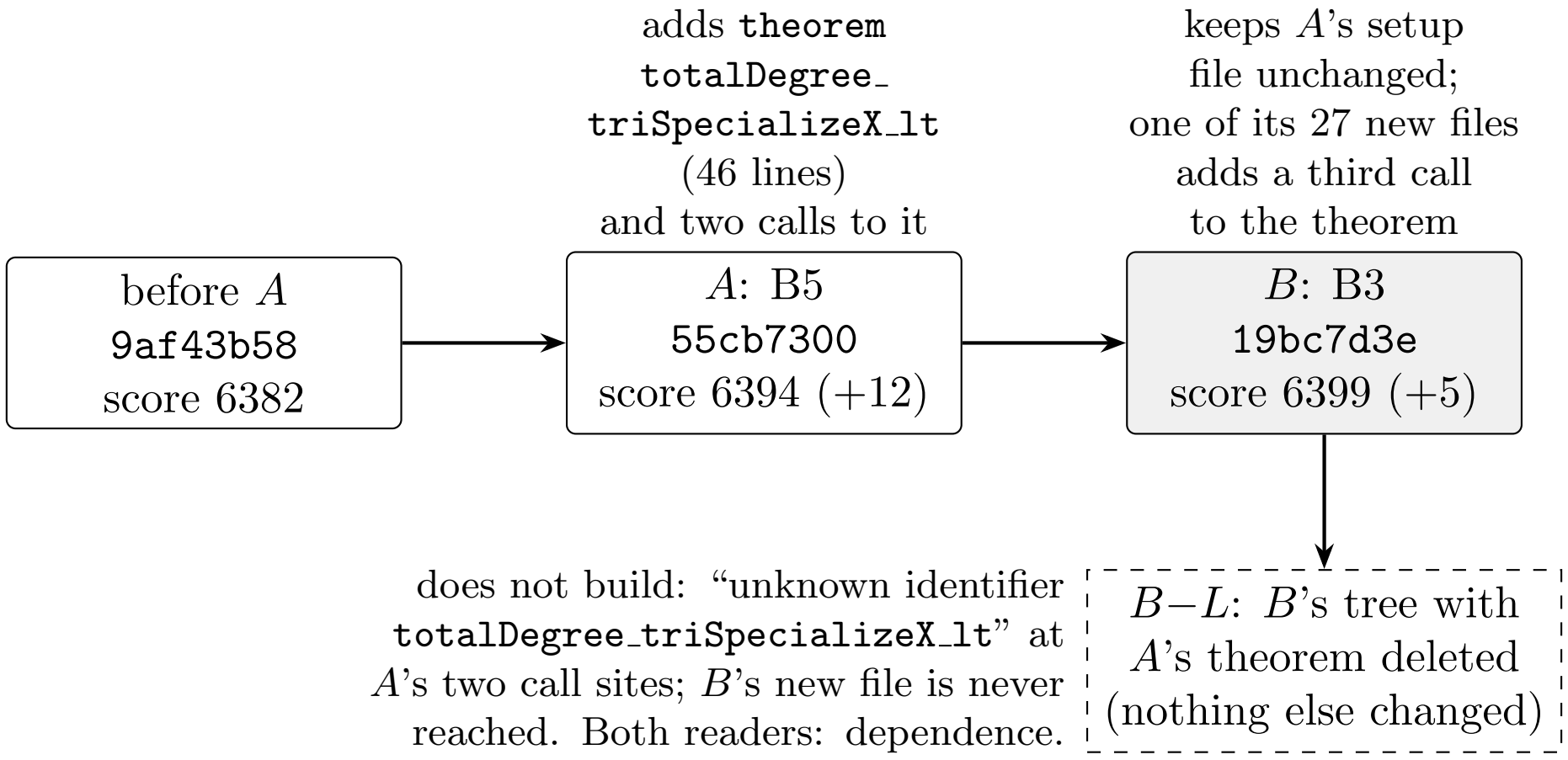
\begin{figure}[H]
\centering
\begin{tikzpicture}[x=1cm,y=1cm,
  commit/.style={draw,rounded corners=2pt,minimum width=3.1cm,minimum height=1.25cm,align=center,font=\small},
  box/.style={draw,dashed,minimum width=3.1cm,minimum height=1.25cm,align=center,font=\small},
  arr/.style={-{Stealth[length=2mm]},thick},
  lab/.style={font=\footnotesize,align=center}]
  \node[commit] (r0) at (0,0) {before $A$\\\texttt{9af43b58}\\score 6382};
  \node[commit] (a)  at (4.4,0) {$A$: B5\\\texttt{55cb7300}\\score 6394 ($+12$)};
  \node[commit,fill=black!6] (b)  at (8.8,0) {$B$: B3\\\texttt{19bc7d3e}\\score 6399 ($+5$)};
  \draw[arr] (r0) -- (a);
  \draw[arr] (a) -- (b);
  \node[lab,anchor=south,text width=3.8cm] at (4.4,0.72)
    {adds \texttt{theorem totalDegree\_}\\\texttt{triSpecializeX\_lt} (46 lines)\\and two calls to it};
  \node[lab,anchor=south,text width=3.8cm] at (8.8,0.72)
    {keeps $A$'s setup file unchanged;\\one of its 27 new files\\adds a third call to the theorem};
  \node[box] (bl) at (8.8,-2.5) {$B{-}L$: $B$'s tree with\\$A$'s theorem deleted\\(nothing else changed)};
  \draw[arr] (b) -- (bl);
  \node[lab,anchor=east,text width=6.4cm,align=right] at (6.9,-2.5)
    {does not build: ``unknown identifier \texttt{totalDegree\_triSpecializeX\_lt}'' at $A$'s
     two call sites; $B$'s new file is never reached. Both readers: dependence.};
\end{tikzpicture}
\caption{What ``rebuilding $B$ without $A$'s declarations'' does, on the pair with the simplest
record (pair-046, Lower track). The diagram shows the one
dependency the deletion test found, not $B$'s whole change: $B$ adds 27 files and edits three
(6{,}713 insertions) on top of $A$'s tree, in which $A$'s setup file with the theorem stays
unchanged. The trees before $A$, of $A$ and of $B$ were rebuilt, not rechecked: the scores are the
challenge's accepted ones. With $A$'s one theorem deleted, $B$'s accepted tree stops at the first
file that names it. The record shows that $B$'s text depends on $A$'s theorem, not that $B$ could
not have proved it alone.}
\label{fig:pair046}
\end{figure}

\paragraph{Method.} Take a pair $(A,B)$ in which $B$'s commit directly follows $A$'s on the same track
and $B$ uses names that $A$ first declared; call that set of declarations $L$. We build $B$'s
accepted tree with $L$ deleted and nothing else changed ($B{-}L$). If $B{-}L$ builds and the checker
accepts it at $B$'s score, $B$ did not need $L$. If it fails, two AI agents, Claude and GPT-6 Astra, read the log independently.
We count a \emph{dependence} only when both agree that $B{-}L$ does not build, that at least one
reported error is an unknown name of $L$, and that every other error lies in a declaration or
command that names a deleted declaration or one that had already failed on it. A build that runs
past one hour is \emph{unresolved}, never a dependence. Figure~\ref{fig:pair046} shows one pair. The build logs and
both readers' verdicts will be included in the release announced in Appendix~\ref{app:records}.
 
\paragraph{What we rebuilt.} The builds ran in a sandbox with no network. We rebuilt $B$'s accepted
tree in every pair (27 of 27 built). For the two Upper pairs and the first five Lower pairs we also
rebuilt $A$'s tree and the tree before $A$, and for the two Upper pairs the checker re-certified all
three trees at the recorded scores. From the sixth Lower pair on we built only $B$'s tree and
$B{-}L$, because one Lower tree takes 2 to 47 minutes to build and its check more than an hour. No
Lower score was rechecked: our one attempt ran the checker for 4.4 hours until it was killed with
the machine's memory nearly full, and the score was neither confirmed nor refuted.

\paragraph{Results (Table~\ref{tab:enablingbuild}).} The saved repository has 27 such pairs (25
Lower, 2 Upper). We attempted 24; in the other 3 our parser could not fix a declaration boundary in
the deletion plan. In all 24, $B$'s accepted tree built. $B{-}L$ failed to build in 22 and ran past the
one-hour limit in 2; none built. In 21 pairs both readers
traced every reported error to the deletion: 16 end with a plain compiler error and 5 on a compiler
resource limit hit while elaborating a declaration that names a deleted one (1 of them the Upper
pair). In 3 pairs we count nothing: the build of $B{-}L$ ran past the hour (2 pairs) or the build of
$A$ alone did (1 pair). No pair built without $L$, and no pair failed for a reason unrelated to the
deletion. $B$'s gains in the 21 pairs are 1 to 32 centibits, 106 in all, against a Lower record that
rose 1{,}515 centibits over 81 steps.

\begin{table}[H]\centering\scriptsize
\caption{Rebuilding $B$ without $A$'s declarations, 27 direct-successor pairs of better.codes (verdicts by two AI agents, Claude and GPT-6 Astra). Solvers are pseudonymised as B1, B2, \ldots{} in order of first appearance. $L$: declarations of $A$ deleted from $B$'s tree (count, lines, how many have a different text in $B$'s tree than in $A$'s). Errors (raw log count): unknown names of $L$ / other errors, each inside a declaration or command that names a deleted declaration or one that failed on it. For the three not-attempted pairs the deletion plan has unparsed names (036: 1 of 6; 385: 1 of 53; 452: 4 of 130); their line and change counts are sums over the parsed names only. Failing blocks: $A$ = every failing declaration is $A$'s own inherited block; mixed = $B$ wrote some. Outcome: the raw build status of $B{-}L$ (\texttt{limit} = the compiler stopped on a resource limit inside such a declaration). Verdict ``dependence'' = $B$'s accepted text does not build without $L$; unresolved = a build ran past the hour; not attempted = parser doubt in the deletion plan. Scores are the record's (centibits, Lower raises, Upper lowers); $B$'s gain is $p(AB)-p(A)$ in the track's direction.}\label{tab:enablingbuild}
\setlength{\tabcolsep}{1.8pt}
\begin{tabular}{llllrrrrrrll}
\hline
pair & $A$ & $B$ & verdict & gain$_B$ & $|L|$ & lines & changed & unk.\ $L$ & other & failing & outcome \\
\hline
005 & B1 & B2 & dependence & 2 & 1 & 2 & 1 & 1 & 0 & A & failed \\
043 & B3 & B4 & dependence & 1 & 5 & 36 & 3 & 4 & 0 & A & failed \\
046 & B5 & B3 & dependence & 5 & 1 & 46 & 0 & 2 & 0 & A & failed \\
066 & B6 & B7 & dependence & 2 & 7 & 1622 & 6 & 5 & 4 & A & failed \\
084 & B8 & B3 & unresolved & 1 & 22 & 577 & 10 & 2 & 1 & A & failed \\
104 & B3 & B6 & dependence & 1 & 7 & 613 & 4 & 11 & 1 & mixed & failed \\
123 & B3 & B9 & dependence & 1 & 10 & 290 & 0 & 10 & 4 & A & failed \\
124 & B3 & B7 & dependence & 1 & 38 & 821 & 6 & 31 & 4 & A & failed \\
150 & B5 & B3 & dependence & 2 & 9 & 46 & 1 & 16 & 8 & mixed & failed \\
158 & B10 & B1 & dependence & 1 & 56 & 487 & 41 & 43 & 17 & A & limit \\
209 & B3 & B11 & dependence & 2 & 1 & 1 & 1 & 32 & 13 & mixed & failed \\
281 & B10 & B3 & dependence & 2 & 102 & 762 & 57 & 14 & 5 & A & limit \\
312 & B12 & B5 & dependence & 1 & 358 & 1887 & 102 & 32 & 16 & mixed & failed \\
317 & B7 & B5 & unresolved & 2 & 34 & 372 & 9 & 63 & 22 & -- & timed out \\
324 & B11 & B6 & dependence & 23 & 90 & 352 & 31 & 74 & 11 & A & failed \\
328 & B2 & B13 & dependence & 1 & 118 & 2245 & 49 & 28 & 13 & A & failed \\
344 & B5 & B3 & dependence & 1 & 48 & 186 & 13 & 37 & 4 & A & failed \\
355 & B14 & B5 & dependence & 32 & 158 & 2491 & 78 & 32 & 10 & mixed & failed \\
359 & B7 & B15 & dependence & 1 & 82 & 857 & 26 & 114 & 16 & mixed & failed \\
385 & B3 & B2 & not attempted & 1 & 53 & $\ge$716 & $\ge$50 & -- & -- & -- & not run \\
411 & B16 & B9 & dependence & 22 & 83 & 516 & 13 & 113 & 77 & A & failed \\
421 & B11 & B5 & unresolved & 18 & 141 & 1285 & 53 & 190 & 40 & -- & timed out \\
452 & B17 & B18 & not attempted & 1 & 130 & $\ge$3779 & $\ge$27 & -- & -- & -- & not run \\
491 & B10 & B5 & dependence & 1 & 612 & 5913 & 373 & 146 & 83 & A & limit \\
516 & B2 & B19 & dependence & 3 & 123 & 1314 & 53 & 46 & 55 & mixed & limit \\
036-U & B20 & B3 & not attempted & 2 & 6 & $\ge$81 & $\ge$2 & -- & -- & -- & not run \\
122-U & B21 & B17 & dependence & 1 & 2 & 4 & 0 & 20 & 10 & mixed & limit \\
\hline
\end{tabular}
\end{table}

\paragraph{What this shows.} $B$'s accepted text depends on $A$'s declarations. It does not show that
$B$ could not have written them: in pair-122 the deleted names are an alias for an external type and a
three-line lemma. In 13 of the 21 pairs every failing declaration is $A$'s own, inherited with the
tree; there the dependence is textual, since $B$ took $A$'s tree and kept what it did not need to
change. In the other 8, declarations that $B$ wrote also fail. The royalty of
Section~\ref{sec:royalty} is defined on ablations that build and score; here the ablation does not
build, so the definition gives no interaction. A fallback that treats a failed build as dependence
pays on $B$'s whole recorded gain, which overstates $A$'s contribution where any solver could have
restated the names.

\subsection{Not every step names a predecessor}\label{app:notenabled}

A declaration \emph{uses} a name when the name occurs as a token in its text and resolves to that
declaration under an approximate reading of namespaces; this is a textual match, not Lean's
elaboration. Under it, 86 of the 101 submissions add or change a declaration that uses a name first
declared by another solver. The other 15 (Lower~3 of~81, Upper~8 of~11, BinaryLower~1 of~4,
BinaryUpper~3 of~5) use none, and their gains are not small: median 32 centibits, maximum 1{,}045,
against a median of 2 for the others. Whether the score of these steps depends on earlier work is
not tested: a declaration can depend on another without naming it, and changes that name nothing of
each other can still interact in the score.

\subsection{The royalty on the measured pairs}\label{app:enablingroyalty}

\paragraph{The fallback, and its two costs.} A sponsor can count a failed build as full dependence,
so that $d$'s whole gain is the interaction. Replayed with the clamped royalty, no window and no cap,
on the 20 pairs of better.codes' largest leaderboard where we confirmed the dependence, against the
budget of its 81 accepted submissions, it moves $1.7\%$ of the budget at $\alpha=\tfrac14$ and
$3.5\%$ at $\alpha=\tfrac12$ (below). It overstates the earlier contribution when
the follower could have rewritten what it used, and a follower can copy the earlier code under its
own names, so that removing $c$ breaks nothing; detecting the copy means judging ancestry again.
This is the hiding of Section~\ref{sec:impossibility}: when restating the earlier code removes the
charge and changes nothing else, a follower pays the royalty only while it costs less than
restating. It need not remove the charge: if a later step by another operator still calls $c$,
removing $c$ still breaks the final artifact. Values for cooperative games under precedence
constraints~\cite{faigle1992precedence} or permission structures~\cite{gilles1992permission} fit the
case where $d$ cannot exist without $c$, but take the dependence as known.

We replay the per-step rule with the follower-financed clamped royalty on the Lower leaderboard, at
one dollar per centibit (budget 1{,}515 dollars for 81 accepted submissions). A royalty is paid only
on the 20 Lower pairs with verdict ``dependence'', and, since $B{-}L$ has no score, we assign each
pair the \emph{failed-build convention} $h=$ $B$'s recorded gain. There is no window, reserve or cap;
an identity that owes on several pairs pays the sum. With $\alpha=1/4$ the royalties move 26.25
dollars of 1{,}515 (1.7\%) among 16 identities; with $\alpha=1/2$, 52.50 dollars (3.5\%). No
identity's royalties exceed its per-step pay. Restricted to the 7 pairs in which declarations that
$B$ wrote fail, 21 dollars move at $\alpha=1/2$. On this record the royalty is a correction at the
margin, not a transfer of the budget. These are assigned, not measured, interactions.

\subsection{Removal elsewhere, and its cost}\label{app:removalelsewhere}

The royalty as defined (Definition~\ref{def:pairroyalty}) says nothing about a pair whose ablation
does not build or run, and a deployment must extend it.

\paragraph{Removal worked where changes were switchable components.} In our training experiment the
three changes ran in all eight on and off combinations, and \citet{hessel2018rainbow} compared
variants of a reinforcement-learning agent ``each corresponding to removing a single component''. On
the speedrun we did not test it (Appendix~\ref{app:speedruncensus}).

\paragraph{Where removal works, noise makes each measurement expensive.} An interaction is a second
difference of four noisy means; at our $\sigma=0.038$, resolving one worth one percent of the
baseline score at $1.96$ standard errors needs 132 runs per pair, about thirteen machine-hours, by a planning
calculation that is not a bound. When we ran one interaction with ten seeds per cell, the prediction
we fixed before the first run, a positive interaction with a $95\%$ interval excluding zero, failed
(Appendix~\ref{app:interactionreal}). When the per-step payments are exact and only the ablations are
noisy, the signed royalty is unbiased on every pair, while the clamped royalty pays
$\alpha\lambda\,\mathbb{E}[\eta^+]$ in expectation on every pair with no interaction, where $\eta$
is the mean-zero measurement error (Proposition~\ref{prop:royaltynoise}).
\section{The four records in detail}\label{app:records}

We computed everything below from the public pages and repositories, which we saved in October
and November 2026. The records show what was accepted, when, by whom and by how much. They do not show what any
step cost, what anyone intended, or whether two changes interacted. Table~\ref{tab:evidence}
summarises the four challenges. ECDSA.Fail, the fifth challenge of Table~\ref{tab:which}, rests on
its paper~\cite{long2026ecdsafail} alone; we hold no record of it. The parsed records, the replay code, the simulation code, the
training-run logs and the experiment specifications will be released with the arXiv version of
this paper.

\paragraph{Progress comes in many small steps, and a few large early ones.} The median speedrun
record improves on the previous one by $1.2\%$, and 32 of 91 steps by less than $1\%$; on
better.codes, 39 of 80 accepted steps raise the bound by $0.01$~bits, the smallest amount the page
displays. A few steps are huge and early: QSB's fifth step, on its second day, gained $158.71\%$.

\paragraph{Later work touches earlier work by others.} In the speedrun, the Muon optimiser (a new
rule for updating the model's weights), introduced in the third record (October 2024; 31.4 to 24.9
minutes), is touched again in 11 later records, 10 of them credited to other handles. This shows
continuity, not that the later change needed the earlier one. On better.codes, in 21 of the 24
rebuilt pairs two AI agents, Claude and GPT-6 Astra, each reading the build logs on its own, traced
every reported error to the deleted declarations (Appendix~\ref{app:enablingbuild}). In 15 of the
101 accepted steps, no added or changed declaration uses a name first declared by another solver
(Appendix~\ref{app:notenabled}).

\begin{table}[!tb]
\centering\footnotesize
\setlength{\tabcolsep}{4pt}
\begin{tabular}{>{\raggedright\arraybackslash}p{0.17\textwidth}>{\raggedright\arraybackslash}p{0.19\textwidth}>{\raggedright\arraybackslash}p{0.17\textwidth}>{\raggedright\arraybackslash}p{0.17\textwidth}>{\raggedright\arraybackslash}p{0.2\textwidth}}
\hline
 & NanoGPT speedrun~\cite{jordan2026speedrun} & better.codes~\cite{bettercodes2026leaderboard} & QSB, pinning~\cite{yukon2026qsbleaderboard} & Muse Spark Open Kernel~\cite{paradigm2026okc} \\
\hline
Score & minutes to 3.28 val.\ loss on 8 H100 & soundness lower bound, bits (Lean-checked) & verified candidates/s on one GPU & kernel boards: speed-up over SGLang on one H200; serving track: throughput and interactivity on 8 H200 \\
Public frontier (O2) & yes: every record is the code to beat & yes: promoted submissions reset the floor & yes & not on the board we inspected: best result per account, no artifact \\
Record & 92 records, 2024-05 to 2026-08, 45 to 0.665 min & 81 promotions, 22 solvers, 2026-08-20 to 10-01, 53.12 to 68.15 bits & 44 promotions, 27 solvers, 2026-09-15 to 10-05, $198$M to $1{,}036$M & kernel board \texttt{muse\_bf16\_proj}: 10 rows captured, more paginated \\
Step size & median $1.2\%$; 32 of 91 under $1\%$; 6 over $20\%$ & median $0.02$ bits; 39 of 80 at the display minimum $0.01$ & median $1.35\%$; 10 of 41 rows under $1\%$; 2 over $20\%$ & --- \\
Noise rule & $p<0.01$ over run logs that the mean loss is $\le3.28$; waived for systems-only speed-ups & exact (proof checker) & one fresh run per candidate on the organiser's GPU, against the best's stored score; hits re-derived on a CPU; noise $\approx1/\sqrt{K}$, about $0.31\%$ & serving track: median of $\ge3$ reruns of the top 25 on held-out data \\
Reward rule & none documented & discretionary (no committed rule) & two-period top-3 + equal-share raffle at $3\%$ (\$20,000) & serving track: top-10 ranked prizes (\$100,000); kernel boards: none \\
\hline
\end{tabular}
\caption{Four challenges, from their public pages on 2026-10-23 and, for the speedrun, the
repository at commit \texttt{4ea6b93}. Step sizes are relative to the previous record (speedrun,
QSB: per leaderboard row; a QSB row marked ``2 runs'' shows the combined gain) or absolute
(better.codes: per promoted submission). The speedrun's two re-timings of record~21 are not steps;
record~22 against the old time shows $-1.9\%$, against the retimed baseline $+0.8\%$; the counts
quoted are the same either way. Contributors: speedrun 52 handles (35 credited once), better.codes
22 solvers (top solver 21 of 81), QSB 27 (top solver 7 of 44).}
\label{tab:evidence}
\end{table}

\paragraph{What else the records show.}
\begin{itemize}\setlength{\itemsep}{1pt}\setlength{\parskip}{0pt}
\item \emph{Large steps come early.} Six speedrun steps exceed $20\%$, five of them among the first
  twelve records. QSB's first accepted step gained $35.37\%$ and its fifth, on the second day,
  $158.71\%$. QSB's organisers wrote that ``early entrants can take very large gains simply by being
  first'' and reset the baseline after week~1 for that reason~\cite{starkware2026qsbrewards}.
\item \emph{Reuse is common.} Besides Muon, the speedrun's value embeddings (record~14) reappear in 7
  later records, its attention windows (record~12) in 5, FP8 arithmetic (record~19) in 6 and the
  bigram hash (record~62) in 3, each time with a handle from outside the original record. The
  repository's rules say: ``Incorporating open PRs into a new record is strongly encouraged.'' We
  matched the descriptions by text, and one AI agent checked each match against the descriptions;
  handles are not verified people.
\item \emph{Bundles and removals.} 36 of the 92 speedrun records describe two or more distinct changes
  in one step (47 by a delimiter heuristic). Under a fixed bonus per submission a bundle earns one
  bonus where a split into separately accepted pieces with the same end points would earn several;
  under the per-step rule both pay the same. We did not measure whether the observed bundles could
  have been split that way. Seven records
  remove or revert a component; our royalty theorems (Section~\ref{sec:royaltytheory}) are for
  records that only add changes.
\item \emph{Runs of minimal steps.} On better.codes, one solver made five consecutive accepted
  submissions of $0.01$~bits each and holds 21 of the 81. Under the per-step rule, five steps of
  $0.01$ pay the same as one step of $0.05$ between the same end points.
\item \emph{Noise is handled by hand-written rules.} The speedrun requires ``enough run logs to attain
  a statistical significance level of $p<0.01$ that their mean val loss is $\le3.28$'', waived for
  speed-ups that do not touch the model. The kernel challenge, under its rules of 23 October 2026, decides its
  serving prizes by the median of at least three reruns of the top~25 entries on held-out data. Its
  rules page changed on 3 November: five reruns paired with stock runs, a geometric mean, and
  winners chosen with a review of the finalists' patches by the SGLang maintainers, so (O3) still
  fails. better.codes has no noise:
  a proof checker decides. QSB measures each candidate once, on the organiser's GPU and a fresh problem
  instance, against the stored score of the current best; its scoring rules put the relative noise of
  one score at about $1/\sqrt{K}$ for $K$ verified hits~\cite{layr2026qsbharness}, which a
  submitter's note on the challenge page computes as $0.31\%$ at the frontier. That note re-measured
  the frontier with no code change and observes that, with the margin then at zero, ``any positive
  draw promotes''~\cite{yukon2026qsbleaderboard}.
\item \emph{What pays.} QSB publishes a two-period top-3 with an equal-share raffle at a $3\%$
  threshold; its winners page lists 22 raffle qualifiers with a claimable share of \$272.72
  each~\cite{starkware2026qsbleaderboard}. We read the schedule after the contest and have no receipts.
  The kernel challenge promises ranked prizes; better.codes pays at its discretion; the speedrun
  documents no payment rule. None of the four uses the per-step rule.
\end{itemize}

\paragraph{What the QSB replay can and cannot show.} We replay the 44 accepted pinning steps
embedded in the saved leaderboard page, in exact arithmetic, with the published schedule and the
same \$20,000. The second track is not on the pages we saved, so we replay pinning alone, and every
statement about the replay is about that track. The replay reproduces 8 of the 20 published raffle
gains of solvers with a pinning step to two decimals and under-counts the rest, as a missing track predicts. Three solvers whose
accounts QSB tags as staff of the two sponsoring firms clear the raffle's threshold on pinning alone
but are not in the published list of qualifiers. QSB's page says that such members ``are not eligible for the rewards''~\cite{starkware2026qsbleaderboard}. We therefore pay only the 24 eligible solvers; the
steps of the other three still move the frontier. Without them, the replay's 18 raffle qualifiers on
pinning alone are all in the published list, and the four published qualifiers it misses have no
pinning step or stay below $3\%$ on pinning alone, as a missing track predicts.
Table~\ref{tab:replayfull} adds two other ways to set the price of the per-step rule to
Table~\ref{tab:replay}.

\begin{table}[!htb]
\centering\footnotesize
\begin{tabular}{lrrrrr}
\hline
Rule (same \$20,000) & Gini & top-3 share & Sybil split & pooling & step split \\
\hline
Published schedule, replayed (top-3 $\times2$ + raffle) & 0.687 & 0.670 & $+4{,}167$ & $+3{,}286$ & 0 \\
Per-step, one $\lambda$ & 0.574 & 0.483 & 0 & 0 & 0 \\
Per-step, $\lambda$ per period (QSB's reset) & 0.781 & 0.671 & 0 & 0 & 0 \\
Per-step, concave $\Phi(x)=1-x_0/x$ & 0.738 & 0.690 & 0 & 0 & 0 \\
Winner-take-all & 0.958 & 1.000 & 0 & 0 & 0 \\
Per-submission bonus (\$487.80) & 0.349 & 0.390 & 0 & 0 & $+488$ \\
\hline
\end{tabular}
\caption{The QSB pinning record (44 steps by 27 solvers) under six rules, paying its 24 eligible
solvers. Gini and top-3 share over the 24 eligible solvers. The last three columns are the largest gain, in dollars, that one solver (or one
pair, for pooling) obtains by re-submitting its own steps under 2 to 4 identities, by two solvers
appearing as one, or by cutting one step in two at the midpoint of its gain; prices and prize
pools fixed in advance from the honest record. No solver is claimed to have done any of this.}
\label{tab:replayfull}
\end{table}

\paragraph{How the price is set matters.} With one price $\lambda$, the per-step rule is less
concentrated than the replayed schedule, by Gini (0.574 against 0.687) and by top-3 share (0.483
against 0.670); its largest payment to one step is \$1,516.07, $7.6\%$ of the budget. The two other
price curves are more concentrated, for opposite reasons. A per-step rule priced per period, our
counterfactual built on QSB's budgets for its two periods (\$2,000 for week~1 and \$18,000 for
weeks~2--3) and not QSB's published ranking, prices a unit of later progress about 15 times a unit of
early progress, so the \$18,000 go to ten late steps by six solvers. The concave $\Phi(x)=1-x_0/x$
prices early progress higher, so the solvers of the two largest early eligible steps are paid
\$9,134.51 and \$3,391.84. Who is eligible matters as much. Counting the three ineligible solvers as
eligible, the one price pays an ineligible solver's $+158.71\%$ step \$8,881.95, 44\% of the budget, and is
more concentrated than the schedule by Gini (0.706 against 0.682). Where progress was concentrated,
so is the pay. On the speedrun and
better.codes, with nominal prices, the per-step rule is less concentrated than the QSB shape
transposed to those budgets (Gini 0.921 against 0.944; 0.873 against 0.923). On the speedrun the
Sybil split pays under the transposed top-3 and raffle; on better.codes the splits we tested gained
nothing (two of our searches there are heuristic lower bounds). Under no per-step variant does any
split pay.

\paragraph{Copying, replayed.}\label{app:copyreplay} We appended one hypothetical step to the end of each record: a
new identity copies the final best and adds the record's smallest real step ($52{,}006$ verified
candidates per second on QSB, $0.24$~s on the speedrun, $1$~centibit on better.codes).
Winner-take-all pays this copier the whole budget on all three records (\$20{,}000 on QSB; \$2{,}665
and \$1{,}515 at the nominal prices), all of it taken from the identity that held the record. The
published QSB schedule, top-3 by total improvement and the raffle pay this minimal copier nothing,
because its total improvement is only its own small step. That zero holds only below the raffle's
threshold: an appended step of $0.42\%$ of QSB's final score reaches the raffle and is paid \$315.79 by the
raffle and by the published schedule, all of it from the other identities' shares (on better.codes,
\$252.50 by the raffle for 159~centibits); whether such a step is cheap to produce is not measured.
The per-submission bonus pays the copier a full bonus (\$487.80
on QSB). The per-step rule pays the value of the step (\$2.48 on QSB at one $\lambda$) and changes
nobody else's pay. On QSB, as everywhere in this appendix, only the 24 eligible solvers and the copier
are paid.

\subsection{Removal on the speedrun: a textual comparison}\label{app:speedruncensus}

The NanoGPT speedrun
is a training challenge. We compared the recovered code of each of its records (the training script
and the files it imports) with that of the last earlier record whose code we could recover, in the
order of the record table, and labelled each line with the record whose comparison added it along
this chain (a line removed and later added again takes the later label). In 80 of the 86 comparisons, the record deletes or replaces at least one line labelled
with a record credited to none of its contributors; the count is also 80 of 86 under two other ways
of labelling the lines, and 76 of 82 if we leave out the four records whose code we could not choose
unambiguously. No record changes only numbers. This is a textual comparison of selected snapshots in
an imposed order: it does not establish ancestry or authorship, and we did not test whether a version
of the code without an earlier change could be built, automatically or by hand.

\subsection{A counterfactual on the speedrun}\label{app:speedruncf}

The speedrun documents no payment rule. We split a budget $B=10{,}000$ over its 91 steps (the two re-timings of
record~21 are not steps; they move the reference time and pay nothing). A step's pay is shared
equally among the handles credited for it; names in the record table that are not credited handles
(6 steps) receive nothing. Record~66 credits no contributor (its contributors cell is ``-''): its pay
goes to one anonymous placeholder, which counts among the 52 recipients (51 credited handles and the
placeholder) for the paid count, Gini and the top-3 share, and is left out of every manipulation. The linear price pays $B/2{,}664.96$ per second saved. The log price pays
a step $\mu\log(t_{\mathrm{before}}/t_{\mathrm{after}})$, with $\mu$ set so that the 91 steps pay $B$.
Top-3 prizes ($3{,}333$, $2{,}000$, $1{,}333$) go to the identities with the largest total
improvement, and an equal-share raffle of $3{,}333$ to every identity whose total improvement is at
least 81 seconds, as in our QSB replay; together they pay $B$. The bonus pays $B/91$ per step.
Manipulations are computed on the record as in the QSB replay: each identity's steps split among
two to four fake identities (exhaustively for identities with at most 10 steps; for the one identity
with 18 steps, a heuristic search, so its gain is a lower bound), one step split at the midpoint of
its gain, and every pair of the 51 credited handles (1{,}275 pairs) merged into one. For winner-take-all, top-3 and the raffle,
which pay identities rather than steps, the shares of the first 10 and last 41 steps apportion an
identity's pay over its steps in proportion to its credited gain.

\begin{table}[H]
\centering\footnotesize
\begin{tabular}{lrrrrrrr}
\hline
Rule & paid & Gini & top-3 & first 10 & last 41 & fake identities & pooling \\
\hline
Per-step, per second & 52 & 0.92 & 0.81 & $85.1\%$ & $3.2\%$ & 0 & 0 \\
Per-step, log & 52 & 0.80 & 0.52 & $43.2\%$ & $26.8\%$ & 0 & 0 \\
Winner-take-all & 1 & 0.98 & 1.00 & $0\%$ & $100\%$ & 0 & 0 \\
Top-3 (pays $\tfrac23B$) & 3 & 0.95 & 1.00 & $99.7\%$ & $0\%$ & 3{,}333 & 1{,}333 \\
Raffle, 81~s (pays $\tfrac13B$) & 4 & 0.92 & 0.75 & $74.8\%$ & $0\%$ & 1{,}071 & 667 \\
Bonus per step & 52 & 0.55 & 0.35 & $11.0\%$ & $45.1\%$ & 0 & 0 \\
\hline
\end{tabular}
\caption{The speedrun's 91 steps under six rules, as a counterfactual with $B=10{,}000$ (the speedrun
documents no payment rule). ``paid'': identities paid, of 52 (51 credited handles and the placeholder of record~66); Gini over
the 52; ``top-3'': share of
the paid amount going to the three best-paid identities; ``first 10'', ``last 41'': share of the
rule's payment for those steps (apportioned by credited gain for the three rank rules); the last two
columns: the largest gain of one identity from splitting its steps among fake identities, and of a
pair of identities from merging into one. Under the log price the computed gains are below
$10^{-12}$ (floating-point rounding). The bonus also pays $B/91\approx110$ for splitting any one step.}
\label{tab:speedruncf}
\end{table}
\section{Related work, extended}\label{app:relatedextra}

\paragraph{Open autoresearch, in more detail.} \citet{long2026ecdsafail} write that ``evaluators and
incentives therefore influence which discoveries are pursued and preserved''; Section~\ref{sec:which}
uses their nonce rerolls as evidence. Gauntlet names peer copying and registering many times as
attacks, and defends against both by giving each peer its own data to train
on~\cite{lidin2025gauntlet}. We study the payment rule itself: what it can guarantee when identities
are free, scores are noisy and nobody rules on a candidate.

\paragraph{Machine-learning contests, in more detail.} In the contest of
\citet{abernethy2011collaborative}, the mechanism publishes the current model; a participant pays to
update it; once the test labels are revealed, each update earns by how much it lowered the loss on
them. The total is the improvement of the final model, and participants post collateral, since an
update that raises the loss loses money. The basic mechanism pays at the end; they also sketch paying
in stages, each on a part of the test set. \citet{waggoner2015market} extend the design to buying
data, with privacy. \citet{harris2019decentralized} propose a version for Ethereum smart contracts,
with a stake on each update, and note that creating false identities then gains nothing. In these designs, an update is scored by its loss on the revealed
labels, so the same update always gets the same score. We ask what such a rule needs when anyone can
submit under free identities, nobody rules on a submission, a step is accepted and paid on scores
measured with noise, and contributors want credit for enabling later work.

\paragraph{Audit contests.} Audit contests meet the problem in practice: Code4rena-style pots cut the
points of a finding by its number of duplicates, and an analysis shows that two accounts submitting a
finding that others also found still earn more than one~\cite{yaudit2026sybil}.

\paragraph{Personhood credentials make identities costly; the per-step rule makes them irrelevant to pay.}
A challenge could require credentials that let a person show online that they are
real~\cite{adler2024personhood}. With exact scores, the per-step rule does not need one
(Proposition~\ref{prop:potential}).

\paragraph{Open participation is older than open frontier challenges.} Volunteer computing lets anyone donate
computing time to a science project: Folding@home simulates proteins on the spare cycles that
volunteers donate~\cite{foldingathome2026about}, and the BOINC platform runs such
projects~\cite{anderson2004boinc}. BOINC treats its participants as untrusted. A project can compute
each unit of work several times on different machines, compare the results and keep one they agree
on, because results are sometimes wrong, ``occasionally from malicious participants''. The same
mechanism makes cheating on credit hard: each correct result is granted only the average or the
minimum of the credit claimed by the correct results. Participants are ``highly motivated by
credit'' and by their rank on leaderboards~\cite{anderson2004boinc}. This is our quorum of verifiers,
with credit for work done rather than pay for progress made. The Polymath Project showed that many
contributors can work together on a hard mathematical problem~\cite{gowers2009polymath}. The Netflix Prize offered \$1 million for a $10\%$ improvement of Netflix's
recommendations, with a public leaderboard. Two teams finished above the threshold with the same
score, and the one that had submitted 20 minutes earlier won. The winner was an alliance: BellKor
merged with BigChaos in 2008 and added Pragmatic Theory in 2009~\cite{bell2010alltogether}. A single
prize to the best entry is the winner-take-all rule of Section~\ref{sec:rule}.

\paragraph{Contests and leaderboards study effort; we study robustness.} Rank-order tournaments
\cite{lazear1981rank}, prize allocation~\cite{moldovanu2001optimal} and dynamic contests
with public leaderboards, with evidence from Kaggle~\cite{lemus2021dynamic} and prizes contingent on
the contest's history~\cite{lemus2025contingent}, study effort when solutions are private. Here
solutions are public and copyable, and the question is robustness. \citet{blum2015ladder} limit
overfitting of a public leaderboard by releasing a new score only on a significant improvement; a
challenge that shows every candidate's score would need such a rule, which we do not specify.

\paragraph{Trustworthy measurement is a layer below the payment rule.} We take the measurements as
reported honestly. Recent designs that produce them rely on one trusted party: the agent-commerce
standard ERC-8183 lets one designated evaluator decide each job~\cite{crapis2026erc8183}, and
\citet{ren2026outerloop} has a trusted kernel
measure each submission of untrusted coding agents on a seed the agent never sees. Outerloop treats
agents as untrusted but not adversarial, assigns credit, not payment, and credits a change on the
measurement that decides acceptance, which, as the paper notes, overstates gains near the noise floor.
\citet{hu2026winners} measure the same effect in self-improvement loops of language models, and find
that scoring on items never used for selection removes the average bias. The per-step rule takes such
measurements as input and asks what to pay.

\paragraph{Charging for each use against cheap identities is an old idea.} \citet{dwork1992pricing}
proposed to fight junk mail by requiring ``a moderately hard, but not intractable'' computation for
each message, ``thus preventing frivolous use''; they preferred this computational cost to a usage
fee. Our fee's payment differs from that of private prediction markets: it is never negative, made on
fresh runs after a test for acceptance, under free identities, and we check a refund on acceptance
and splitting. In stochastic contests, \citet{whitmeyer2023submission} shows that a positive
submission cost helps a designer who wants the best entry to be as good as possible. The other way to
charge, a negative payment backed by a stake, is the design of \citet{abernethy2011collaborative} and
\citet{harris2019decentralized}, and Numerai deploys it~\cite{numerai2026staking}; in our simulated
challenge, a stake covering three standard deviations of the payment would be about twice what an
accepted step earns (Appendix~\ref{app:noisyscorer}).

\paragraph{Blockchain incentives are the nearest deployed designs.} Mining-pool reward
functions~\cite{schrijvers2017pool}, selfish mining~\cite{eyal2018majority}, proportional reward
splitting~\cite{aumayr2025proportional} and front-running~\cite{daian2020flash} are the nearest
deployed incentive designs; a verifier who submits a renamed copy of a revealed change would be
front-running. Bitcoin's mining puzzle~\cite{miller2015nonoutsourceable} has open entry
and no referee, but no public frontier: a solution seeds the next puzzle without improving a shared
artifact, so copying, splitting and merging have no analogue in it.

\begin{thebibliography}{76}
\providecommand{\natexlab}[1]{#1}
\providecommand{\url}[1]{\texttt{#1}}
\expandafter\ifx\csname urlstyle\endcsname\relax
  \providecommand{\doi}[1]{doi: #1}\else
  \providecommand{\doi}{doi: \begingroup \urlstyle{rm}\Url}\fi

\bibitem[Abernethy and Frongillo(2011)]{abernethy2011collaborative}
Jacob Abernethy and Rafael~M. Frongillo.
\newblock A collaborative mechanism for crowdsourcing prediction problems.
\newblock In \emph{Advances in Neural Information Processing Systems 24
  (NeurIPS 2011)}. Curran Associates, 2011.
\newblock URL
  \url{https://proceedings.neurips.cc/paper/2011/hash/4edaa105d5f53590338791951e38c3ad-Abstract.html}.

\bibitem[Abernethy et~al.(2011)Abernethy, Chen, and
  Wortman~Vaughan]{abernethy2011optimization}
Jacob Abernethy, Yiling Chen, and Jennifer Wortman~Vaughan.
\newblock An optimization-based framework for automated market-making.
\newblock In \emph{Proceedings of the 12th ACM Conference on Electronic
  Commerce (EC '11)}. ACM, 2011.
\newblock \doi{10.1145/1993574.1993621}.

\bibitem[Adler et~al.(2024)Adler, Hitzig, Jain, et~al.]{adler2024personhood}
Steven Adler, Zo{\"e} Hitzig, Shrey Jain, et~al.
\newblock Personhood credentials: Artificial intelligence and the value of
  privacy-preserving tools to distinguish who is real online.
\newblock arXiv preprint arXiv:2408.07892, 2024.

\bibitem[Agarwal et~al.(2019)Agarwal, Dahleh, and
  Sarkar]{agarwal2019marketplace}
Anish Agarwal, Munther Dahleh, and Tuhin Sarkar.
\newblock A marketplace for data: An algorithmic solution.
\newblock In \emph{Proceedings of the 2019 ACM Conference on Economics and
  Computation (EC '19)}, pages 701--726, 2019.
\newblock \doi{10.1145/3328526.3329589}.

\bibitem[Anderson(2004)]{anderson2004boinc}
David~P. Anderson.
\newblock {BOINC}: A system for public-resource computing and storage.
\newblock In \emph{Fifth IEEE/ACM International Workshop on Grid Computing
  (GRID 2004)}, pages 4--10, 2004.
\newblock \doi{10.1109/GRID.2004.14}.

\bibitem[{ARC Prize Inc}(2024)]{arcprize2024rules}
{ARC Prize Inc}.
\newblock {ARC} prize 2024: competition rules.
\newblock Kaggle competition page, rules updated 2024-09-18, 2024.
\newblock URL \url{https://www.kaggle.com/c/arc-prize-2024/rules}.
\newblock Accessed 2026-10-23.

\bibitem[Aumayr et~al.(2026)Aumayr, Avarikioti, Karakostas, Kreder, and
  Shastry]{aumayr2025proportional}
Lukas Aumayr, Zeta Avarikioti, Dimitris Karakostas, Karl Kreder, and Shreekara
  Shastry.
\newblock Optimal reward allocation via proportional splitting.
\newblock In \emph{8th Conference on Advances in Financial Technologies (AFT
  2026)}, volume 395 of \emph{LIPIcs}, pages 30:1--30:23. Schloss Dagstuhl --
  Leibniz-Zentrum f{\"u}r Informatik, 2026.
\newblock \doi{10.4230/LIPIcs.AFT.2026.30}.

\bibitem[Babaioff et~al.(2012)Babaioff, Dobzinski, Oren, and
  Zohar]{babaioff2012redballoons}
Moshe Babaioff, Shahar Dobzinski, Sigal Oren, and Aviv Zohar.
\newblock On bitcoin and red balloons.
\newblock In \emph{Proceedings of the 13th ACM Conference on Electronic
  Commerce (EC '12)}, pages 56--73, 2012.
\newblock \doi{10.1145/2229012.2229022}.
\newblock Proof comparison checked in arXiv:1111.2626v2.

\bibitem[Banerjee et~al.(2014)Banerjee, Goel, and
  Krishnaswamy]{banerjee2014reincentivizing}
Siddhartha Banerjee, Ashish Goel, and Anilesh~Kollagunta Krishnaswamy.
\newblock Re-incentivizing discovery: Mechanisms for partial-progress sharing
  in research.
\newblock In \emph{Proceedings of the Fifteenth ACM Conference on Economics and
  Computation (EC '14)}, pages 149--166, 2014.
\newblock \doi{10.1145/2600057.2602888}.

\bibitem[Bell et~al.(2010)Bell, Koren, and Volinsky]{bell2010alltogether}
Robert~M. Bell, Yehuda Koren, and Chris Volinsky.
\newblock All together now: A perspective on the {Netflix} prize.
\newblock \emph{CHANCE}, 23\penalty0 (1):\penalty0 24--29, 2010.
\newblock \doi{10.1080/09332480.2010.10739787}.

\bibitem[Blum and Hardt(2015)]{blum2015ladder}
Avrim Blum and Moritz Hardt.
\newblock The ladder: A reliable leaderboard for machine learning competitions.
\newblock In \emph{Proceedings of the 32nd International Conference on Machine
  Learning (ICML 2015)}, volume~37 of \emph{Proceedings of Machine Learning
  Research}, pages 1006--1014, 2015.

\bibitem[Burnat and Davidson(2026)]{burnat2026quotient}
Florian A.~D. Burnat and Brittany~I. Davidson.
\newblock Quotient semivalues for false-name-resistant data attribution.
\newblock arXiv preprint arXiv:2605.07663, 2026.

\bibitem[Cheng and Friedman(2005)]{cheng2005sybilproof}
Alice Cheng and Eric Friedman.
\newblock Sybilproof reputation mechanisms.
\newblock In \emph{Proceedings of the 2005 ACM SIGCOMM Workshop on Economics of
  Peer-to-Peer Systems (P2PECON '05)}, 2005.
\newblock \doi{10.1145/1080192.1080202}.

\bibitem[Cox(1975)]{cox1975datasplitting}
D.~R. Cox.
\newblock A note on data-splitting for the evaluation of significance levels.
\newblock \emph{Biometrika}, 62\penalty0 (2):\penalty0 441--444, 1975.
\newblock \doi{10.1093/biomet/62.2.441}.

\bibitem[Crapis et~al.(2026)Crapis, Lim, Tay, and Chooi]{crapis2026erc8183}
Davide Crapis, Bryan Lim, Weixiong Tay, and Zuhwa Chooi.
\newblock {ERC-8183}: Agentic commerce.
\newblock Ethereum Improvement Proposals, no. 8183, draft, created 2026-02-25,
  2026.
\newblock URL \url{https://eips.ethereum.org/EIPS/eip-8183}.
\newblock Accessed 2026-10-23.

\bibitem[Daian et~al.(2020)Daian, Goldfeder, Kell, Li, Zhao, Bentov,
  Breidenbach, and Juels]{daian2020flash}
Philip Daian, Steven Goldfeder, Tyler Kell, Yunqi Li, Xueyuan Zhao, Iddo
  Bentov, Lorenz Breidenbach, and Ari Juels.
\newblock Flash boys 2.0: Frontrunning in decentralized exchanges, miner
  extractable value, and consensus instability.
\newblock In \emph{2020 IEEE Symposium on Security and Privacy (SP)}, pages
  910--927, 2020.
\newblock \doi{10.1109/SP40000.2020.00040}.

\bibitem[Douceur and Moscibroda(2007)]{douceur2007lottery}
John~R. Douceur and Thomas Moscibroda.
\newblock Lottery trees: Motivational deployment of networked systems.
\newblock In \emph{Proceedings of the 2007 Conference on Applications,
  Technologies, Architectures, and Protocols for Computer Communications
  (SIGCOMM '07)}, pages 121--132, 2007.
\newblock \doi{10.1145/1282380.1282395}.

\bibitem[{Drastic Watermelon}(2025)]{yaudit2026sybil}
{Drastic Watermelon}.
\newblock The sybilooor's guide to audit competition math.
\newblock yAudit blog, September 2025.
\newblock URL \url{https://blog.yaudit.dev/contest_sybil}.
\newblock Accessed 2026-11-10.

\bibitem[Dwork and Naor(1993)]{dwork1992pricing}
Cynthia Dwork and Moni Naor.
\newblock Pricing via processing or combatting junk mail.
\newblock In \emph{Advances in Cryptology -- CRYPTO '92}, volume 740 of
  \emph{Lecture Notes in Computer Science}, pages 139--147. Springer, 1993.
\newblock \doi{10.1007/3-540-48071-4_10}.

\bibitem[{Eigen Labs}(2026)]{layr2026qsbharness}
{Eigen Labs}.
\newblock {QSB} grinding benchmark: the {Quantum Safe Bitcoin} challenge
  harness.
\newblock GitHub repository Layr-Labs/quantum-safe-bitcoin-challenge, README
  and spec/SCORING.md at commit 86f6d52 (2026-09-14), 2026.
\newblock URL
  \url{https://github.com/Layr-Labs/quantum-safe-bitcoin-challenge}.
\newblock Accessed 2026-11-10.

\bibitem[Emek et~al.(2011)Emek, Karidi, Tennenholtz, and
  Zohar]{emek2011multilevel}
Yuval Emek, Ron Karidi, Moshe Tennenholtz, and Aviv Zohar.
\newblock Mechanisms for multi-level marketing.
\newblock In \emph{Proceedings of the 12th ACM Conference on Electronic
  Commerce (EC '11)}, pages 209--218, 2011.
\newblock \doi{10.1145/1993574.1993606}.

\bibitem[{Ethereum Foundation and Eigen
  Labs}(2026)]{bettercodes2026leaderboard}
{Ethereum Foundation and Eigen Labs}.
\newblock better.codes: the {Proximity Prize} soundness leaderboard.
\newblock Web page, 2026.
\newblock URL \url{https://better.codes/}.
\newblock Accessed 2026-10-23.

\bibitem[{Ethereum Foundation Formal Verification
  team}(2026)]{ef2026bettercodes}
{Ethereum Foundation Formal Verification team}.
\newblock Raising machine-checked security benchmarks to advance hash-based
  {SNARKs} through agentic collaboration.
\newblock Ethereum Foundation Blog, 20 August 2026, 2026.
\newblock URL
  \url{https://blog.ethereum.org/2026/08/20/better-codes-challenge}.
\newblock Accessed 2026-10-23.

\bibitem[Eyal and Sirer(2018)]{eyal2018majority}
Ittay Eyal and Emin~G{\"u}n Sirer.
\newblock Majority is not enough: Bitcoin mining is vulnerable.
\newblock \emph{Communications of the ACM}, 61\penalty0 (7):\penalty0 95--102,
  2018.
\newblock \doi{10.1145/3212998}.

\bibitem[Faigle and Kern(1992)]{faigle1992precedence}
Ulrich Faigle and Walter Kern.
\newblock The {Shapley} value for cooperative games under precedence
  constraints.
\newblock \emph{International Journal of Game Theory}, 21\penalty0
  (3):\penalty0 249--266, 1992.
\newblock \doi{10.1007/BF01258278}.

\bibitem[{Folding@home}(2026)]{foldingathome2026about}
{Folding@home}.
\newblock About {Folding@home}.
\newblock Web page, 2026.
\newblock URL \url{https://foldingathome.org/about/}.
\newblock Accessed 2026-11-06.

\bibitem[Frongillo and Waggoner(2018{\natexlab{a}})]{frongillo2018axiomatic}
Rafael Frongillo and Bo~Waggoner.
\newblock An axiomatic study of scoring rule markets.
\newblock In \emph{9th Innovations in Theoretical Computer Science Conference
  (ITCS 2018)}, volume~94 of \emph{LIPIcs}, pages 15:1--15:20. Schloss Dagstuhl
  -- Leibniz-Zentrum f{\"u}r Informatik, 2018{\natexlab{a}}.
\newblock \doi{10.4230/LIPIcs.ITCS.2018.15}.

\bibitem[Frongillo and Waggoner(2018{\natexlab{b}})]{frongillo2018bounded}
Rafael Frongillo and Bo~Waggoner.
\newblock Bounded-loss private prediction markets.
\newblock In \emph{Advances in Neural Information Processing Systems 31
  (NeurIPS 2018)}. Curran Associates, 2018{\natexlab{b}}.
\newblock URL
  \url{https://proceedings.neurips.cc/paper_files/paper/2018/hash/b2dc449578a4744a1684d3b0ea933702-Abstract.html}.

\bibitem[Gilles et~al.(1992)Gilles, Owen, and van~den
  Brink]{gilles1992permission}
Robert~P. Gilles, Guillermo Owen, and Ren{\'e} van~den Brink.
\newblock Games with permission structures: The conjunctive approach.
\newblock \emph{International Journal of Game Theory}, 20\penalty0
  (3):\penalty0 277--293, 1992.
\newblock \doi{10.1007/BF01253782}.

\bibitem[Gowers and Nielsen(2009)]{gowers2009polymath}
Timothy Gowers and Michael Nielsen.
\newblock Massively collaborative mathematics.
\newblock \emph{Nature}, 461\penalty0 (7266):\penalty0 879--881, 2009.
\newblock \doi{10.1038/461879a}.

\bibitem[Grabisch(1997)]{grabisch1997kadditive}
Michel Grabisch.
\newblock $k$-order additive discrete fuzzy measures and their representation.
\newblock \emph{Fuzzy Sets and Systems}, 92\penalty0 (2):\penalty0 167--189,
  1997.
\newblock \doi{10.1016/S0165-0114(97)00168-1}.
\newblock Definition of $k$-additivity as restated in Tehrani et al., Mach.\
  Learn.\ 2012.

\bibitem[Grabisch and Roubens(1999)]{grabisch1999interaction}
Michel Grabisch and Marc Roubens.
\newblock An axiomatic approach to the concept of interaction among players in
  cooperative games.
\newblock \emph{International Journal of Game Theory}, 28\penalty0
  (4):\penalty0 547--565, 1999.
\newblock \doi{10.1007/s001820050125}.
\newblock Abstract consulted.

\bibitem[Green and Scotchmer(1995)]{green1995division}
Jerry~R. Green and Suzanne Scotchmer.
\newblock On the division of profit in sequential innovation.
\newblock \emph{The RAND Journal of Economics}, 26\penalty0 (1), 1995.
\newblock \doi{10.2307/2556033}.

\bibitem[Hanson(2003)]{hanson2003combinatorial}
Robin Hanson.
\newblock Combinatorial information market design.
\newblock \emph{Information Systems Frontiers}, 5\penalty0 (1):\penalty0
  107--119, 2003.
\newblock \doi{10.1023/A:1022058209073}.

\bibitem[Hanson(2007)]{hanson2007logarithmic}
Robin Hanson.
\newblock Logarithmic market scoring rules for modular combinatorial
  information aggregation.
\newblock \emph{Journal of Prediction Markets}, 1\penalty0 (1):\penalty0 3--15,
  2007.

\bibitem[Harris and Waggoner(2019)]{harris2019decentralized}
Justin~D. Harris and Bo~Waggoner.
\newblock Decentralized and collaborative {AI} on blockchain.
\newblock In \emph{2019 IEEE International Conference on Blockchain
  (Blockchain)}, pages 368--375. IEEE, 2019.
\newblock \doi{10.1109/Blockchain.2019.00057}.

\bibitem[Hessel et~al.(2018)Hessel, Modayil, van Hasselt, Schaul, Ostrovski,
  Dabney, Horgan, Piot, Azar, and Silver]{hessel2018rainbow}
Matteo Hessel, Joseph Modayil, Hado van Hasselt, Tom Schaul, Georg Ostrovski,
  Will Dabney, Dan Horgan, Bilal Piot, Mohammad Azar, and David Silver.
\newblock Rainbow: Combining improvements in deep reinforcement learning.
\newblock In \emph{Proceedings of the AAAI Conference on Artificial
  Intelligence}, volume~32, 2018.
\newblock \doi{10.1609/aaai.v32i1.11796}.

\bibitem[Hu and Tang(2026)]{hu2026winners}
Litao Hu and Yutong Tang.
\newblock The winner's curse in {LLM} self-improvement loops: Selection noise,
  lock-in, and acceptance rules.
\newblock arXiv preprint 2610.09239, 2026.
\newblock URL \url{https://arxiv.org/abs/2610.09239}.

\bibitem[Jordan and contributors(2026)]{jordan2026speedrun}
Keller Jordan and contributors.
\newblock modded-nanogpt: the {NanoGPT} speedrun.
\newblock GitHub repository, README ``World record history'' and ``Rules'',
  master at commit 4ea6b93 (2026-09-28), 2026.
\newblock URL \url{https://github.com/KellerJordan/modded-nanogpt}.
\newblock Accessed 2026-10-23.

\bibitem[Ju et~al.(2007)Ju, Miyagawa, and Sakai]{ju2007nonmanipulable}
Biung-Ghi Ju, Eiichi Miyagawa, and Toyotaka Sakai.
\newblock Non-manipulable division rules in claim problems and generalizations.
\newblock \emph{Journal of Economic Theory}, 132\penalty0 (1):\penalty0 1--26,
  2007.
\newblock \doi{10.1016/j.jet.2005.08.003}.

\bibitem[Karpathy(2026)]{karpathy2026autoresearch}
Andrej Karpathy.
\newblock autoresearch: {AI} agents running research on single-{GPU} nanochat
  training automatically.
\newblock GitHub repository, master at commit 228791f (2026-03-26), 2026.
\newblock URL \url{https://github.com/karpathy/autoresearch}.
\newblock Accessed 2026-10-23.

\bibitem[Kleinberg and Oren(2022)]{kleinberg2022credit}
Jon Kleinberg and Sigal Oren.
\newblock Mechanisms for (mis)allocating scientific credit.
\newblock \emph{Algorithmica}, 84\penalty0 (2):\penalty0 344--378, 2022.
\newblock \doi{10.1007/s00453-021-00902-y}.

\bibitem[Knoop(2024)]{knoop2024arcwinners}
Mike Knoop.
\newblock {ARC} prize 2024 winners \& technical report.
\newblock ARC Prize blog, 6 December 2024, 2024.
\newblock URL
  \url{https://arcprize.org/blog/arc-prize-2024-winners-technical-report}.
\newblock Accessed 2026-10-23.

\bibitem[Knudsen and {\O}sterdal(2012)]{knudsen2012merging}
Peter~H. Knudsen and Lars~Peter {\O}sterdal.
\newblock Merging and splitting in cooperative games: Some (im)possibility
  results.
\newblock \emph{International Journal of Game Theory}, 41\penalty0
  (4):\penalty0 763--774, 2012.
\newblock \doi{10.1007/s00182-012-0337-7}.

\bibitem[Lazear and Rosen(1981)]{lazear1981rank}
Edward~P. Lazear and Sherwin Rosen.
\newblock Rank-order tournaments as optimum labor contracts.
\newblock \emph{Journal of Political Economy}, 89\penalty0 (5):\penalty0
  841--864, 1981.
\newblock \doi{10.1086/261010}.

\bibitem[Lee and Shen(2018)]{lee2018winners}
Minyong~R. Lee and Milan Shen.
\newblock Winner's curse: Bias estimation for total effects of features in
  online controlled experiments.
\newblock In \emph{Proceedings of the 24th ACM SIGKDD International Conference
  on Knowledge Discovery \& Data Mining (KDD '18)}, pages 491--499. ACM, 2018.
\newblock \doi{10.1145/3219819.3219905}.

\bibitem[Lehrer(1988)]{lehrer1988banzhaf}
Ehud Lehrer.
\newblock An axiomatization of the {Banzhaf} value.
\newblock \emph{International Journal of Game Theory}, 17\penalty0
  (2):\penalty0 89--99, 1988.
\newblock \doi{10.1007/BF01254541}.

\bibitem[Lemus and Marshall(2021)]{lemus2021dynamic}
Jorge Lemus and Guillermo Marshall.
\newblock Dynamic tournament design: Evidence from prediction contests.
\newblock \emph{Journal of Political Economy}, 129\penalty0 (2):\penalty0
  383--420, 2021.
\newblock \doi{10.1086/711762}.

\bibitem[Lemus and Marshall(2025)]{lemus2025contingent}
Jorge Lemus and Guillermo Marshall.
\newblock Contingent prizes in dynamic contests.
\newblock \emph{The RAND Journal of Economics}, 56\penalty0 (4):\penalty0
  476--493, 2025.
\newblock \doi{10.1111/1756-2171.70019}.
\newblock Content checked in the working-paper version of 2024-03-01.

\bibitem[Lidin et~al.(2025)Lidin, Sarfi, Pappas, Dare, Belilovsky, and
  Steeves]{lidin2025gauntlet}
Joel Lidin, Amir Sarfi, Evangelos Pappas, Samuel Dare, Eugene Belilovsky, and
  Jacob Steeves.
\newblock Incentivizing permissionless distributed learning of {LLMs}.
\newblock arXiv preprint arXiv:2505.21684; a shorter version appeared at DAI
  2025, 2025.

\bibitem[Long et~al.(2026)Long, Pender, Huang, et~al.]{long2026ecdsafail}
Jieyi Long, Theodore Pender, Zhao Huang, et~al.
\newblock {ECDSA.Fail}: Open autoresearch for optimizing elliptic-curve point
  addition in {Shor's} algorithm.
\newblock arXiv preprint arXiv:2609.09582, 2026.

\bibitem[Lv and Moscibroda(2016)]{lv2016incentivetree}
Yuezhou Lv and Thomas Moscibroda.
\newblock Fair and resilient incentive tree mechanisms.
\newblock \emph{Distributed Computing}, 29\penalty0 (1):\penalty0 1--16, 2016.
\newblock \doi{10.1007/s00446-015-0250-y}.

\bibitem[Miller et~al.(2015)Miller, Kosba, Katz, and
  Shi]{miller2015nonoutsourceable}
Andrew Miller, Ahmed Kosba, Jonathan Katz, and Elaine Shi.
\newblock Nonoutsourceable scratch-off puzzles to discourage {Bitcoin} mining
  coalitions.
\newblock In \emph{Proceedings of the 22nd {ACM} {SIGSAC} Conference on
  Computer and Communications Security ({CCS} '15)}, 2015.
\newblock \doi{10.1145/2810103.2813621}.

\bibitem[Moldovanu and Sela(2001)]{moldovanu2001optimal}
Benny Moldovanu and Aner Sela.
\newblock The optimal allocation of prizes in contests.
\newblock \emph{American Economic Review}, 91\penalty0 (3):\penalty0 542--558,
  2001.
\newblock \doi{10.1257/aer.91.3.542}.

\bibitem[{mutable-state-inc}(2026)]{mutablestate2026athome}
{mutable-state-inc}.
\newblock autoresearch-at-home: a collaborative, {SETI@home}-style fork of
  autoresearch.
\newblock GitHub repository, 2026.
\newblock URL \url{https://github.com/mutable-state-inc/autoresearch-at-home}.
\newblock Accessed 2026-10-23.

\bibitem[{Numerai}(2026)]{numerai2026staking}
{Numerai}.
\newblock Numerai tournament documentation: staking.
\newblock Web page, 2026.
\newblock URL \url{https://docs.numer.ai/numerai-tournament/staking}.
\newblock Accessed 2026-10-29.

\bibitem[Ohrimenko et~al.(2019)Ohrimenko, Tople, and
  Tschiatschek]{ohrimenko2019collaborative}
Olga Ohrimenko, Shruti Tople, and Sebastian Tschiatschek.
\newblock Collaborative machine learning markets with data-replication-robust
  payments.
\newblock arXiv preprint 1911.09052, 2019.
\newblock URL \url{https://arxiv.org/abs/1911.09052}.

\bibitem[Ohta et~al.(2008)Ohta, Conitzer, Satoh, Iwasaki, and
  Yokoo]{ohta2008anonymity}
Naoki Ohta, Vincent Conitzer, Yasufumi Satoh, Atsushi Iwasaki, and Makoto
  Yokoo.
\newblock Anonymity-proof {Shapley} value: Extending {Shapley} value for
  coalitional games in open environments.
\newblock In \emph{Proceedings of the 7th International Conference on
  Autonomous Agents and Multiagent Systems (AAMAS 2008)}, pages 927--934, 2008.
\newblock URL
  \url{https://www.ifaamas.org/Proceedings/aamas08/proceedings/pdf/paper/AAMAS08_0775.pdf}.

\bibitem[{Paradigm Operations LP}(2026)]{paradigm2026okc}
{Paradigm Operations LP}.
\newblock Muse spark open kernel challenge: rules and official contest terms.
\newblock Web page and terms PDF, 2026.
\newblock URL \url{https://openkernelchallenge.dev/rules}.
\newblock Accessed 2026-10-23.

\bibitem[{Proximity Prize maintainers}(2026)]{proximityprize2026repo}
{Proximity Prize maintainers}.
\newblock proximity-prize/proximity-prize: Lean repository of the {Proximity
  Prize} challenge.
\newblock GitHub repository, Apache License 2.0, 2026.
\newblock URL \url{https://github.com/proximity-prize/proximity-prize}.
\newblock Accessed 2026-10-26.

\bibitem[Ren(2026)]{ren2026outerloop}
Mengye Ren.
\newblock Outerloop: Open and trustworthy multi-agent autoresearch.
\newblock Paper on the project web page, dated 1 November 2026, 2026.
\newblock URL \url{https://outerloop.science/papers/outerloop-2026.pdf}.
\newblock Accessed 2026-11-03.

\bibitem[Schrijvers et~al.(2017)Schrijvers, Bonneau, Boneh, and
  Roughgarden]{schrijvers2017pool}
Okke Schrijvers, Joseph Bonneau, Dan Boneh, and Tim Roughgarden.
\newblock Incentive compatibility of bitcoin mining pool reward functions.
\newblock In \emph{Financial Cryptography and Data Security (FC 2016)}, Lecture
  Notes in Computer Science, pages 477--498, 2017.
\newblock \doi{10.1007/978-3-662-54970-4_28}.

\bibitem[Seuken and Parkes(2014)]{seuken2014sybilproof}
Sven Seuken and David~C. Parkes.
\newblock Sybil-proof accounting mechanisms with transitive trust.
\newblock In \emph{Proceedings of the 13th International Conference on
  Autonomous Agents and Multiagent Systems (AAMAS 2014)}, pages 205--212, 2014.

\bibitem[Shapley(1953)]{shapley1953value}
Lloyd~S. Shapley.
\newblock A value for $n$-person games.
\newblock In \emph{Contributions to the Theory of Games, Volume II}, volume~28
  of \emph{Annals of Mathematics Studies}, pages 307--317. Princeton University
  Press, 1953.
\newblock \doi{10.1515/9781400881970-018}.

\bibitem[Smith and Winkler(2006)]{smith2006optimizer}
James~E. Smith and Robert~L. Winkler.
\newblock The optimizer's curse: Skepticism and postdecision surprise in
  decision analysis.
\newblock \emph{Management Science}, 52\penalty0 (3):\penalty0 311--322, 2006.
\newblock \doi{10.1287/mnsc.1050.0451}.

\bibitem[Somekh-Baruch et~al.(2018)Somekh-Baruch, Leshem, and
  Saligrama]{somekhbaruch2018nonexistence}
Anelia Somekh-Baruch, Amir Leshem, and Venkatesh Saligrama.
\newblock On the non-existence of unbiased estimators in constrained estimation
  problems.
\newblock \emph{IEEE Transactions on Information Theory}, 64\penalty0
  (8):\penalty0 5549--5554, 2018.
\newblock \doi{10.1109/TIT.2017.2731845}.

\bibitem[Stannat et~al.(2021)Stannat, Ileri, Gijswijt, and
  Pouwelse]{stannat2021sybilproofness}
Alexander Stannat, Can~Umut Ileri, Dion Gijswijt, and Johan Pouwelse.
\newblock Achieving sybil-proofness in distributed work systems.
\newblock In \emph{Proceedings of the 20th International Conference on
  Autonomous Agents and Multiagent Systems (AAMAS 2021)}, 2021.
\newblock URL \url{http://ifaamas.org/Proceedings/aamas2021/pdfs/p1263.pdf}.

\bibitem[{StarkWare and Yukon
  Research}(2026{\natexlab{a}})]{starkware2026qsbleaderboard}
{StarkWare and Yukon Research}.
\newblock Quantum-safe bitcoin optimization challenge: final results and
  leaderboard data.
\newblock Web page and its leaderboard data,
  \url{https://quantum.starkware.co/claim/api/qsb/leaderboard},
  2026{\natexlab{a}}.
\newblock URL \url{https://quantum.starkware.co/claim/winners}.
\newblock Accessed 2026-11-08.

\bibitem[{StarkWare and Yukon
  Research}(2026{\natexlab{b}})]{starkware2026qsbrewards}
{StarkWare and Yukon Research}.
\newblock Quantum-safe bitcoin optimization challenge: solver rewards.
\newblock Web page, challenge closed 2026-10-07, 2026{\natexlab{b}}.
\newblock URL \url{https://quantum.starkware.co/claim}.
\newblock Accessed 2026-10-23.

\bibitem[{Stiftung Ethereum}(2026)]{ef2026programterms}
{Stiftung Ethereum}.
\newblock The {Ethereum Foundation} program terms \& conditions (better codes
  research rewards program).
\newblock Web page, 2026.
\newblock URL \url{https://better.codes/program-terms}.
\newblock Accessed 2026-10-23.

\bibitem[{trevin-creator}(2026)]{trevin2026autoresearchmlx}
{trevin-creator}.
\newblock autoresearch-mlx: Apple silicon ({MLX}) port of karpathy's
  autoresearch.
\newblock GitHub repository, commit 766a25f (2026-07-02), 2026.
\newblock URL \url{https://github.com/trevin-creator/autoresearch-mlx}.
\newblock Accessed 2026-10-23.

\bibitem[Waggoner et~al.(2015)Waggoner, Frongillo, and
  Abernethy]{waggoner2015market}
Bo~Waggoner, Rafael Frongillo, and Jacob Abernethy.
\newblock A market framework for eliciting private data.
\newblock In \emph{Advances in Neural Information Processing Systems 28
  (NeurIPS 2015)}. Curran Associates, 2015.
\newblock URL
  \url{https://papers.nips.cc/paper/2015/hash/7af6266cc52234b5aa339b16695f7fc4-Abstract.html}.

\bibitem[Whitmeyer(2023)]{whitmeyer2023submission}
Mark Whitmeyer.
\newblock Submission costs in risk-taking contests.
\newblock \emph{Games and Economic Behavior}, 142:\penalty0 101--112, 2023.
\newblock \doi{10.1016/j.geb.2023.07.014}.

\bibitem[Yokoo et~al.(2004)Yokoo, Sakurai, and Matsubara]{yokoo2004falsename}
Makoto Yokoo, Yuko Sakurai, and Shigeo Matsubara.
\newblock The effect of false-name bids in combinatorial auctions: New fraud in
  internet auctions.
\newblock \emph{Games and Economic Behavior}, 46\penalty0 (1):\penalty0
  174--188, 2004.
\newblock \doi{10.1016/S0899-8256(03)00045-9}.

\bibitem[Yokoo et~al.(2005)Yokoo, Conitzer, Sandholm, Ohta, and
  Iwasaki]{yokoo2005coalitional}
Makoto Yokoo, Vincent Conitzer, Tuomas Sandholm, Naoki Ohta, and Atsushi
  Iwasaki.
\newblock Coalitional games in open anonymous environments.
\newblock In \emph{Proceedings of the Twentieth National Conference on
  Artificial Intelligence (AAAI-05)}, pages 509--514. AAAI Press, 2005.
\newblock URL
  \url{https://aaai.org/papers/00509-aaai05-080-coalitional-games-in-open-anonymous-environments/}.

\bibitem[{Yukon Research and StarkWare}(2026)]{yukon2026qsbleaderboard}
{Yukon Research and StarkWare}.
\newblock Quantum safe bitcoin challenge: pinning leaderboard.
\newblock Web page (challenge closed 2026-10-07), 2026.
\newblock URL \url{https://qsb.fast/}.
\newblock Accessed 2026-10-23.

\end{thebibliography}
 \end{document}